\documentclass[12pt]{article}
\usepackage{amsmath,amsfonts}
\usepackage{multirow}
\usepackage{amssymb}
\usepackage[colorlinks=true,urlcolor=blue]{hyperref}
\usepackage{graphicx}
\usepackage{ulem}
\usepackage{cancel}
\usepackage{enumitem}
\usepackage{fancyvrb}
\hypersetup{            
backref = true,          
pagebackref = true,              
hyperindex = true,     
colorlinks = true,          
breaklinks = true,          
urlcolor = blue,            
linkcolor = blue,          
bookmarks = true,         
bookmarksopen = true,      
citecolor=red,
}

\usepackage{hyperref}
\usepackage{breakurl}
\hypersetup{breaklinks=true}
\def\bepro{\begin{proposition}}
\def\enpro{\end{proposition}}
\newtheorem{theorem}{Theorem}[section]

\newtheorem{proposition}[theorem]{Proposition}

\newcommand{\qeed}{\hfill\textrm{QED}\break\null}

\advance \textheight by \topskip
\makeatletter

\def\cc#1{\kern .7em\hfill #1 \hfill\kern .7em}

\usepackage{mathtools}

\newcommand{\beqa}{\begin{eqnarray}}
\newcommand{\eeqa}{\end{eqnarray}}
\newcommand{\bpm}{\begin{pmatrix}}
\newcommand{\epm}{\end{pmatrix}}

\newcommand{\nn}{\nonumber}

\newcommand{\LL}{L^2(\mathrm{SL}(2,\mathbb R))}
\newcommand{\SL}{\mathrm{SL}(2,\mathbb R)}
\newcommand{\g}{\mathfrak{g}}

\newcommand{\II}{\sum_{\Lambda,m,n} \hskip -.6 truecm \int{}\;}
\newcommand{\Kg}{\widehat{\mathfrak g}(\mathrm{SL}(2,\mathbb R))}
\newcommand{\s}{\mathfrak{sl}(2,\mathbb R)}

\def\d{\mathrm{d}}

\numberwithin{equation}{section}
\title{An infinite--rank Lie algebra associated to \\
SL$(2,\mathbb R)$ and
SL$(2,\mathbb R)/U(1)$}
\begin{document}
\maketitle
\begin{center}

{\large Rutwig Campoamor-Stursberg$^{1\ast}$, Alessio Marrani $^{2\dagger}$,} \\and {\large Michel Rausch de Traubenberg$^{3\ddagger}$,}\\
\bigskip
\bigskip

$^1$  Instituto de Matem\'atica Interdisciplinar and Dp.to de Geometr\'\i a y Topolog\'\i a, UCM, E-28040 Madrid, Spain\\

$^2$Instituto de F\'isica Te\'orica, Dep.to de F\'isica, Universidad de Murcia, Campus de Espinardo,
E-30100, Spain \\

$^3$ Universit\'e de Strasbourg, CNRS, IPHC UMR7178, F-67037 Strasbourg Cedex, France\\

\end{center}

\noindent $^\ast$ rutwig@ucm.es

\noindent $^\dagger$ alessio.marrani@um.es

\noindent $^\ddagger$ Michel.Rausch@iphc.cnrs.fr

\bigskip
\bigskip

\begin{abstract}
We construct a generalised notion of Kac-Moody algebras using smooth maps from the non-compact manifolds
${\cal M}=$SL$(2,\mathbb R)$ and ${\cal M}=$ SL$(2,\mathbb R)/U(1)$ to a finite-dimensional simple Lie group $G$.
This construction is achieved through two equivalent ways: by means of the Plancherel Theorem and by  identifying a Hilbert basis within $L^2(\mathcal{M})$.
We analyse the  existence of central extensions   and identify those in duality with Hermitean operators on $\cal M$.
By inspecting the Clebsch-Gordan coefficients of $\mathfrak{sl}(2,\mathbb{R})$, we derive the Lie brackets characterising the corresponding generalised Kac-Moody algebras. The root structure of these algebras is identified, and it is shown that an infinite
number of simultaneously commuting operators can be defined.
Furthermore, we briefly touch upon applications of these algebras within the realm of supergravity, particularly in scenarios where the scalar fields  coordinatize the non-compact manifold $\text{SL}(2,\mathbb{R})/U(1)$.

\end{abstract}
\newpage
\tableofcontents
\newpage

\section{Introduction}

As an algebraic object, the theory of Kac--Moody algebras emerges naturally from the Chevalley--Serre relations for complex semisimple Lie algebras \cite{CH}, by dropping the condition on the non-degeneracy of the Cartan matrix. This leads to a new ample class of (generally infinite-dimensional) algebras that extends semisimple algebras, by simultaneously preserving some of their salient properties \cite{Kac, Kac2, Moo}. Kac--Moody algebras can further be extended, either using an axiomatic approach,
leading to generalized intersection matrix Lie algebras, quasi-simple Lie algebras or generalised Kac--Moody algebras, among others (see \cite{SLO,KT,Frap} and references therein), or using analytical tools, by considering affine extensions of the loop algebra of smooth maps from the one-dimensional sphere $\mathbb{S}^{1}$ into a simple Lie group, leading to  the notion of affine Lie algebras. Actually this construction motivated the use of Kac--Moody  algebras in physical applications, such as Kaluza-Klein or string theories \cite{go, dd},  in which a deep relation with other important objects as the Virasoro algebra was soon observed. Motivated by this fact, and having in mind applications to Kaluza-Klein compactifications of higher dimensional spaces, in \cite{rmm,rmm2} a notion of  generalised Kac--Moody algebras based on spaces of differentiable maps from compact manifolds to compact Lie groups was proposed, providing also a detailed construction of the  corresponding generalized Kac--Moody algebras. These algebras have indeed also been
considered by various authors in the case of specific manifolds, such as the two-sphere \cite{bars} --to our knowledge this is the first reference where such algebras were studied-- or the $n-$tori \cite{bars, KT, MRT, jap}.

The main objective of this work is to extend the results of \cite{rmm} to the case of the non-compact Lie group $SL(2,\mathbb{R})$ and  of the  related symmetric coset space $SL(2,\mathbb{R})/U(1)$. Being a non-compact group, the harmonic analysis turns out to be much more complicated, and the approach must be modified accordingly to extract suitable bases of the corresponding Hilbert spaces.

This paper is structured as follows: In Section 2 we review the square-integrable functions on the non-compact Lie group $SL(2,\mathbb{R})$, with special emphasis on the matrix elements of the discrete and principal continuous series, in order to expand square-integrable functions on $SL(2,\mathbb{R})$ using the Plancherel formula. Section 3 is devoted to the construction of a Hilbert basis of $L^2\big(SL(2,\mathbb{R})\big)$. This is done identifying all normalisable eigenfunctions, and considering a specific decomposition of $L^2\big(SL(2,\mathbb{R})\big)$ into complementary orthogonal spaces. While a basis for the set of square-integrable functions expanded in the discrete series follows easily from the analysis of matrix elements, for the part corresponding to the continuous principal series the procedure does not work, due to non-normalisability. This technical difficulty is surmounted using an ansatz proposed to the authors by
Viktor Losert, that allows us to extract a basis (called Losert basis) which enables us to express the matrix elements of the principal continuous series in terms  of these elements, as well as the converse. In Section 4, the Clebsch-Gordan coefficients are studied,  and it is shown that the use of the Losert basis is critical for the analysis of the product of matrix elements corresponding to the continuous principal series. In Section 5, an infinite-dimensional algebra ``\`a la Kac--Moody" (denoted by $\Kg$) associated to $SL(2,\mathbb{R})$ as manifold is defined,  thus extending the method proposed in \cite{rmm} to the non-compact case. It is shown that two  (equivalent) strategies  can be followed:  namely, one can either apply the Plancherel theorem, or use the aforementioned Losert basis. In analogy with the compact case, it is shown that this algebra can also be written in terms of the usual current algebra with a Schwinger term.  The  algebra $\Kg$ also admits a root system, and it turns out that there exists an infinite
number of simultaneously commuting operators, implying that the
algebra $\Kg$ is of infinite rank. In Section 6, motivated by applications to supergravity, the analysis is extended to the  non-compact, Riemannian,  locally symmetric coset space $SL(2,\mathbb{R})/U(1)$,
and the corresponding infinite dimensional Lie algebra $\widehat{\g}($SL$(2,\mathbb R)/U(1))$ is defined. Finally, in Section 7 some conclusions and future perspectives are drawn.

\section{Square-integrable functions on SL$(2,\mathbb R)$}

The purpose of this section is to study square-integrable functions on $\SL$ as a manifold. As is well known, the Lie group $\SL$ is non-compact with maximal compact subgroup SO$(2)\subset \SL$. Due to the non-compactness of $\SL$, harmonic analysis on $\LL$ is much more delicate than on compact Lie groups $G_c$, for which the matrix elements of all unitary representations, once correctly normalised, constitute an orthonormal Hilbert basis of the Hilbert space $L^2(G_c)$, according to the  Peter-Weyl theorem \cite{PW}. For the non-compact case $G_{nc}$, an analogous result is provided by the Plancherel theorem (see {\it e. g.} \cite{sch}). In contrast to the Peter-Weyl theorem, when identifying unitary representations of $G_{nc}$, we have to take into account substantial differences, beginning with the infinite dimension of unitary representations of non-compact groups. Further, for non-compact Lie groups two types of representations are conceivable -- the continuous series and the discrete series. While the former series is not normalisable, the latter is. Non-compact Lie groups always have a continuous series, whereas a discrete series exists {\it iff} the rank of $G_{nc}$ equals the rank of its maximal compact subgroup.
For the case under consideration, namely for $\SL$,  we have
\beqa
\text{rk}(\SL) =\text{rk}(SO(2))=1 \ , \nn
\eeqa
thus the group $\SL$ admits both a discrete and continuous series.

\subsection{Unitary representations of $\s$}
 The unitary representations of $\s$ (the Lie algebra of $\SL$) were classified by
Bargmann in \cite{bar} (see also \cite{vk,ggv} and references therein). We briefly recall the main results of Bargmann. Let
$K_0,K_\pm$ be the generators of $\s$ with Lie brackets
\beqa
\label{eq:LB}
\big[K_0,K_\pm\big]=\pm K_\pm\ , \ \
\big[K_+,K_-\big]=-2 K_0\
\eeqa
and Casimir operator given by
\beqa
Q=K_0^2 -\frac 12\big(K_+ K_-+ K_- K_+\big) \ . \nn
\eeqa

 Unitary representations of  $\s$ (beyond the trivial representation) are subdivided into four classes. The discrete series consists of   two types, one corresponding to representations that are bounded from below and one to representations that are bounded from above. The continuous series also splits into two classes: a principal and a supplementary series. The discrete series are semi-infinite, while the continuous representations are unbounded and hence infinite in both directions.

 The discrete series are characterised by the discrete number  $\lambda \in \mathbb N$ or $\lambda \in \mathbb N + 1/2$, where unitarity of the representation implies that $\lambda>0$. Whenever $\lambda$ is an integer (resp. half-integer) number, the representation is called bosonic (resp. fermionic). For representations  bounded from below, we have ${\cal D}^+_\lambda=\{\big|\; \lambda,+,n\big>, n\ge \lambda>0\}$
  and the action by the $\s$ generators
  \beqa
  \label{eq:D+}
K_{0} \big|\lambda,+,n\big>&=&n\big|\lambda,+,n\big>\ ,  \notag   \\
K_{+}\big|\lambda,+,n\big> &=&\sqrt{(n+\lambda)(n+1-\lambda)}\big|\lambda,+,n+1\big>\ ,  \\
K_{-}\big|\lambda,+,n\big> &=&\sqrt{(n-1+\lambda)(n-\lambda)}\big|\lambda,+,n-1\big>\ , \nn\\
Q\big|\lambda,+,n\big> &=&\lambda \left( \lambda -1\right) \big|\lambda,+,n\big>\ .\notag
\eeqa
 For representations  bounded from  above, we have ${\cal D}^-_\lambda=\{\big|\lambda,-,n\big>, n\le- \lambda\le 0\} $ and the action
 \beqa
   \label{eq:D-}
K_{0} \big|\lambda,-,n\big>&=&n\big|\lambda,-,n\big>;  \notag   \\
K_{+}\big|\lambda,-,n\big> &=&-\sqrt{(-n-\lambda)(-n-1+\lambda)} \big|\lambda,-,n+1\big> \\
  K_{-}\big|\lambda,-,n\big> &=& -\sqrt{(-n+1-\lambda)(-n+\lambda)}\big|\lambda,-,n-1\big>; \nn\\
Q\big|\lambda,-,n\big> &=&\lambda \left( \lambda -1\right) \big|\lambda,-,n\big>.\notag
\eeqa
For the discrete series, the eigenvalues of the Casimir operator $Q$ are  obviously discrete numbers.

Continuous representations,  subdivided into two types, are characterised by a continuous spectrum of the Casimir operator. The principal continuous series can be either of bosonic or of fermionic type ${\cal C}^{i \sigma,\epsilon}=\{\big|\sigma, \epsilon,n\big>, n \in \mathbb Z + \epsilon\}$ with $\epsilon=0$ (resp. $\epsilon=1/2$) for bosonic  (resp. fermionic) representations. Unitarity implies $\sigma \ne 0$. Since  the substitution
$\sigma \to -\sigma$ leads to an equivalent representation, we restrict our analysis to the case $\sigma>0$:
\beqa
\label{eq:Cont}
K_0 \big|\sigma, \epsilon,n\big>&=& n\big|\sigma, \epsilon,n\big>\ , \nn\\
K_+ \big|\sigma, \epsilon,n\big>&=& \sqrt{(n+\frac 12+\frac i2 \sigma)(n+\frac12-\frac i2\sigma)}\big|\sigma, \epsilon,n+1\big>\ , \\
K_- \big|\sigma, \epsilon,n\big>&=& \sqrt{(n-\frac 12+\frac i2 \sigma)(n-\frac12-\frac i2\sigma)}\big|\sigma, \epsilon,n-1\big>\ , \nn\\
Q\big|\sigma, \epsilon,n\big>&=& \left(\frac 12 +i\sigma\right)\left(\frac 12 +i\sigma -1\right)\big|\sigma, \epsilon,n\big>= -\left(\frac 14 +\sigma^2\right)\big|\sigma, \epsilon,n\big> \ . \nn
\eeqa
Observe that,  since  $\sigma>0$, the eigenvalue of the Casimir operator  is upper bounded by $q<-1/4$.

\medskip
 The supplementary continuous series, on the contrary, is only of bosonic type  ${\cal C}^{ \sigma}=\{\big|\sigma,n\big>, n \in \mathbb Z \}$,
\beqa
K_0 \big|\sigma,n\big>&=& n\big|\sigma,n\big>\ , \nn\\
K_+ \big|\sigma,n\big>&=& \sqrt{(n+\frac 12+\frac12 \sigma)(n+\frac12-\frac12\sigma)}\big|\sigma,n+1\big>\ , \nn \\
K_- \big|\sigma,n\big>&=& \sqrt{(n-\frac 12+\frac12 \sigma)(n-\frac12-\frac12\sigma)}\big|\sigma,n-1\big>\ , \nn\\
Q\big|\sigma,n\big>&=& \left(\frac 12 +\sigma\right)\left(\frac 12 +\sigma -1\right)\big|\sigma,n\big>=\left(-\frac 14 +\sigma^2\right)\big|\sigma,n\big> \ . \nn
\eeqa
The unitary condition $0< \sigma^2 <1/4$ is equivalent to the condition $-1/4<q<0$,  where $q$ denotes the eigenvalue of the Casimir operator $Q$.

\subsection{Matrix elements}\label{sec:Mat}

The Plancherel theorem, which allows to express square-integrable functions on the manifold $\SL$, is based upon the matrix elements of the (two)  discrete series, as well as on those of the principal continuous series,  with the supplementary continuous series playing no r\^ole in the harmonic analysis on $\SL$. For further analysis, we reproduce here the results
 as obtained by Bargmann in \cite{bar}, using a slightly different method.

The group $\SL \cong SU(1,1)$ is defined by the set of $2\times 2$ complex matrices
\beqa
SU(1,1) = \Bigg\{U = \begin{pmatrix} z_1 & z_2 \\
 \bar z_2   & \bar z_1 \end{pmatrix}  \ , \ \ z_1, z_2 \in \mathbb C:  \ \ |z_1|^2-|z_2|^2 =1 \Bigg\} \ . \nn
\eeqa
Thus, as a manifold, we have
\beqa
SU(1,1) =\mathbb{H}_{2,2}=\Big\{z_{1},z_{2}\in \mathbb{C}: |z_{1}|^{2}-|z_{2}|^{2}=1%
\Big\}\ ,\nn
\eeqa
where the hyperboloid  $\mathbb{H}_{2,2}$ can be parameterised as follows:
\begin{equation}
\label{eq:param}
z_{1}=\cosh \rho e^{i\varphi _{1}}\ ,\ \ z_{2}=\sinh \rho e^{i\varphi _{2}}\
,\ \ \rho \geq 0\ ,0\leq \varphi _{1},\varphi _{2}<2\pi
\end{equation}%
We then endow $\mathbb{H}_{2,2}$  with the scalar product\footnote{%
 This choice of scalar product corresponds to the one-sheeted hyperboloid,
and thus it fixes the global properties of the group manifold under
consideration (see \textit{e.g.} the $p=1$ case in Eq. (5.33) of \cite%
{Practitioner}).}
\begin{equation}
\label{eq:sp}
(f,g)=\frac{1}{ 4\pi^{2}}\int\limits_{0}^{+\infty }\cosh \rho \sinh
\rho \text{d}\rho \int\limits_{0}^{2\pi }\text{d}\varphi
_{1}\int\limits_{0}^{2\pi }\text{d}\varphi _{2}\bar{f}(\rho ,\varphi
_{1},\varphi _{2})g(\rho ,\varphi _{1},\varphi _{2})\ .
\end{equation}%

Associated to the parameterisation \eqref{eq:param},
we can define  the generators of $\s$ for  the left action as
\begin{eqnarray}
\label{eq:L}
L_{+} &=&\frac{1}{2}e^{i(\varphi _{1}-\varphi _{2})}\Big[i\tanh \rho
\;\partial _{1}+\partial _{\rho }-i\coth \rho \;\partial _{2}\Big]
\\
L_{-} &=&\frac{1}{2}e^{i(\varphi _{2}-\varphi _{1})}\Big[i\tanh \rho
\;\partial _{1}-\partial _{\rho }-i\coth \rho \;\partial _{2}\Big]
\nn\\
L_{0} &=&\frac{i}{2}\big(\partial _{2}-\partial _{1})\ ,
\nn
\end{eqnarray}%
and, for  the  right-action
\begin{eqnarray}
\label{eq:R}
R_{+} &=&\frac{1}{2}e^{i(\varphi _{1}+\varphi _{2})}\Big[-i\tanh
\rho \;\partial _{1}-\partial _{\rho }-i\coth \rho \;\partial _{2}\Big]
\nn \\
R_{-} &=&\frac{1}{2}e^{-i(\varphi _{2}+\varphi _{1})}\Big[-i\tanh
\rho \;\partial _{1}+\partial _{\rho }-i\coth \rho \;\partial _{2}\Big]
 \\
R_{0} &=&-\frac{i}{2}\big(\partial _{2}+\partial _{1})\ .
\nn
\end{eqnarray}%
  The Casimir operator takes the form:
\beqa
Q &=&\frac{1}{4}\frac{(2\cosh ^{2}\rho -1) }{\cosh \rho \sinh \rho }\partial
_{\rho }+\frac{1}{4}\partial _{\rho }^{2}-\frac{1}{4}\frac{1}{\cosh ^{2}\rho
}\partial _{1}^{2}+\frac{1}{4}\frac{1}{\sinh ^{2}\rho }\partial _{2}^{2}\\[0.3cm]
  &=& \frac{1}{2}\coth(2 \rho)\partial_{\rho }+\frac{1}{4}\partial _{\rho }^{2}-\frac{1-\tanh^{2}\rho}{4}\partial _{1}^{2}+\frac{\coth^2\rho-1}{4}\partial _{2}^{2} \ .
\notag
\eeqa

Of course, the operators for the left (resp. right) action satisfy equation \eqref{eq:LB}, and operators of the left action commute with operators of the right action.\\

To obtain the matrix elements of $\SL-$representations, we identify the functions on $\mathbb H_{2,2}$ which are simultaneously eigenfunctions of the commuting operators $L_0,R_0$ and $Q$, with appropriate eigenvalues.
In fact, we are looking for solution(s) of the differential equations
\beqa
\label{eq:SolQ}
L_0 F_{n,q,m}(\rho, \varphi_1, \varphi_2)&=& n F_{n,q,m}(\rho, \varphi_1, \varphi_2)\nn\\
R_0 F_{n,q,m}(\rho, \varphi_1, \varphi_2)&=& m F_{n,q,m}(\rho, \varphi_1, \varphi_2)\\
Q  F_{n,q,m}(\rho, \varphi_1, \varphi_2)&=& q F_{n,q,m}(\rho, \varphi_1, \varphi_2 ) \ ,\nn
\eeqa
where we have  the constraints
\begin{enumerate}[noitemsep]
\item $q= \lambda(\lambda-1), m,n\ge \lambda>1/2$ for representations bounded from below  $\mathcal{D}_{\lambda }^{+}$;
\item $q= \lambda(\lambda-1), m,n\le- \lambda<-1/2$ for representations bounded from above  $\mathcal{D}_{\lambda }^{-}$;\footnote{The solution $\lambda=1/2$ is excluded for the discrete series, as in this case the matrix elements are not square-integrable (see below and \eqref{eq:cv})}
\item $q= -\frac 14- \sigma^2, m,n \in \mathbb Z + \epsilon, \sigma>0$ (with $\epsilon=0,\frac 12$)  for the principal continuous series  $\mathcal{C}^{i\sigma ,\epsilon }$.
\end{enumerate}
\medskip

 As the supplementary series is irrelevant for the harmonic analysis on $SL(2,\mathbb{R})$, from now onwards we only consider the discrete  and
continuous principal series\footnote{%
 In fact, the matrix elements of the supplementary continuous series are obtained via
the substitution $i\sigma \rightarrow \sigma $ (only for the bosonic case).}. Observe also that, differently from the compact case $SU(2)$, the trivial representation does not lead to a normalised matrix element. This last observation will be relevant in the sequel.
The solution of the two first equations in \eqref{eq:SolQ} is obvious, and is given by
\beqa
F_{n,q,m}(\rho, \varphi_1, \varphi_2)=e^{i(n+m)\varphi_1 + i (m-n)\varphi_2} f_{m,q,n}(\rho) \ ,\nn
\eeqa
with $f_{m,q,n}$ an undetermined function to be identified by the last equation in \eqref{eq:SolQ}.
If we set now
\beqa
\label{eq:anz+}
f_{n,q,m}(\rho)= \cosh^{-m-n}\rho \sinh^{m-n}\rho \;\Xi_{n,q,m}(\rho),
\eeqa
the third equation in \eqref{eq:SolQ}   reduces to the second-order differential equation
\beqa
\Bigg(\frac 14 \frac{\text{d}^2}{\text{d} \rho^2}  +
 \frac{(1-2n)}{2}\coth \rho \frac{\text{d}}{\text{d} \rho}+ \frac 14 \frac{(2m+2n-1)}{\cosh \rho\sinh \rho }
 \frac{\text{d}}{\text{d} \rho} +n(n-1)-q\Bigg) \Xi_{n,q,m}(\rho)=0 \ .\nn
\eeqa
 By means of the additional change of variables  $z=-\sinh^2\rho$,  the latter equation is easily seen to be a hypergeometric equation (see Eq.[\ref{eq:hyper}])  with parameters
\beqa
\alpha= \frac12\big(-2n+1 + \sqrt{4q+1})\ ,\nn
\beta= \frac12\big(-2n+1 - \sqrt{4q+1})\ ,\nn
\gamma=m-n+1 \ . \nn
\eeqa
 We observe that, whenever $\gamma$ is an integer, in order to avoid negative values of $\gamma$,  the solution reduces to (compare with the two solutions given in Appendix \ref{app:hyper})
\beqa
\label{eq:S++}
&F_{n,q,m}(\rho, \varphi_1, \varphi_2)=A_{n,q,m}e^{i(n+m)\varphi_1 + i (m-n)\varphi_2}\cosh^{-m-n}\rho \sinh^{m-n}\rho \nn\\
&{}_2F_1\Big( \frac12\big(-2n+1 + \sqrt{4q+1}), \frac12\big(-2n+1 - \sqrt{4q+1}); 1+m-n; -\sinh^2 \rho\Big)
\eeqa
 for $m\ge n$, and to
\beqa
\label{eq:S+-}
&F_{n,q,m}(\rho, \varphi_1, \varphi_2)=A'_{n,q,m}e^{i(n+m)\varphi_1 + i (m-n)\varphi_2}\cosh^{-m-n}\rho \sinh^{-m+n}\rho \nn\\
&{}_2F_1\Big( \frac12\big(-2m+1 + \sqrt{4q+1}), \frac12\big(-2m+1 - \sqrt{4q+1}); 1+n-m; -\sinh^2 \rho\Big)
\eeqa
 for $n\ge m$.
The coefficients $A_{n,q,m}$ and $A'_{n,q,m}$  must still be computed.

\smallskip

Similarly, if we set
\beqa
\label{eq:anz-}
f_{n,q,m}(\rho)= \cosh^{m+n}\rho \sinh^{m-n}\rho \;\Xi_{n,q,m}(\rho)
\eeqa
we obtain  the solution
\beqa
\label{eq:S-+}
&F_{n,q,m}(\rho, \varphi_1, \varphi_2)=B_{n,q,m}e^{i(n+m)\varphi_1 + i (m-n)\varphi_2}\cosh^{m+n}\rho \sinh^{m-n}\rho \nn\\
&{}_2F_1\Big( \frac12\big(2m+1 + \sqrt{4q+1}), \frac12\big(2m+1 - \sqrt{4q+1}); 1+m-n; -\sinh^2 \rho\Big)
\eeqa
  for $m\ge n$, and
\beqa
\label{eq:S--}
&F_{n,q,m}(\rho, \varphi_1, \varphi_2)=B'_{n,q,m}e^{i(n+m)\varphi_1 + i (m-n)\varphi_2}\cosh^{m+n}\rho \sinh^{-m+n}\rho \nn\\
&{}_2F_1\Big( \frac12\big(2n+1 + \sqrt{4q+1}), \frac12\big(2n+1 - \sqrt{4q+1}); 1+n-m; -\sinh^2 \rho\Big)
\eeqa
for $n\ge m$.

\medskip

We  now turn to the identification of the coefficients $A_{nqm}, A'_{nqm}, B_{nqm}, B'_{nqm}$.  We start with the
ansatz \eqref{eq:anz+} and the corresponding solutions \eqref{eq:S++} or \eqref{eq:S+-}. In the case of the discrete series bounded from below, the matrix elements of the action of the $\s$ generators on $\psi^+_{n,\lambda,m}$
must satisfy \eqref{eq:D+}, with the generators of the  left action given by \eqref{eq:L}
and the generators of the right action given by \eqref{eq:R}
(in the case of the right action, we have to replace $n$ by $m$). Considering $m\ge n$ and the solution given in Eq. [\ref{eq:S++}] with $q=\lambda(\lambda-1)$,  we use \eqref{eq:Lhyper} and impose the constraint \eqref{eq:D+}.  This leads to the expression
\beqa
\psi_{n\lambda m}^{+}(\rho,\varphi_1,\varphi_2)  &=&
  \frac {C} {(m-n)!} \sqrt{\frac{(m-\lambda)!(m+\lambda-1)!}
    {(n-\lambda)!(n+\lambda-1)!}} e^{i(m+n)\varphi_1 + i(m-n)\varphi_2} \cosh^{-m-n}\rho \sinh^{m-n}\rho \times\nn
    \\
     && {}_2F_1(-n+\lambda,-n-\lambda+1;1+m-n;-\sinh^2\rho)\nn \ ,
\eeqa
where $C$ is a coefficient  that must be determined. Using again the results of Appendix \ref{app:hyper}, we impose that $\|\psi^+_{n,\lambda,m}\|=1$ (see \eqref{eq:sp}). From \eqref{eq:cv}
\beqa
\psi^+_{n,\lambda,m}  \in \LL \ \ \Longleftrightarrow\ \  \lambda > \frac12 \ .\nn
\eeqa
Finally, using \eqref{eq:poly} and \eqref{eq:CH}, we deduce that
\beqa
C=\sqrt{2(2\lambda-1)} \ .\nn
\eeqa

Proceeding along the same lines for the case $n\ge m$, we obtain the matrix elements of the discrete series bounded from below
($\lambda>1/2, n,m \ge \lambda$):
\beqa
\label{eq:matD+}
\psi_{n\lambda m}^{+}(\rho,\varphi_1,\varphi_2)  &=& \left\{\begin{array}{l}
  \frac {\sqrt{ 2(2\lambda-1)}} {(m-n)!} \sqrt{\frac{(m-\lambda)!(m+\lambda-1)!}
    {(n-\lambda)!(n+\lambda-1)!}} e^{i(m+n)\varphi_1 + i(m-n)\varphi_2} \cosh^{-m-n}\rho \sinh^{m-n}\rho \times
    \\
      {}_2F_1(-n+\lambda,-n-\lambda+1;1+m-n;-\sinh^2\rho) \ , \hskip 1.truecm m\ge n\\ \\
  \frac {\sqrt{ 2(2\lambda-1)}} {(n-m)!} \sqrt{\frac{(n-\lambda)!(n+\lambda-1)!}
    {(m-\lambda)!(m+\lambda-1)!}}
     e^{i(m+n)\varphi_1 + i(m-n)\varphi_2} \cosh^{-m-n}\rho \sinh^{-m+n}\rho \times
    \\
    (-1)^{m-n}  {}_2F_1(-m+\lambda,-m-\lambda+1;1-m+n;-\sinh^2\rho)\ , \hskip .8truecm  n\ge m
\end{array}
\right.
\eeqa

\medskip
To obtain the matrix elements of the representations bounded from above,  we  have to deal with the ansatz \eqref{eq:anz-} instead, to avoid convergence problems (see \eqref{eq:ab}). The same computation as before leads us to  ($\lambda>1/2, n,m \le -\lambda$):
\beqa
\label{eq:matD-}
\psi_{n\lambda m}^{-}(\rho,\varphi_1,\varphi_2)  &=& \left\{\begin{array}{l}
 \frac {\sqrt{2(2\lambda-1)}} {(m-n)!} \sqrt{\frac{(-n-\lambda)!(-n+\lambda-1)!}
    {(-m-\lambda)!(-m+\lambda-1)!}} e^{i(m+n)\varphi_1 + i(m-n)\varphi_2} \cosh^{m+n}\rho \sinh^{m-n}\rho
    \times\\
     (-1)^{m-n}  {}_2F_1(m+\lambda,m-\lambda+1;1+m-n;-\sinh^2\rho)\ , \hskip .8truecm m\ge n\\ \\
  \frac {\sqrt{2(2\lambda-1)}} {(n-m)!} \sqrt{\frac{(-m-\lambda)!(-m+\lambda-1)!}
    {(-n-\lambda)!(-n+\lambda-1)!}} e^{i(m+n)\varphi_1 + i(m-n)\varphi_2} \cosh^{m+n}\rho \sinh^{-m+n}\rho
     \times\\
      {}_2F_1(n+\lambda,n-\lambda+1;1-m+n;-\sinh^2\rho)\ , \hskip 1.truecm n\ge m
\end{array}
\right.
\eeqa
Note that the matrix elements of discrete series have a different normalisation than those in \cite{bar}, since they are
normalised with respect to the scalar product \eqref{eq:sp}
\beqa
\label{eq:Dorth}
(\psi_{n\lambda m}^\eta, \psi_{n'\lambda' m'}^{\eta'}) = \delta_{nn'} \delta_{mm'} \delta_{\lambda \lambda'} \delta_{\eta \eta'}  \ , \
\eeqa
where $\eta, \eta' = +,-$ for discrete series bounded from below and from above, respectively.

Further we have
\beqa
\psi_{n\lambda m}^\eta(0,0,0)=\sqrt{2(2\lambda-1)}\delta_{mn} \ . \nn
\eeqa

 Repeating the procedure for the principal continuous series ($\epsilon=0,1/2, n,m \in \mathbb Z + \epsilon, \sigma>0$), a routine computation shows that the matrix elements are given by
\beqa
\label{eq:matC}
&&\hskip -1.truecm \left.
\begin{array}{lll}
  \psi_{ni\sigma m}^{\epsilon }(\rho,\varphi_1,\varphi_2)&=&
  \frac 1 {(m-n)!} \sqrt{\frac{\Gamma(m+\frac 12 +\frac 12 i\sigma)\Gamma(m+\frac 12 -\frac 12 i\sigma)}
                              {\Gamma(n+\frac 12 +\frac 12 i\sigma)\Gamma(n+\frac 12 -\frac 12 i\sigma)}}
  e^{i(m+n)\varphi_1 +i(m-n)\varphi_2}\cosh^{m+n}\rho \sinh^{m-n}\rho  \times\\
  && {}_2F_1(m+\frac12 + \frac12 i\sigma,m+\frac12 - \frac12 i\sigma;m-n+1; -\sinh^2 \rho)\\ \\
 &=&
  \frac 1 {(m-n)!} \sqrt{\frac{\Gamma(m+\frac 12 +\frac 12 i\sigma)\Gamma(m+\frac 12 -\frac 12 i\sigma)}
                              {\Gamma(n+\frac 12 +\frac 12 i\sigma)\Gamma(n+\frac 12 -\frac 12 i\sigma)}}
  e^{i(m+n)\varphi_1 +i(m-n)\varphi_2} \cosh^{-m-n}\rho \sinh^{m-n}\rho \times  \\
 &&  {}_2F_1(-n+\frac12 + \frac12 i\sigma,-n+\frac12 - \frac12 i\sigma;m-n+1; -\sinh^2 \rho)
\end{array}\right\}\nn \\
&&\hskip 6.truecm  m\ge n\nn\\ \\
&&\hskip -1.truecm \left.
\begin{array}{lll}
  \psi_{n i\sigma m}^{\epsilon }(\rho,\varphi_1,\varphi_2)&=&
  \frac {1  } {(n-m)!} \sqrt{\frac{\Gamma(n+\frac 12 +\frac 12 i\sigma)\Gamma(n+\frac 12 -\frac 12 i\sigma)}
                              {\Gamma(m+\frac 12 +\frac 12 i\sigma)\Gamma(m+\frac 12 -\frac 12 i\sigma)}}
  e^{i(m+n)\varphi_1 +i(m-n)\varphi_2}\cosh^{m+n}\rho \sinh^{-m+n}\rho \times \\
  &&(-1)^{m-n} {}_2F_1(n+\frac12 + \frac12 i\sigma,n+\frac12 - \frac12 i\sigma;-m+n+1; -\sinh^2 \rho)\\ \\
 &=&
  \frac {1} {(n-m)!} \sqrt{\frac{\Gamma(n-\frac 12 +\frac 12 i\sigma)\Gamma(n-\frac 12 -\frac 12 i\sigma)}
                              {\Gamma(m-\frac 12 +\frac 12 i\sigma)\Gamma(m-\frac 12 -\frac 12 i\sigma)}}
  e^{i(m+n)\varphi_1 +i(m-n)\varphi_2} \cosh^{-m-n}\rho \sinh^{n-m}\rho  \times\\
 && (-1)^{m-n} {}_2F_1(-m+\frac12 + \frac12 i\sigma,-m+\frac12 - \frac12 i\sigma;-m+n+1; -\sinh^2 \rho)
\end{array}\right\} \nn\\
&&\hskip 6.truecm  n\ge m\nn\nn
\eeqa
As the eigenvalue of the Casimir operator is continuous, the matrix elements are not normalisable and we have
(see Section \ref{sec:Plan})
\beqa
\label{eq:Corth}
(\psi^{\epsilon }_{ni \sigma m}, \psi^{\epsilon' }_{n'i \sigma' m'})=\frac{1}{\sigma\tanh\pi(\sigma+i\epsilon)} \delta_{\epsilon \epsilon'}
\delta_{mm'} \delta_{n n'} \delta(\sigma-\sigma')\ ,
\eeqa
(with $\tanh\pi(\sigma+\frac i 2)) =\coth\pi\sigma\nn$ \cite{vk}). We further have
\beqa
\psi^{\epsilon }_{ni \sigma m}(0,0,0)=\delta_{mn}\ . \nn
\eeqa

To end this section,  we observe that for the discrete series ($\eta=\pm 1$) the following relation holds:
\beqa
\label{eq:cc1}
\overline{\psi^{m\lambda n}_{\eta}}(\rho,\varphi_1,\varphi_2) = \psi^{-\eta}_{-n\lambda -m}(\rho,\varphi_1,\varphi_2) \ ,
\eeqa
 meaning that  the complex conjugate of matrix elements of a representation bounded from below are matrix elements of a representation bounded from above  (and the other way around). In the case of the principal continuous series, we have
\beqa
\label{eq:cc2}
\overline{\psi^{m i\sigma n}_{\epsilon}}(\rho,\varphi_1,\varphi_2) =\psi_{-m i\sigma -n}^{\epsilon}(\rho,\varphi_1,\varphi_2) \ .
\eeqa
To prove this identity, we have used \footnote{%
Because $\Gamma (n)=(n-1)!$ for $n\in \mathbb{N}\setminus\left\{ 0\right\} $.} $\Gamma(\alpha+m)=(\alpha+m-1) \cdots \alpha \Gamma(\alpha)$ to simplify the coefficient
which appears in the square root
in \eqref{eq:matC}. For instance, if $m\ge n$

\beqa
\label{eq:n-n}
\frac{\Gamma(m+\frac 12 +\frac 12 i\sigma)\Gamma(m+\frac 12 -\frac 12 i\sigma)}
                              {\Gamma(n+\frac 12 +\frac 12 i\sigma)\Gamma(n+\frac 12 -\frac 12 i\sigma)}
                              =\Big|m-\frac12+\frac12 i\sigma\Big|^2\;\cdots\;
                              \Big|n+\frac12+\frac12 i\sigma\Big|^2
\eeqa

\subsection{Plancherel Theorem}\label{sec:Plan}

The Peter-Weyl theorem is a generalisation of the Fourier series on compact Lie groups. Similarly, the Plancherel theorem is a generalisation of the Fourier transform to the case of non-compact Lie groups \cite{sch}.
In particular, the Plancherel theorem enables us to obtain a decomposition of square-integrable functions on non-compact Lie groups.

 We briefly recall  how these results hold for $L^2(\mathbb R)$. In this case,  the real line $\mathbb R$
is considered as a group of translations. Irreducible representations are given by
  \beqa
  \label{eq:qm1}
 f_p(x)= \frac 1{\sqrt{2 \pi}}e^{ipx} \ ,\ \  p \in \mathbb R
\eeqa
and correspond to plane waves.
The representations ${\cal D}_p=\{f_p\}, p\in\mathbb R$ are analogous to a continuous representation in the $\SL$ case, and we have
\beqa
\label{eq:qm2}
(f_p,f_q) = \delta(p-q) \ ,
\eeqa
with $(\cdot,\cdot)$ the usual scalar product on $L^2(\mathbb R)$.
Since there is no discrete series, for any $f\in L^2(\mathbb R)$ we have
\beqa
f(x) =  \int\limits_{-\infty}^{+\infty} \text{d} p \tilde f(p)  f_p(x)\nn
\eeqa
which is the Fourier decomposition, that turns out to be the Plancherel theorem associated to the group of translations $\mathbb R$.

When the Lie group is $\SL$, the decomposition of functions on $\LL$ involves the matrix elements of the two discrete
series, together with the matrix elements of the continuous principal series. Let $f \in \LL$.  Then, we have (see \cite{hc} and \cite{vil}, pp. 336-337)
\beqa
\label{eq:Plan}
f(\rho,\varphi_1,\varphi_2) &=&\sum \limits_{\lambda>\frac12}\;
\sum \limits_{m,n\ge \lambda}\; f_{+  }^{n\lambda m}\; \psi_{n\lambda m}^{+ }(\rho,\varphi_1,\varphi_2) +
  \sum \limits_{\lambda>\frac 12}\;
 \sum \limits_{m,n\le - \lambda}\; f_{-}^{n\lambda m}\; \psi_{n\lambda m}^{- }(\rho,\varphi_1,\varphi_2)\nn\\
 && + \int \limits_0^{+\infty} \text{d} \sigma\; \sigma \tanh \pi \sigma \sum \limits_{m,n \in \mathbb Z}
 f_0^{nm}(\sigma) \psi^{0 }_{n i \sigma m}(\rho,\varphi_1,\varphi_2)\\
&&  +\int \limits_0^{+\infty} \text{d} \sigma\; \sigma \coth \pi \sigma \sum \limits_{m,n \in \mathbb Z + \frac 12}
 f_\frac 12^{nm}(\sigma) \psi^{\frac 12 }_{n i \sigma m}(\rho,\varphi_1,\varphi_2)\nn
 \eeqa
 The components of $f$ are given by the scalar products \eqref{eq:sp}
 \beqa
 \label{eq:PC}
 f_{\pm }^{n \lambda m}&=&(\psi^{\pm  }_{n\lambda m},f) \\
 f^{\epsilon}_{nm}(\sigma)&=& (\psi^{\epsilon }_{ni\sigma m},f) \ , \nn
 \eeqa
 which in fact legitimate \eqref{eq:Corth} {\it a posteriori}.

 In fact, \eqref{eq:Plan} and \eqref{eq:PC}  (as well as the standard relations
 of quantum mechanics \eqref{eq:qm1} and \eqref{eq:qm2}) must be treated with care. Indeed, the relation \eqref{eq:Plan} is {\it a priori}
 defined only for  the set of  functions ${C}^\infty_c($SL$(2,\mathbb R))$ on SL$(2,\mathbb R)$
which vanish outside a compact set
$c\subset$\;SL$(2,\mathbb R)$
and which are infinitely differentiable \cite{hc}. Nevertheless, it is possible to legitimate
\eqref{eq:Plan} and \eqref{eq:PC} (and similarly \eqref{eq:qm1} and \eqref{eq:qm2}) by considering a Gel'fand triple
which is introduced to deal with Hermitean operators which have continuous spectra and
thus have `eigenvectors' which are not normalisable. More precisely, a Gel'fand triple of a Hilbert space ${\cal H}$
(see for instance \cite{Ba-Ra}) is given
by a dense subspace ${\cal S}$ of smooth functions of  ${\cal H}$ and its dual  ${\cal S}'$ (in Quantum Mechanics, ${\cal S}$ corresponds to
the space of Schwartz functions) such that
\beqa
{\cal S} \stackrel{\mathclap{\tiny\mbox{J}}}{\subset} {\cal H} \stackrel{\mathclap{\tiny\mbox{K}}}{\subset} {\cal S}' \ , \nn
\eeqa
where $J:{\cal S} \rightarrow {\cal H}$ is an injective bounded operator with dense image and $K$ is the composition of the canonical isomorphism
${\cal H}\simeq {\cal H}^{\prime}$ determined by the inner product (given by the Riesz theorem) and the dual $J^{\prime}:{\cal H}^{\prime}\rightarrow {\cal S}' $ of $J$.
If we assume that the functions  are in ${\cal S}$, then both \eqref{eq:Plan}
and \eqref{eq:PC} make sense.  Note also that the matrix elements of the continuous principal series
$\psi^\epsilon_{m i\sigma n}$ belong to ${\cal S}'$; see for instance \cite{Ba-Ra}, Theorem 1,  p. 426, and Eqs.[34-37] p. 429, or \cite{Va} Chap. 8 (where in
particular the set of Schwartz functions for $\LL$, which are rapidly decreasing in the $\rho-$direction are defined)\footnote{ Schwartz functions were defined
for any semisimple Lie group by Harish-Chandra (see \cite{Va} and references therein).}. Furthermore Eq.[\ref{eq:Corth}] is
also a consequence of the formalism (see \cite{Ba-Ra},  p. 427, Eq.[21]). Thus,
we will hitherto consider functions which are compatible with  \eqref{eq:Plan} and \eqref{eq:PC}.

Since the matrix elements enable us to expand square-integrable functions on $\SL$,
we also have
(see \cite{Ba-Ra} p. 429)
 \beqa
 \label{eq:delP}
&\sum \limits_{\lambda>\frac12}\; \sum \limits_{m,n\ge \lambda}\;\psi^+_{n \lambda m}(\rho,\varphi_1,\varphi_2)
 \overline{\psi}_+^{n \lambda m}(\rho',\varphi'_1,\varphi'_2) +
\sum \limits_{\lambda>\frac12}\; \sum \limits_{m,n\le - \lambda}\;\psi^-_{n \lambda m}(\rho,\varphi_1,\varphi_2)
 \overline{\psi}_-^{n \lambda m}(\rho',\varphi'_1,\varphi'_2)
\nn\\
&+
  \int \limits_0^{+\infty} \text{d} \sigma\; \sigma \tanh \pi \sigma \sum \limits_{m,n \in \mathbb Z}
 \psi^0_{n i \sigma m}(\rho,\varphi_1,\varphi_2) \overline{\psi}_0^{n i \sigma m}(\rho',\varphi'_1,\varphi'_2) \\\
 &+
  \int \limits_0^{+\infty} \text{d} \sigma\; \sigma \coth \pi \sigma \sum \limits_{m,n \in \mathbb Z}
 \psi^\frac12_{n i \sigma m}(\rho,\varphi_1,\varphi_2) \overline{\psi}_\frac12^{n i \sigma m}(\rho',\varphi'_1,\varphi'_2) \nn\\
& = \frac 12\delta(\sinh^2 \rho -\sinh^2\rho') \delta(\varphi_1-\varphi'_1)\delta(\varphi_2-\varphi'_2)\nn
\eeqa

Introducing  the symbol
 $\sum \hskip -.4 truecm \int{}$ to reproduce the summation over all discrete and continuous series, we can denote  \eqref{eq:Plan}  as
\beqa
\label{eq:IP}
f(\rho,\varphi_1,\varphi_2) = \sum_{\Lambda,m,n} \hskip -.6 truecm \int{} f^{n \Lambda m} \psi_{n\Lambda,m}(\rho,\varphi_1,\varphi_2)
\eeqa
with $\Lambda=(+,\lambda), (-,\lambda), (0,i\sigma), (1/2,i\sigma)$.  As expected, in the case of the principal series, the integral involves $\sigma \tanh \pi (\sigma + i\epsilon)$, thus justifying (\eqref{eq:Corth}) \textit{a posteriori}.

\section{Hilbert basis of $\LL$}

 In the preceding section we have seen how the Plancherel theorem allows us to expand square-integrable functions on  $\SL$ by means of the matrix elements of  the discrete and  principal continuous series.  As the latter series  is not normalisable,
this decomposition involves an integral, whilst since the former are normalised
the part involving discrete series involves a
sum. Thus the set of matrix elements of discrete series and of principal continuous series doesn't
constitute  a Hilbert basis of $\LL$
It is however well known   (see \textit{e.g.} \cite{RS, GP}) that any Hilbert space admits a Hilbert basis, {\it i.e.}, a complete,
countable set of orthonormal vectors which enables to express any element of the
Hilbert space as a convergent infinite sum
over this countable set of basis vectors.
In that context, the matrix elements of the discrete series bounded
from below and from above constitute a countable set of orthonormal functions of $\LL$, but this set is not complete. Differently, the set of  matrix elements of the continuous principal series is neither countable, nor
normalisable (see \eqref{eq:Corth}) and consequently the matrix elements identified in Section
\ref{sec:Mat} cannot constitute a Hilbert basis of $\LL$.
 It is the purpose of this section to identify a Hilbert basis of $\LL$.

\subsection{Normalisable eigenfunctions of the three commuting operators $L_0,R_0$ and $Q$}\label{sec:H1}
In Section \ref{sec:Mat} we have identified the matrix elements of representations of $SL(2,\mathbb{R})$. The goal of this section is to identify all normalised functions satisfying \eqref{eq:SolQ} with $m,n \in \mathbb Z$ or $m,n\in \mathbb Z+\frac12$, by looking to the two ansatz \eqref{eq:anz+} and \eqref{eq:anz-} as well as their corresponding eigenfunctions \eqref{eq:S++}, \eqref{eq:S+-}, \eqref{eq:S-+} and \eqref{eq:S--}
respectively.
All solutions are expressed in terms of hypergeometric functions. In order to have a normalisable solution, we must have a hypergeometric
polynomial. This implies, in particular, that in ${}_2F_1(\alpha,\beta;\gamma,z)$ either $\alpha$ or $\beta$ has to be a negative integer (see (\eqref{eq:poly}) and the comment above it).
This is possible {\it iff}
\beqa
4q+1=N^2\ , \ \ N \in \mathbb N\ .\nn
\eeqa
Setting $\mu=\frac{N+1}2$,
we obtain
\beqa
q=\mu(\mu-1) \ .\nn
\eeqa

\subsubsection{ Case $m,n\ge0$}
 Let us first consider  $m\ge n$.
In this case, we obtain the solution \eqref{eq:S++} with the  first ansatz \eqref{eq:anz+}:
\beqa
\Psi(\rho,\varphi_1,\varphi_2)&=&e^{i(m+n)\varphi_1+i(m-n)\varphi_2} \cosh^{-m-n}\rho \sinh^{m-n}\rho\nn\\
&&{}_2F_1(-n+\mu,-n-\mu+1;1+m-n;-\sinh^2\rho) \ .\label{eq:this}
\eeqa
 In these conditions, one of the three following possibilities holds:
\begin{enumerate}[noitemsep]
\item  $m\ge n\ge \mu$.
This corresponds  to matrix elements of the discrete series  bounded from below
${\cal D}^+_\mu$ with $\mu>1/2$.
\item  $m\ge \mu> n$. Then ${}_2F_1(-n+\mu,-n-\mu+1;1+m-n;-\sinh^2\rho)$ is a degree $(n+\mu-1)$ polynomial in $\sinh^2\rho$, which is not
normalisable by \eqref{eq:convP}.
\item  $\mu>m\ge n$.  As in the preceding case, the solution is not normalisable.
\end{enumerate}
The case where $n\ge m$ is similar,  and we are led  to the same conclusion.
Hence, when $m,n>1/2$, the only normalisable solutions correspond
to the matrix elements of the discrete series bounded from  below ${\cal D}^+_\mu, \mu>\frac12$.

\subsubsection{ Case  $m,n\le 0$}
 Similarly, as happens for  $m,n\ge 0$, using ansatz \eqref{eq:anz-}, we obtain that the only normalisable solutions correspond
to the matrix elements of the discrete series bounded from  above ${\cal D}^-_\mu, \mu>\frac12$ with $n,m<-1/2$.

\subsubsection{ Case  $m\ge 0$ and $n\le 0$}
 Suppose first that $m\ge -n$ holds. The solution is obtained again with the first ansatz, and it is given by \eqref{eq:this}.

 Note that ${}_2F_1(-n+\mu,-n-\mu+1;1+m-n;-\sinh^2\rho)$ is  a degree $n+\mu-1$ polynomial in $\sinh^2\rho$, which is not
normalisable by \eqref{eq:convP}.

\smallskip

If $-n\ge m$, the second ansatz \eqref{eq:anz-} leads to
\beqa
\Psi(\rho,\varphi_1,\varphi_2)&=&e^{i(m+n)\varphi_1+i(m-n)\varphi_2} \cosh^{m+n}\rho \sinh^{m-n}\rho  \times\nn\\
&&{}_2F_1(m+\mu,m-\mu+1;1+m-n;-\sinh^2\rho) \ .\nn
\eeqa
If $m>\mu-1, {}_2F_1(m+\mu,m-\mu+1;1+m-n;-\sinh^2\rho)$ is not a polynomial, thus we do not have a square-integrable function.
If $m<\mu-1\le0$, then ${}_2F_1(m+\mu,m-\mu+1;1+m-n;-\sinh^2\rho)$ is a polynomial of degree $\mu-1-m$  in $\sinh^2\rho$, which is not
normalisable by \eqref{eq:convP}.\\

As the case $n\ge 0, m\le n$ is similar  to the previous one, we  conclude that the only normalisable eigenfunctions of the operators $L_0,R_0$ and $Q$ are the matrix elements
of the discrete series.

\subsection{Hilbert basis of $\LL$}\label{sec:Hilb}
 Once the normalisable eigenfunctions have been identified, we proceed with the identification of a Hilbert basis of $\LL$. We have seen that the only
normalisable eigenfuctions of the Cartan of the left and right actions and the Casimir operator are the matrix
elements of the discrete series. This is in accordance with the property that the only irreducible unitary representations of $\SL$ which
are normalisable are the discrete series \cite{bar}. This of course can be compared with usual
eigenvalues/eigenfunctions problem of the momentum in the case of $L^2(\mathbb R)$, which does not admit normalised eigenfuctions.
However, as it is the case for $L^2(\mathbb R)$, it is fully possible to identify a Hilbert basis of $\LL$. To start with
this identification we introduce the notations
\beqa
\LL= \LL^d \oplus \LL^{d^\perp} \nn
\eeqa
where $\LL^d$ constitutes the set of square-integrable functions expanded with the discrete series. This corresponds to
functions in \eqref{eq:Plan}, with only the first line  being non-vanishing. Stated differently, this denotes
the part corresponding to the discrete series. The second term, $\LL^{d^\perp}$ corresponds to functions
in \eqref{eq:Plan}, where the second and third lines  of \eqref{eq:Plan} are non-vanishing.  This denotes
the part corresponding to the principal continuous series. \\

 In Sec. \ref{sec:Mat}, a Hilbert basis of $\LL^d$  has been identified. We now  inspect  a Hilbert basis of  $\LL^{d^\perp}$.
This identification is mainly due to Viktor Losert  \cite{los} [V. Losert, Institut f\"ur Mathematik, Universit\"at Wien,  viktor.losert@univie.ac.at].
We reproduce here the  construction  of the Hilbert basis of $\LL^{d^\perp}$ with his kind authorisation.
We define the eigenspaces of the operators $L_0$ and $R_0$
\beqa
W_{mn} = \Bigg\{ F \in \LL\ , \ \ F(\rho,\varphi_1,\varphi_2)=e^{i(m+n)\varphi_1 + i(m-n)\varphi_2} f(\rho) \Bigg\} \ .\nn
\eeqa
Correspondingly, we define
\beqa
\label{eq:wmn}
w_{mn} &=& \Bigg\{f(\rho)= e^{-i(m+n)\varphi_1 - i(m-n)\varphi_2}F(\rho,\varphi_1, \varphi_2) : F \in W_{mn}\Bigg\}
\eeqa
Our strategy is to identify a Hilbert basis of $L^2(w_{mn})$ for  fixed values of $m,n$.
We first focus on the discrete series.  As shown in Appendix \ref{app:J},  the matrix elements of the discrete series can be expressed in terms of  Jacobi polynomials.
For the discrete series bounded from below (resp. above) and a given value of
$\lambda>1/2$, we have $m,n\ge \lambda  >1/2$ (resp. $m,n<-\lambda  <-1/2$).
That is, for given values of $m,n$ the degree of the polynomial
$k=$min$(|n|,|m|)-\lambda$
is $ 0\le k<$ min$(|m|,|n|) -  1/2$, since for bosons $m,n$ are integer
and for fermions $m,n$ are half integer, this latter relation writes equivalently as
 $ 0\le k<$ min$(|m|,|n|) - \epsilon$
with $\epsilon=0$ (resp. $\epsilon=1/2$) for bosons (resp. fermions).
Using the
relationship \eqref{eq:HJ}, the Hilbert basis ${\cal B}_d$  of $L^{2}\left( SL(2,\mathbb{R})\right) ^{d}$ is given by:
\beqa
\label{eq:Losert}
    {\cal B}_{d}&&= \Bigg\{ \phi_{m,n,k}(\rho,\varphi_1,\varphi_1)=e^{i (m+n)\varphi_1 + i(m-n)\varphi_2 }e_{m,n,k} (\cosh 2\rho)\ ,\\
 && \hskip1.truecm  \epsilon=0, \frac12; n,m \in \mathbb Z+\epsilon; |n|, |m| > \frac12; nm>0; 0 \le k < \text{min}(|n|,|m|)-\epsilon
    \Bigg\}, \nn
    \eeqa
 where $e_{m,n,k}$ is defined as follows:
        If $n\ge m>1/2$,
\beqa
\label{eq:ed}
e_{m,n,k}(x)&=& \sqrt{2^{2m-1}\frac{(2m-2k-1)k!(m+n-k-1)!}{(2m-k-1)!(n-m+k)!} } \times\\
&&(x-1)^\frac{n-m}2 (x+1)^{-\frac{m+n}2} P_k^{(n-m,-n-m)}(x) \ , 0\le k< m-\epsilon,\nn
\eeqa
where  $x=\cosh 2 \rho$.
     For $m\ge n > 1/2$,  $e_{m,n,k}=(-1)^{m-n}e_{n,m,k}$,  while for $m,n<-\frac 12$, we have $e_{m,n,k}=e_{-m,-n,k}$.\\

Now we turn to the Hilbert basis  ${\cal B}^{d^\perp}$ of $\LL^{d^\perp}$. We define
\beqa
\label{eq:Hil2}
{\cal B}^{d^\perp} &=&\Big\{ \phi_{m,n,k}(\rho,\varphi_1,\varphi_1)=e^{i (m+n)\varphi_1 + i(m-n)\varphi_2 }e_{m,n,k} (\cosh 2\rho)\ ,\nn\\
&& \hskip 1.5truecm
 \epsilon=0,\frac 12; m,n \in \mathbb Z+\epsilon; k\ge k_{\text{min}  }
    \Big\}
\eeqa
where
\beqa
\label{eq:kmin}
k_{\text{min}}= \left\{
\begin{array}{ll}
\text{min}(|n|,|m|) -\epsilon&\text{if} \ \ |n|, |m| > \frac12;  nm>0\\
0&\text{elsewhere}
\end{array}
\right.
\eeqa
 In $\LL^{d^\perp}$ we only have polynomials of degree $d\ge$min$(|n|,|m|)-\epsilon$ when $mn>0$ and $|m|,|n|>\frac 12$,
  as
polynomials of degree $d<$min$(|n|,|m|)-\epsilon$ for the corresponding values of $m,n$ are contained in $\LL^d$.
However, in all other cases, we have to consider polynomials of degree $d\ge 0$.
\\

Consider first the case $mn\ge 0$ and $|m|,|n|>1/2$.
If $n\ge m>1/2$, we set
    \beqa
    \label{eq:e}
e_{m,n,k}(x) =  \sqrt{2^{2k + 2 \epsilon+1}\frac{(2k +n-m + 2\epsilon+1) (k +n-m + 2\epsilon) !k! }{(k+2\epsilon)! (n-m+k)!} } \times\nn\\
    (x-1)^{\frac{n-m}2} (x+1)^{\frac{m-n}2 -k-\epsilon -1}
    P_k^{(n-m,m-n-2k-2\epsilon-1)}(x) \ , k\ge m -\epsilon
    \eeqa
 Then, for  $m\ge n > 1/2$, we have $e_{m,n,k}=e_{n,m,k}$, and for $m,n<-\frac 12$,  $e_{m,n,k}=e_{-m,-n,k}$.

 In the remaining cases, we define $e_{m,n,k}$ as in \eqref{eq:e} for $n>m$,
and  $e_{m,n,k}$ as in \eqref{eq:e} for $m>n$, permuting $n$ and $m$,
but now with $k\ge 0$.
Of course, because of \eqref{eq:N4}, the functions are normalised.

 We now make an  observation that will be important in the sequel. In the Hilbert basis constructed
so far, when the condition
\beqa
\label{eq:cond}
mn>0, |m|,|n|>1/2
\eeqa
holds the vectors $e_{m,n,k}$ belongs to ${\cal B}^d$ if
$k<$min$(|n|,|m|)-\epsilon$ and belongs to ${\cal B}^{d^\perp}$ if $k\ge$min$(|n|,$ $|m|)-\epsilon$. On the other hands, when
the condition \eqref{eq:cond} doesn't hold the vectors $e_{m,n,k}$  belongs to ${\cal B}^{b^\perp}$ for all integer values  of $k$ (see \eqref{eq:Losert} and \eqref{eq:kmin}).
Thus we have for any values of $m$ and $n$:
$w_{mn}=\{e_{m,n,k}\ , k \in \mathbb N\}.$ We should stress that the elements of ${\cal B}^d$ are normalised with respect to \eqref{eq:N3}, whereas the elements of ${\cal B}^{d^\perp}$ are normalised with respect to
\eqref{eq:N4}.
\\

We now prove that for  any fixed value of $m,n$,  the set $\{e_{m,n,k}, k\in \mathbb N\}$ is a Hilbert basis of
$L^2(w_{mn}) \cong L^2([1,+\infty[)$. Following Losert, we define the isomorphism \eqref{eq:bij} between $[-1,1]$ and $[1,+\infty [$ and set
\beqa
\begin{array}{lcll}
U_{mn}:&L^2([1,+\infty[)&\to&L^2([-1,1], (1-x)^{2\epsilon} (1+x)^{n-m})\\
&f(x)&\mapsto&(U_{mn}f)(x)=2(1-x)^{-1-\epsilon}(1+x)^{\frac{m-n}2} f\Big(\frac{3+x}{1-x}\Big),
\end{array}\nn
\eeqa
where $ (1-x)^{2\epsilon} (1+x)^{n-m}$ is the weight for the scalar product in $L^2([-1,1])$. Indeed, we have
\beqa
\|U_{mn} f\|^2_{[-1,1]}=\int\limits_{-1}^1 \text{d}x (1-x)^{2\epsilon}(1+x)^{n-m} |U_{mn}f(x)|^2 = \int
\limits_{1}^{+\infty} \text{d}x|f(x)|^2 = \|f\|^2_{[1,+\infty[} . \nn
\eeqa

We consider  $n\ge m>1/2$ and show that $\{e_{m,n,k}, k \in \mathbb N\}$ is a Hilbert basis of $L^2([1,+\infty[)$ through
the isomorphism $U_{mn}$. By  \eqref{eq:this2}, the image of  \eqref{eq:e} is
\beqa
U_{mn} e_{m,n,k}(y) =(-1)^{k}\sqrt{2^{m-n-2\epsilon-1}\frac{(2k+n-m+2\epsilon+1)\;k!\;(k+n-m+2\epsilon)!}{(k+2\epsilon)!\;(n-m+k)!}}
P_k^{(2\epsilon,n-m)}(y) \ , \nn
\eeqa
which is a polynomial of degree $k\ge m-\epsilon$,
whilst the image of \eqref{eq:ed}  (up to a multiplicative constant)
\beqa
\label{eq:zz}
U_{mn} e_{m,n,k}=(1-x)^{m-k-1-\epsilon} P_k^{(2m-2k-1,n-m)}(x) \ ,
\eeqa
is a polynomial of degree $m-1-\epsilon<m-\epsilon$. Note that the assumption on $k$ implies that $m-k-1-\epsilon$
is a non-negative integer.
We now prove that $\{U_{mn}e_{m,n,k}\}$ is a Hilbert basis of $L^2([1,-1])$.

We first show that the set $\{U_{mn}e_{m,n,k}, k\in \mathbb N\}$ is  orthonormal. We already know that $\|e_{m,n,k}\|=1$. We have
to prove their orthogonality.
Three cases must be considered:
\begin{enumerate} [noitemsep]
\item $k,k'<m-\epsilon, k\ne k'$. By \eqref{eq:N3} we have $e_{m,n,k}\perp e_{m,n,k'}$.
\item $k,k'\ge m-\epsilon, k\ne k'$. Then $U_{mn}e_{m,n,k} \perp U_{mn} e_{m,n,k'}$ holds in $L^2([-1,1])$ with
weight $ (1-x)^{2\epsilon}(1+x)^{m-n}$, hence $e_{m,n,k}\perp e_{m,n,k'}$ in
 $L^2([1,+\infty[)$ because $U_{mn}$ is an isometry.
 \item $k<m-\epsilon, k'\ge m-\epsilon$. Since $U_{mn}e_{m,n,k}$ is a polynomial of degree $<m-\epsilon$ and $U_{mn}e_{m,n,k'}$ is of degree
 $\ge m-\epsilon$, we have $U_{mn}e_{m,n,l} \perp U_{mn} e_{m,n,k'}$ and, again,  $e_{m,n,k}\perp e_{m,n,k'}$
 because $U_{mn}$ is an isometry.
\end{enumerate}

 Now, as for the discrete series (bounded from below) the set $\{U_{mn}e_{m,n,k}, 0\le k<m-\epsilon\}$ generates the space of polynomials of degree $m-\epsilon-1$ ($m-\epsilon$ independent polynomials; {\it cf.} \eqref{eq:zz}),
  the image $\{U_{mn}e_{m,n,k}, k\ge0\}$ generates the space of all polynomials, which is dense in $L^2([-1,1])$ (with weight $(1-x)^{2\epsilon} (1+x)^{n-m}$), from which, by inverting the isomorphism $U_{mn}$, follows that  $\{e_{m,n,k}, k\ge 0\}$ is
a Hilbert basis of $L^2(w_{mn})\cong L^2([1,+\infty[)$.\\

This  argument extends to any values of $m,n$.  As a consequence
\beqa
\label{eq:BHil}
{\cal B} =\Big\{ \Phi_{m,n,k}(\rho,\varphi_1,\varphi_2)=e^{i(m+n)\varphi_1 + i (m-n)\varphi_2} e_{m,n,k}(\cosh 2 \rho)\ ,
\epsilon=0, \frac12; m,n \in \mathbb Z + \epsilon\ ; k\in \mathbb N \Big\},\nn\\
\eeqa
where $e_{m,n,k}$ are defined above (see \eqref{eq:ed} and \eqref{eq:e})
constitutes, as stated above, a Hilbert basis of $\LL$, which is called from now on the Losert basis.
 To recap we thus have
\begin{enumerate}[noitemsep]
\item If $m,n> \frac 12\phantom{-} $ then $W_{mn} \cap \LL^d \subset \LL^{(d,+)}$;
\item If $m,n< -\frac 12 $ then $W_{mn} \cap \LL^d \subset \LL^{(d,-)}$;
\item If $m,n \in \mathbb Z$ then  $W_{mn} \cap \LL^{d^\perp} \subset \LL^{(d^\perp,0)}$;
\item If $m,n \in \mathbb Z+\frac12$ then  $W_{mn} \cap \LL^{d^\perp} \subset \LL^{(d^\perp, \frac 12)}$;
\end{enumerate}
where $\LL^{d} = \LL^{(d, +)}\oplus  \LL^{(d, -)}$ with $\LL^{(d, +)}$ (resp. $\LL^{(d, -)}$) corresponding
to the discrete series bounded from below (resp. above), and $\LL^{d^\perp}= \LL^{(d^\perp,0)}
\oplus \LL^{(d^\perp,\frac12)}$ with $\LL^{(d^\perp,0)}$ (resp. $\LL^{(d^\perp,\frac 12)}$) corresponding to the
part corresponding to the principal continuous series of bosonic (resp. fermionic) nature.\\

The construction above leads to
\beqa
\label{eq:cc3}
\overline{\Phi^{m,n,k}}(\rho, \varphi_1,\varphi_2)=\Phi_{-m,-n,k}(\rho, \varphi_1,\varphi_2) \ ,
\eeqa
and to the completeness relation
\beqa
\label{eq:delL}
\delta(\cosh 2 \rho -\cosh 2 \rho') \delta(\varphi_1-\varphi_1') \delta(\varphi_2-\varphi_2')
=\sum \limits_{\epsilon=0,\frac12} \sum \limits _{m,n \in \mathbb Z + \epsilon} \sum \limits_{k\ge0}
\overline{\Phi^{m,n,k}}(\rho, \varphi_1,\varphi_2)\Phi_{m,n,k}(\rho', \varphi'_1,\varphi'_2)\nn\\
\eeqa

We introduce the following notations
\beqa
\Phi_{m,n,k}^d(\rho,\varphi_1,\varphi_2) &=& e^{i(m+n)\varphi_1+i(m-n)\varphi_2} e^d_{m,n,k}(\cosh 2 \rho) \in \LL^d\nn\\
\Phi_{m,n,k}^p(\rho,\varphi_1,\varphi_2) &=& e^{i(m+n)\varphi_1+i(m-n)\varphi_2} e^p_{m,n,k}(\cosh 2 \rho) \in \LL^{d^\perp}\nn\
\eeqa
when we want to distinguish basic elements of the discrete series from the  (principal) continuous series.

Let
\beqa
\psi^{\epsilon}_{n i\sigma m}(\rho,\varphi_1,\varphi_2)= e^{i(m+n)\varphi_1+i(m-n)\varphi_2}  \chi^{\epsilon}_{n i\sigma m}(\rho)\nn\ ,
\eeqa
be
 the matrix elements of the (principal) continuous series \eqref{eq:matC}.

The asymptotic  (\textit{i.e.}, for $\rho \rightarrow \infty $) behaviour of the functions $e_{m,n,k}$  for the part corresponding to the discrete series $e^d_{m,n,k}$ (recall that $mn>0, |m|,|n|>1/2$ and  $k= \text{min}(|m|,|n|) -\lambda$) and for the part orthogonal to the discrete series  $e^p_{m,n,k}$ is
\beqa
\label{eq:asymp}
e^d_{m,n,k}(x) &\sim& C_{mnk} \;x^{-\text{min}(|m|,|n|) +k} \ , \ \ 0\le k < \text{min}(|m|,|n|) -\epsilon\nn\\
 &\sim&  C_{mnk} \;x^{-\lambda}\ , \hskip 2.1truecm \lambda\ge \epsilon+1\nn\\
  &\precsim&   C_{mnk}\; x^{-\epsilon -1}\\
  e^{ p}_{m,n,k}(x) &\sim& C_{mnk}  x^{-\epsilon-1}\nn
  \eeqa
where $x = \cosh 2 \rho\sim + \infty$, and $C_{mnk}$ is a suitable constant which can be deduced from \eqref{eq:ed} or  \eqref{eq:e} (see also below).
Since   $x^{-\epsilon -1} \sim e^{-2(1+\epsilon)\rho}, \rho>0$, $e^d_{m,n,k}$ are rapidly decreasing and thus  belong to the Schwartz space \cite{Va}. \footnote{
The principal continuous series are oscillatory in the $\sigma-$direction
around $e^{-\rho}$ and thus are not square-integrable \cite{bar}. The oscillatory character  means that
one can construct wave packets from continuous principal series which are in the Schwartz space \cite{Va}. Differently, the matrix elements of the trivial representation and of the supplementary series
are also bounded \cite{bar} but cannot be used to construct functions in the Schwartz space.
This legitimates in some sense the Plancherel formula \eqref{eq:delP}.}
 Similarly $e^p_{m,n,k}$ also belong to the Schwartz space. Observe also that we have asymptotically $\rho \sim + \infty$:
\beqa
\label{eq:asym2}
e_{m,n,k}(x)e_{m',n',k'}(x) &\precsim&  C_{mnk}C_{m^{\prime }n^{\prime }k^{\prime }}\;  x^{-2\epsilon-2}
\eeqa
thus the product of two elements of the Losert basis is a Schwartz function. This observation
will be crucial for the computation of the Clebsch-Gordan coefficients in Sec.
\ref{sec:CG} or for the definition of the infinite dimensional algebra associated to SL$(2,\mathbb R)$ in Sec.
\ref{sec:infa}.

Applying the Plancherel theorem we have
\beqa
\label{eq:PL}
e_{m,n,k}^p(\cosh 2 \rho) = \int \limits_0^{+\infty} \text{d} \sigma \; \sigma\tanh \pi(\sigma+i\epsilon) f_{mnk}(\sigma) \chi^{\epsilon}_{n i\sigma m}(\rho)\ , \ \ k \ge k_{\text{min}}
\eeqa
and using \eqref{eq:PC}, we have (with the scalar product \eqref{eq:sp} restricted to $\rho$)
\beqa
f_{mnk}(\sigma)= (\chi^\epsilon_{ni\sigma m},e^p_{mnk}) \ .\nn
\eeqa
Thus relation \eqref{eq:PL} can be inverted
\beqa
\label{eq:LP}
\chi^\epsilon_{n i \sigma m}(\rho) = \sum \limits_{k\ge k_{\text{min}}} \overline{ f^{mnk}}(\sigma)e^p_{m,n,k}(\cosh 2\rho) \ ,
\eeqa
where $k_{\text{min}}$ is defined in \eqref{eq:kmin}.
 We are thus able
to express the matrix elements of the  (principal) continuous series in terms of the Losert basis, and conversely.\\

\subsection{Action of the $\mathfrak{sl}(2,\mathbb R)-$generators on $\LL^{d^\perp}$}
 The  $\mathfrak{sl}(2,\mathbb R)$ generators act on $\LL$, thus $\LL$ itself is a representation of $\SL$. Furthermore,  by the action of $\mathfrak{sl}(2,\mathbb R)$, $\LL^d$ (which corresponds to the matrix elements of the two discrete series, bounded from below and from
 above) transforms onto itself, implying that this space is a sub-representation, {\it i.e.}, an invariant subspace of $\LL$. As a consequence,
 $\LL^{d^\perp}$ is also a representation of $\SL$.
To analyse the action of the $\mathfrak{sl}(2,\mathbb R)$  generators on $\LL^{d^\perp}$, we start and
express all basis elements of the $\LL^{d^\perp}-$part in terms of hypergeometric functions using \eqref{eq:HJ} to get
\beqa
\label{eq:Cmnk}
\Phi_{mnk}(\rho,\varphi_1,\varphi_2)=
  C_{mnk}\; e^{i(m+n) \varphi_1 + i (m-n) \varphi_2}
\cosh\rho^{-|m-n| -2k-2\epsilon -2}\sinh \rho^{|n-m|}\times\\\
\hskip 1.truecm {}_2 F_1(-k,-k -2\epsilon;|n-m|+1; -\sinh^2\rho)\nn
\eeqa
where
\beqa
C_{mnk} = \frac1 {|m-n|!} \sqrt{\frac 12 \frac{(k+|m-n|)!(k+|m-n|+2\epsilon)!
(2k+|m-n|+2\epsilon+1)}
{k! (k+2\epsilon)!}}\nn \ .
\eeqa
Then, the left action
of $\mathfrak{sl}(2,\mathbb R)$ \eqref{eq:L} leads to
\beqa
&&\hskip 1.5truecm L_+(\Phi_{m,n,k})= C_{mnk} e^{i(m+n+1)\varphi_1 + i(m-n-1) \varphi_2} \cosh \rho^{n-m-2\epsilon -2k-1}\sinh\rho^{m-n-1}\times\nn\\
&&\Bigg[\Big(-(k+n+\epsilon+1) + (k+m+\epsilon+1) \frac 1{\cosh\rho^2}\Big)
   {}_2 F_1(-k,-k-2\epsilon;m-n+1;-\sinh^2\rho)\nn\\
         &&\hskip .4truecm -\frac{k(k+2\epsilon)}{m-n+1} \sinh^2\rho {}_2 F_1(-k+1,-k-2\epsilon+1;m-n+2;-\sinh^2\rho)\Bigg]\ , m \ge n\nn\\\nn\\
&&\hskip 1.5truecm L_+(\Phi_{m,n,k})= C_{nmk}  e^{i(m+n+1)\varphi_1 + i(m-n-1) \varphi_2} \cosh \rho^{m-n-2\epsilon -2k-1}\sinh\rho^{n-m-1}\times\nn\\
  &&\Bigg[-(\epsilon+k+n+1)\frac{\sinh^2\rho}{\cosh^2\rho} {}_2 F_1(-k,-k-2\epsilon;n-m+1;-\sinh^2\rho)\nn\\
  &&\hskip .4truecm -\frac{k(k+2\epsilon)}{n-m+1} \sinh^2\rho {}_2 F_1(-k+1,-k-2\epsilon+1;n-m+2;-\sinh^2\rho)\Bigg]\ , n \ge m\nn\\ \nn
\eeqa

and
\beqa
&&\hskip 1.5truecm L_-(\Phi_{m,n,k})= C_{mnk} e^{i(m+n-1)\varphi_1 + i(m-n+1) \varphi_2} \cosh \rho^{n-m-2\epsilon-2k -3}\sinh\rho^{m-n+1}\times\nn\\
  &&\Bigg[(k-n+\epsilon+1)  {}_2 F_1(-k,-k-2\epsilon;m-n+1;-\sinh^2\rho)\nn\\
  &&\hskip .4truecm +\frac{k(k+2\epsilon)}{m-n+1} \cosh^2\rho {}_2 F_1(-k+1,-k-2\epsilon+1;m-n+2;-\sinh^2\rho)\Bigg]\ , m \ge n\nn\\
&& \hskip 1.5truecm L_-(\Phi_{m,n,k})= C_{nmk} e^{i(m+n-1)\varphi_1 + i(m-n+1) \varphi_2} \cosh \rho^{m-n-2\epsilon-2k -3}\sinh\rho^{n-m+1}\times\nn\\
&&\Bigg[\Big( -\frac{(\epsilon +k-m+1)}{\sinh^2 \rho}+ (\epsilon+k-n+1)\frac{\cosh^2 \rho}{\sinh^2 \rho}\Big)
   {}_2 F_1(-k,-k-2\epsilon;n-m+1;-\sinh^2\rho)\nn\\
  &&\hskip .4truecm +\frac{k(k+2\epsilon)}{n-m+1} \cosh^2\rho {}_2 F_1(-k+1,-k-2\epsilon+1;n-m+2;-\sinh^2\rho)\Bigg]\ , n \ge M\nn
\eeqa
and  the the right action
of $\mathfrak{sl}(2,\mathbb R)$ of \eqref{eq:R}  reduces to
\beqa
&&\hskip 1.5truecm R_+(\Phi_{m,n,k})= C_{mnk} e^{i(m+n+1)\varphi_1 + i(m-n+1) \varphi_2} \cosh \rho^{n-m-2\epsilon-2k -3}\sinh\rho^{m-n+1}\times\nn\\
  &&\Bigg[(k+m+\epsilon+1)  {}_2 F_1(-k,-k-2\epsilon;m-n+1;-\sinh^2\rho)\nn\\
  &&\hskip .4truecm +\frac{k(k+2\epsilon)}{m-n+1} \cosh^2\rho {}_2 F_1(-k+1,-k-2\epsilon+1;m-n+2;-\sinh^2\rho)\Bigg]\ , m \ge n\nn\\\nn\\
&& \hskip 1.5truecm R_+(\Phi_{m,n,k})= C_{nmk} e^{i(m+n+1)\varphi_1 + i(m-n+1) \varphi_2} \cosh \rho^{m-n-2\epsilon -2k-3}\sinh\rho^{n-m+1}\times\nn\\
&&\Bigg[\Big((k+m+\epsilon+1)\frac{\cosh^2 \rho}{\sinh^2\rho} -\frac{(k+n+1 +\epsilon)}{\sinh^2\rho}
\Big)
 {}_2 F_1(-k,-k-2\epsilon;n-m+1;-\sinh^2\rho)\nn\\
  &&\hskip .4truecm +\frac{k(k+2\epsilon)}{n-m+1} \cosh^2\rho {}_2 F_1(-k+1,-k-2\epsilon+1;n-m+2;-\sinh^2\rho)\Bigg]\ , n \ge m\nn
\eeqa
and
\beqa
&& \hskip 1.5truecm R_-(\Phi_{m,n,k})= C_{mnk} e^{i(m+n-1)\varphi_1 + i(m-n-1) \varphi_2} \cosh \rho^{n-m-2\epsilon-2k -1}\sinh\rho^{m-n-1}\times\nn\\
&&\Bigg[\Big(-(k-m+\epsilon+1)
  +(\epsilon+k-n+1)\frac 1 {\cosh^2 \rho}
  {}_2 F_1(-k,-k-2\epsilon;m-n+1;-\sinh^2\rho)\nn\\
  &&\hskip .4truecm -\frac{k(k+2\epsilon)}{m-n+1} \sinh^2\rho {}_2 F_1(-k+1,-k-2\epsilon+1;m-n+2;-\sinh^2\rho)\Bigg]\ , m \ge n\nn\\\nn\\
&& \hskip 1.5truecm R_-(\Phi_{m,n,k})= C_{nmk} e^{i(m+n-1)\varphi_1 + i(m-n-1) \varphi_2} \cosh \rho^{m-n-2\epsilon -2k-1}\sinh\rho^{n-m-1}\times\nn\\
  &&\Bigg[-(k-m+\epsilon+1)\frac {\sinh^2 \rho}{\cosh^2 \rho}    {}_2 F_1(-k,-k-2\epsilon;n-m+1;-\sinh^2\rho)\nn\\
  &&\hskip .4truecm -\frac{k(k+2\epsilon)}{n-m+1} \sinh^2\rho {}_2 F_1(-k+1,-k-2\epsilon+1;n-m+2;-\sinh^2\rho)\Bigg]\ , n \ge m\nn
\eeqa
In the expression above all RHS's belong to $W_{mn+1}$ (resp. $W_{mn-1}, W_{m+1n}$ or  $W_{m-1n}$). Thus all RHS's can be
expended in the basis $\{\Phi_{m,n+1,,k}, k\ge k_{\text{min}}\}$ (resp. $\{\Phi_{m,n-1,k}, k\ge k_{\text{min}}\}$,
$\{\Phi_{m+1,n,k}, k\ge k_{\text{min}}\}$ or $\{\Phi_{m-1,n,k}, k\ge k_{\text{min}}\}$).
Analysing the action of the $\mathfrak{sl}(2,\mathbb R)$ generators, one can guess the following
expressions for the left action
\beqa
\label{eq:c1}
L_0(\Phi_{m,n,k})&=&n\Phi_{m,n,k}\nn\\
 L_+(\Phi_{m,n,k})&=&
\left\{
\begin{array}{ll}
 \alpha_{1L}{}_{m,n,k} \Phi_{m,n+1,k+1} + \beta_{1L}{}_{m,n,k}  \Phi_{m,n+1,k}& m>n\\
 \alpha_{2L} {}_{m,n,k}\Phi_{m,n+1,k-1} + \beta_{2L}{}_{m,n,k}  \Phi_{m,n+1,k}& n\ge m
 \end{array}\right.\nn\\\\
L_-(\Phi_{m,n,k})&=&
\left\{
\begin{array}{ll}
 \gamma_{1L}{}_{m,n,k} \Phi_{m,n-1,k-1} + \delta_{1L}{}_{m,n,k}  \Phi_{m,n-1,k}& m\ge n\\
 \gamma_{2L}{}_{m,n,k} \Phi_{m,n-1,k+1} + \delta_{2L}{}_{m,n,k}  \Phi_{m,n-1,k}& n> m
 \end{array}\right.\nn
\eeqa
and for the right action
\beqa
\label{eq:c2}
R_0(\Phi_{m,n,k})&=&m\Phi_{m,n,k}\nn\\
 R_+(\Phi_{m,n,k})&=&
\left\{
\begin{array}{ll}
 \alpha_{1R}{}_{m,n,k} \Phi_{m+1,n,k-1} + \beta_{1R}{}_{m,n,k}  \Phi_{m+1,n,k}& m \ge n\\
 \alpha_{2R}{}_{m,n,k} \Phi_{m+1,n,k+1} + \beta_{2R}{}_{m,n,k}  \Phi_{m+1,n,k}& n> m
 \end{array}\right.\nn\\ \\
R_-(\Phi_{m,n,k})&=&
\left\{
\begin{array}{ll}
 \gamma_{1R}{}_{m,n,k} \Phi_{m-1,n,k+1} + \delta_{1R}{}_{m,n,k}  \Phi_{m-1,n,k}& m> n\\
 \gamma_{2R} {}_{m,n,k}\Phi_{m-1,n,k-1} + \delta_{2R}{}_{m,n,k}  \Phi_{m-1,n,k}& n\ge m
 \end{array}\right.\nn
\eeqa
Then multiplying the RHS by an appropriate
power of $\cosh \rho$ and of $\sinh \rho$, the equations \eqref{eq:c1} and \eqref{eq:c2} reduce to polynomial identities. It is then
straightforward but tedious to identify all the coefficients $\alpha,\beta,\gamma,\delta$:

\beqa
\label{eq:ccc1}
L_0(\Phi_{m,n,k})&=&n\Phi_{m,n,k}\nn\\
 L_+(\Phi_{m,n,k})&=&
\left\{
\begin{array}{ll}
\phantom{+} \sqrt {{\frac { \left( \epsilon+k+m+1 \right) ^{2} \left( k+1 \right)
 \left( k+2\,\epsilon+1 \right) }{ \left( 2\,k+m-n+2\,\epsilon+1
       \right) \; \left( 2\,k+m-n+2\,\epsilon+2 \right) }}} \Phi_{m,n+1,k+1}&
\multirow{2}{*}{$m>n$}
 \nn\\
 + \sqrt {{\frac { \left( \epsilon+k-n \right) ^{2} \left( m-n+k \right)
 \left( k+m-n+2\,\epsilon \right) }{ \left( 2\,k+m-n+2\,\epsilon+1
 \right)  \left( -n+m+2\,k+2\,\epsilon \right) }}}\;
  \Phi_{m,n+1,k-1}&\\
-\sqrt {{\frac { \left( k+2\,\epsilon \right)  \left( \epsilon+k-m
 \right) ^{2}k}{ \left( 2\,k+n-m+2\,\epsilon+1 \right)  \left( -m+n+2
      \,k+2\,\epsilon \right) }}} \;\Phi_{m,n+1,k-1} &
\multirow{2}{*}{$n\ge m$}
\nn\\
 -\sqrt {{\frac { \left( k+1-m+n \right)  \left( \epsilon+k+n+1 \right)
^{2} \left( k+2\,\epsilon+1-m+n \right) }{ \left( 2\,k+n-m+2\,\epsilon
+1 \right)  \left( -m+n+2\,k+2\,\epsilon+2 \right) }}} \; \Phi_{m,n+1,k}&
 \end{array}\right.\nn\\\\
L_-(\Phi_{m,n,k})&=&
\left\{
\begin{array}{ll}
\phantom{+} \sqrt {{\frac { \left( k+2\,\epsilon \right)  \left( \epsilon+k+m
 \right) ^{2}k}{ \left( 2\,k+m-n+2\,\epsilon+1 \right)  \left( -n+m+2
       \,k+2\,\epsilon \right) }}} \;\Phi_{m,n-1,k-1}&
\multirow{2}{*}{$m\ge n$}\\
 + \sqrt {{\frac { \left( m-n+k+1 \right)  \left( k-n+\epsilon+1 \right)
^{2} \left( k+m-n+2\,\epsilon+1 \right) }{ \left( 2\,k+m-n+2\,\epsilon
+1 \right)  \left( 2\,k+m-n+2\,\epsilon+2 \right) }}}  \Phi_{m,n-1,k}&\\
  -\sqrt {{\frac { \left( -m+\epsilon+k+1 \right) ^{2} \left( k+1
 \right)  \left( k+2\,\epsilon+1 \right) }{ \left( 2\,k+n-m+2\,
        \epsilon+1 \right)  \left( -m+n+2\,k+2\,\epsilon+2 \right) }}}\Phi_{m,n-1,k+1}
&\multirow{2}{*}{$n>m$}\\
  -\sqrt {{\frac { \left( -m+\epsilon+k+1 \right) ^{2} \left( k+1
 \right)  \left( k+2\,\epsilon+1 \right) }{ \left( 2\,k+n-m+2\,
\epsilon+1 \right)  \left( -m+n+2\,k+2\,\epsilon+2 \right) }}}  \; \Phi_{m,n-1,k}&
 \end{array}\right.\nn
\eeqa
and
\beqa
\label{eq:ccc2}
R_0(\Phi_{m,n,k})&=&m\Phi_{m,n,k}\nn\\
 R_+(\Phi_{m,n,k})&=&
\left\{
\begin{array}{ll}
 \phantom{+} \sqrt {{\frac { \left( k+2\,\epsilon \right)  \left( \epsilon+k-n
 \right) ^{2}k}{ \left( 2\,k+m-n+2\,\epsilon+1 \right)  \left( -n+m+2
\,k+2\,\epsilon \right) }}}
  \;\Phi_{m+1,n,k-1}&\multirow{2}{*}{$ m \ge n$} \nn\\
+ \sqrt {{\frac { \left( m-n+k+1 \right)  \left( \epsilon+k+m+1 \right)
^{2} \left( k+m-n+2\,\epsilon+1 \right) }{ \left( 2\,k+m-n+2\,\epsilon
+1 \right)  \left( 2\,k+m-n+2\,\epsilon+2 \right) }}}  \; \Phi_{m  +1,n,k}&
 \\
 -\sqrt {{\frac { \left( \epsilon+k+n+1 \right) ^{2} \left( k+1 \right)
 \left( k+2\,\epsilon+1 \right) }{ \left( 2\,k+n-m+2\,\epsilon+1
       \right)  \left( -m+n+2\,k+2\,\epsilon+2 \right) }}}
\; \Phi_{m+1,n,k+1}&\multirow{2}{*}{$n> m$}\nn\\
 -\sqrt {{\frac { \left( \epsilon+k-m \right) ^{2} \left( n-m+k \right)
 \left( k+2\,\epsilon-m+n \right) }{ \left( 2\,k+n-m+2\,\epsilon+1
 \right)  \left( -m+n+2\,k+2\,\epsilon \right) }}}  \;\Phi_{m  +1,n,k}&
 \end{array}\right.\nn\\ \\
R_-(\Phi_{m,n,k})&=&
\left\{
\begin{array}{ll}
\phantom{+}\sqrt {{\frac { \left( k-n+\epsilon+1 \right) ^{2} \left( k+1 \right)
 \left( k+2\,\epsilon+1 \right) }{ \left( 2\,k+m-n+2\,\epsilon+1
      \right)  \left( 2\,k+m-n+2\,\epsilon+2 \right) }}} \; \Phi_{m-1,n,k+1}
 &\multirow{2}{*}{$m > n$}
\nn\\
+ \sqrt {{\frac { \left( \epsilon+k+m \right) ^{2} \left( m-n+k \right)
 \left( k+m-n+2\,\epsilon \right) }{ \left( 2\,k+m-n+2\,\epsilon+1
 \right)  \left( -n+m+2\,k+2\,\epsilon \right) }}} \;  \Phi_{m  -1,n,k}&\nn\\
-\sqrt {{\frac { \left( k+2\,\epsilon \right)  \left( \epsilon+k+n
 \right) ^{2}k}{ \left( 2\,k+n-m+2\,\epsilon+1 \right)  \left( -m+n+2
      \,k+2\,\epsilon \right) }}} \;\Phi_{m-1,n,k-1} &
\multirow{2}{*}{$n\ge m$}
\nn\\
 -\sqrt {{\frac { \left( k+1-m+n \right)  \left( -m+\epsilon+k+1
 \right) ^{2} \left( k+2\,\epsilon+1-m+n \right) }{ \left( 2\,k+n-m+2
\,\epsilon+1 \right)  \left( -m+n+2\,k+2\,\epsilon+2 \right) }}}\; \Phi_{m -1,n,k}&
 \end{array}\right.\nn
\eeqa
Note however that in order to simplify the computation, we have firstly computed the action of $L_\pm, R_\pm$ on unormalised
fuctions ({\it i.e}, without the normalisation factor $C_{mnk}$ in \eqref{eq:Cmnk})  and then properly rescaled the obtained coefficients.

The coefficients $\alpha,\beta,\gamma,\delta$ share many symmetry properties.
The symmetry between left and right actions reduces to:
\beqa
&&\alpha_{1L}{}_{m,n,k}=-\alpha_{2R}{}_{n,m,k}\ ,\ \  \alpha_{2L}{}_{m,n,k}= -\alpha_{1R}{}_{n,m,k}\ ,\nn\\
&&\beta_{1L}{}_{m,n,k}=-\beta_{2R}{}_{n,m,k}\ ,\ \
\beta_{2L}{}_{m,n,k}=-\beta_{1R}{}_{n,m,k}\ , \nn\\
&& \gamma_{1L}{}_{m,n,k}=-\gamma_{2R}{}_{n,m,k}\ ,\ \  \gamma_{2L}{}_{m,n,k}=-\gamma_{1R}{}_{n,m,k}\ ,\nn\\
&&\delta_{1L}{}_{m,n,k}= -\delta_{2R}{}_{n,m,k}\ ,\ \  \delta_{2L}{}_{m,n,k}=-\delta_{1R}{}_{n,m,k}\ .
\nn
\eeqa
This shows the isomorphism between left and right actions. We also have the symmetry $(m,n)\to (-m,-n)$:
\beqa
&&\alpha_{1L}{}_{m,n,k}=-\gamma_{2L}{}_{-m,-n,k} \ ,\ \  \beta_{1L}{}_{m,n,k}= -\delta_{2L}{}_{-m,-n,k}\ ,\nn\\
&&\alpha_{2L}{}_{m,n,k}=-\gamma_{1L}{}_{-m,-n,k}\ ,\ \
\beta_{2L}{}_{m,n,k}=-\delta_{1L}{}_{-m,-n,k}\nn\\
&&\alpha_{1R}{}_{m,n,k}=-\gamma_{2R}{}_{-m,-n,k} \ ,\ \  \beta_{1R}{}_{m,n,k}= -\delta_{2R}{}_{-m,-n,k}\ ,\nn\\
&&\alpha_{2R}{}_{m,n,k}=-\gamma_{1R}{}_{-m,-n,k}\ ,\ \
\beta_{2R}{}_{m,n,k}=-\delta_{1R}{}_{-m,-n,k}\nn\ .
\eeqa
 Finally, we observe
  that  the fact that the representation is Hermitean, implies $L_+^\dag= L_-$ and
  $R_+^\dag= R_-$ thus we have
\beqa
&&\alpha_{1L}{}_{m,n,k}= \gamma_{1L}{}_{m,n+1,k+1}\ , \ \  \beta_{1L}{}_{m,n,k}=  \delta_{1L}{}_{m,n+1,k}\ ,
\nn\\
&&\alpha_{2L}{}_{m,n,k}= \gamma_{2L}{}_{m,n+1,k-1}\ ,\ \  \beta_{2L}{}_{m,n,k}= \delta_{2L}{}_{m,n+1,k}\nn\\
&&\alpha_{1R}{}_{m,n,k}= \gamma_{1R}{}_{m+1,n,k-1}\ ,\ \   \beta_{1R}{}_{m,n,k}=  \delta_{1R}{}_{m+1,n,k}\ ,
\nn\\
&&\alpha_{2R}{}_{m,n,k}= \gamma_{2R}{}_{m+1,n,k+1}\ ,\ \  \beta_{2R}{}_{m,n,k}= \delta_{2R}{}_{m+1,n,k}\nn
\eeqa
  It is worthy to be observed that the coefficients are such that under the action
  of $L_\pm$ or $R_\pm$ we always have $k\ge k_{\text{min}}$. Indeed, for $m>n >1/2$ we have
$
L_+ \Phi_{m,n,n-\epsilon} = \alpha_{1L}  \Phi_{m,n+1,n+1-\epsilon}
$
since in this case, the coefficient $\beta_{1L}$ vanishes as it should. Similar relations hold in the other cases.
This implies
in particular that if $\Phi_{m,n,k}$ belongs to $\LL^{d^\perp}$ then
$L_\pm(\Phi_{m,n,k}), R_\pm(\Phi_{m,n,k})$ belong  to $\LL^{d^\perp}$, as well.

The  action of the Casimir operator on $\Phi_{m,n,k}$ can be easily
deduced from Eqs.[\ref{eq:ccc1}-\ref{eq:ccc2}]
to give
\beqa
Q \Phi_{mnk} &=&  -\frac { \left( k-n+\epsilon+1 \right)  \left( \epsilon+k+m+1
 \right) }{2\,k+m-n+2\,\epsilon+2} \times \nn\\
 && \sqrt {{\frac { \left( m-n+k+1
 \right)  \left( k+m-n+2\,\epsilon+1 \right)  \left( k+1 \right)
 \left( k+2\,\epsilon+1 \right) }{ \left( 2\,k+m-n+2\,\epsilon+1
      \right)  \left( -n+m+2\,k+2\,\epsilon+3 \right) }}}
\; \Phi_{mnk+1} \nn\\
&&\Bigg(n(n+1)-{\frac { \left( \epsilon+k+m+1 \right) ^{2} \left( k+1 \right)
 \left( k+2\,\epsilon+1 \right) }{ \left( 2\,k+m-n+2\,\epsilon+1
 \right)  \left( 2\,k+m-n+2\,\epsilon+2 \right) }}\nn\\
 &&\hskip 1.7 truecm -\;{\frac { \left( \epsilon+k-n \right) ^{2} \left( m-n+k \right)
 \left( k+m-n+2\,\epsilon \right) }{ \left( 2\,k+m-n+2\,\epsilon+1
 \right)  \left( -n+m+2\,k+2\,\epsilon \right) }}\Bigg)
\Phi_{mnk}\nn\\
&&
 -\frac { \left( \epsilon+k+m \right)  \left( \epsilon+k-n \right) }{-
     n+m+2\,k+2\,\epsilon}\times\nn\\
   &&\sqrt {{\frac { \left( k+2\,\epsilon \right) k
 \left( m-n+k \right)  \left( k+m-n+2\,\epsilon \right) }{ \left( 2\,k
+m-n+2\,\epsilon+1 \right)  \left( -n+m+2\,k+2\,\epsilon-1 \right) }}}
\;
\Phi_{mnk-1}\ ,\nn
\eeqa
 The action above of the generators      of
the left and right action of $\mathfrak{sl}(2,\mathbb R)$ on  $\LL^{d^\perp}$
clearly shows that this representation
is not irreducible. Furthermore, none of the elements of  $\LL^{d^\perp}$  can be identified with an eigenvector of the Casimir operator.
This is in accordance with  Section \ref{sec:H1},  where we have shown that the only normalisable simultaneous eigenvectors of $L_0, R_0$ and $Q$ are the matrix elements of the discrete series.\\

\section{The Clebsch-Gordan coefficients} \label{sec:CG} 

In \cite{hb1}, Holman and Biedenharn studied the coupling of two discrete series
${\cal D}^+_{\lambda_1} \otimes {\cal D}_{\lambda_2}^\pm$ and asked whether this tensor
product reproduces the representation ${\cal D}_\Lambda$,   where $\Lambda$ is either equal to
$\Lambda=\lambda\in \mathbb N \setminus\{0\}  + \epsilon$  with $\epsilon=0,1/2$ (discrete series),
 to $\Lambda=\frac 12 +i\sigma, \sigma>0$ (principal continuous series), or  to $\Lambda=\frac 12 +\sigma$, $0<\sigma<1/2$) (supplementary continuous series), and such that the eigenvalues of the Casimir operator are given by $q=\Lambda(\Lambda-1)$.
 The fact that one obtains the continuous series from the product ${\cal D}^+_{\lambda_1} \otimes {\cal D}_{\lambda_2}^-$ seems at  first glance surprising. However, if we remember that the matrix elements
 $\psi^+_{n_1\lambda_1m_1} \in {\cal D}_{\lambda_1}^+$ (resp. $\psi^-_{n_2\lambda_2m_2} \in {\cal D}_{\lambda_2}^-$)
 are such that $m_1,n_1 \ge \lambda_1$ (resp. $m_1,n_1 \le -\lambda_2)$, it conceivable to have
 $(m_1 +n_1)(m_2+n_2)<0$, and thus in this case $\psi^+_{n_1\lambda_1m_1}\psi^-_{n_2\lambda_2m_2}$ decomposes
 into continuous series.
Thus, one can write for $M\ge \Lambda$
\beqa
\label{eq:C}
\big|\Lambda,M\big>= \sum_{m_1\ge \lambda_1}
\scriptsize
\begin{pmatrix}\lambda_1&\lambda_2&\Lambda\\
m_1&M-m_1&M \end{pmatrix}
 \mid \lambda_1,+,m_1\rangle \otimes \mid\lambda_2,\pm, M-m_1\rangle
\eeqa
where  $\begin{pmatrix}\lambda_1&\lambda_2&\Lambda\\
m_1&M-m_1&M \end{pmatrix}$ are the Clebsch-Gordan coefficients, and $\big|\lambda_1,+,m_1\big> \in {\cal D}^+_{\lambda_1}$ and  $\big|\lambda_2,\pm,M-m_1\big> \in {\cal D}^\pm_{\lambda_2}$. The existence of solutions to Eq. [\ref{eq:C}] for a given value of $\Lambda$ is proved using a recursion relation for the Clebsch-Gordan coefficients derived from
 a realisation of the discrete series by means of two commuting harmonic oscillators \cite{hb1}. The case where $\Lambda$ corresponds to a continuous representation is studied by analytic continuation.
In \cite{hb2}, this analysis was extended to study the coupling of the discrete series with the continuous series and the coupling of two continuous series. The main results are the following:

\begin{enumerate} [noitemsep]
\item Product of two series bounded from below:
  \beqa
  \label{eq:++}
      {\cal D}^+_{\lambda_1} \otimes {\cal D}^+_{\lambda_2}
      =\bigoplus\limits_{\lambda\ge \lambda_1+\lambda_2} {\cal D}_\lambda^+ \ .
      \eeqa
    \item Product of two series bounded from above:
 \beqa
 \label{eq:--}
      {\cal D}^-_{\lambda_1} \otimes {\cal D}^-_{\lambda_2}
      =\bigoplus\limits_{\lambda\ge \lambda_1+\lambda_2} {\cal D}_\lambda^- \ .
      \eeqa
    \item Product of a series bounded from  below with a series bounded from above:
      \begin{enumerate}
      \item If $\lambda_1>\lambda_2$,
\beqa
\label{eq:+-1}
    {\cal D}^+_{\lambda_1} \otimes {\cal D}^-_{\lambda_2} \supset
    \bigoplus\limits_{0<\lambda\le \lambda_1-\lambda_2} {\cal D}_\lambda^+ \ .
    \eeqa
  \item     If $\lambda_2>\lambda_1$,
\beqa
\label{eq:+-2}
    {\cal D}^+_{\lambda_1} \otimes {\cal D}^-_{\lambda_2} \supset
    \bigoplus\limits_{0<\lambda\le \lambda_2-\lambda_1} {\cal D}_\lambda^- \ .
    \eeqa
  \item If $\lambda_1=\lambda_2$,
 \beqa
    {\cal D}^+_{\lambda_1} \otimes {\cal D}^-_{\lambda_1} \supset
     {\cal D}_0,\nn
     \eeqa
 where $ {\cal D}_0$ is  the trivial representation.
\item   For any value $q=-\frac 14 -\sigma^2$ of the Casimir operator corresponding to a principal continuous series we have
  \beqa
  \label{eq:+-3}
 {\cal D}^+_{\lambda_1} \otimes {\cal D}^-_{\lambda_2} \supset {\cal C}^{i\sigma,\epsilon}
 \eeqa
 where $\epsilon=0$ if $\lambda_1,\lambda_2$ are both integers or half-integers, and
 $\epsilon=1/2$ if one is an integer, whilst the other is half-integer.
\item   For any value $q=-\frac 14 +\sigma^2$ of the Casimir operator corresponding to a supplementary continuous series we have
  \beqa
 {\cal D}^+_{\lambda_1} \otimes {\cal D}^-_{\lambda_2} \supset {\cal C}^{\sigma} \nn
 \eeqa
 where obviously $\lambda_1,\lambda_2$ are both integers or half-integers  (in fact, only bosonic supplementary continuous series exist).
  \end{enumerate}

\item Product of a discrete series and a principal continuous series ${\cal D}^\pm_\lambda \otimes {\cal C}^{i \sigma,\epsilon}$:
\begin{enumerate}
\item For any  $\lambda' \in \mathbb N \setminus\{0\}$,
  \beqa
  \label{eq:DC1}
      {\cal D}^\pm_\lambda \otimes {\cal C}^{i \sigma,\epsilon}\supset{\cal D}^+_{\lambda'}\nn\\
       {\cal D}^\pm_\lambda \otimes {\cal C}^{i \sigma,\epsilon}\supset{\cal D}^-_{\lambda'}
  \eeqa
\item For any $\sigma'>0$,
  \beqa
   \label{eq:DC2}
{\cal D}^\pm_\lambda \otimes {\cal C}^{i \sigma,\epsilon}\supset  {\cal C}^{i \sigma',\epsilon'}
\eeqa
\end{enumerate}
where both sides are bosons or fermions.
\item Product of two continuous principal series
  \begin{enumerate}
  \item For any values of $\lambda \in \mathbb N \setminus\{0\}$,
    \beqa
     \label{eq:CC1}
{\cal C}^{i\sigma,\epsilon} \otimes {\cal C}^{i\sigma',\epsilon'} \supset {\cal D}^\pm_\lambda \ .
\eeqa
\item For any values of $\sigma''>0$,
  \beqa
  \label{eq:CC2}
{\cal C}^{i\sigma,\epsilon} \otimes {\cal C}^{i\sigma',\epsilon'} \supset 2 {\cal C}^{ i \sigma'',[\epsilon+\epsilon']} \ ,
\eeqa
(with $[1/2+1/2]=0$ and $[0+\epsilon]=\epsilon$).
 Since the recursion formul\ae \ for the Clebsch-Gordan coefficients lead
 to two independent solutions denoted by $\Big( \hskip 1.truecm \Big)_1$ and $\Big( \hskip 1.truecm \Big)_2$, the
 representation ${\cal C}^{i \sigma'',[\epsilon+\epsilon'] }$ appears with multiplicity two.
    \end{enumerate}
In both cases,  an expression for the Clebsch-Gordan coefficients is given in \cite{hb2}.

\end{enumerate}

 In this analysis we are not considering any factor of the tensor product to be a supplementary
continuous series because these series play no r\^ole in the harmonic analysis on $\SL$.

\subsection{Product of matrix elements of SL$(2,\mathbb R)$}

We now apply the results of \cite{hb1,hb2} to the product of matrix elements of representations of the discrete series
and the continuous principal series.
 We have seen in the previous section that the tensor product of discrete series involves the
trivial representation, discrete series
and (principal and supplementary) continuous series. However,
when considering the product of matrix elements, it turns out that the supplementary series and the
trivial representation never appear.  This is on the one hand a consequence of
the fact that the trivial representation is not normalisable, and on the other hand a consequence of the Plancherel theorem and of the Losert basis (for
the analysis of the consequences of Losert basis see Sec. \ref{sec:Los}).\\

Consider first the product of two representations bounded from below/above.
Let $\psi^\pm_{m\lambda n}$ be a matrix element of the discrete series  ${\cal D}_\lambda^\pm$. Define
\beqa
\label{eq:mat}
f_{a,b,N}(\rho)=\psi^\pm_{m \lambda n}(\rho,\varphi_1,\varphi_2) e^{-i(m+n)\varphi_1 -i(m-n)\varphi_2}=\cosh^a \rho \sinh^b \rho P_N(-\sinh^2 \rho)
\eeqa
see \eqref{eq:wmn},
where $P_N$ is a degree $N$ polynomial in $-\sinh^2 \rho$.
 We have
\beqa
\label{eq:convab}
a+b+2N<-1 \ \ \text{and} \ \ b\ge 0
\eeqa
(see Eqs. [\ref{eq:matD+},\ref{eq:matD-}]).
By \eqref{eq:convP}, as expected, $f_{a,b,N} \in \LL$. If we now consider the product of two matrix elements of the discrete series, we get a term like $f_{a,b,N}(\rho)f_{a',b',N'}(\rho)$.
Since $f_{a,b,N}$ and  $f_{a',b',N'}$ correspond to matrix elements of the discrete series, the coefficients $(a,b,N)$ and $(a',b',N')$ clearly satisfy equation \eqref{eq:convab}.
As a consequence, the coefficients $(a+a',b+b', N+N')$ also satisfy \eqref{eq:convab}.
This means
that $f_{a,b,N}f_{a',b',N'}$ is square-integrable. \footnote{Note that, in general, the product of two square-integrable functions is not a square-integrable function. Here the specific form of the matrix elements implies that the product of two matrix elements of the discrete series is
square-integrable.} The behaviour is even better  since the product $e^d_{m,n,k} e^d_{m',n',k'}$ is a Schwartz function (see \eqref{eq:asym2}) thus
can be expanded using the Plancherel theorem. To proceed further, recall that for the matrix elements of the discrete series \eqref{eq:mat}, we have
 $mn>0$ and $|m|,|n|>1/2$.  Thus $f_{a,b,N} f_{a',b',N'} \in \LL^d$. This means that the  product of matrix elements of two representations bounded from above (resp. below) decomposes  from the Plancherel Theorem into a sum of matrix elements of
 representations bounded from below (resp. above). Finally, if
 $f_{a,b,N}$ (resp. $f_{a',b',N'}$) corresponds to the matrix elements of
 a representation ${\cal D}_\lambda^{\eta}$ (resp. ${\cal D}_{\lambda'}^{\eta}$) with $\eta=\pm$, the matrix elements of  ${\cal D}_\lambda^{\eta}$
 (resp.  ${\cal D}_{\lambda'}^{\eta}$) belong to $w_{mn}$ with $\eta m, \eta n \ge \lambda$ (resp.  $\eta m, \eta n \ge \lambda'$)
 --- see \eqref{eq:wmn}  and comments below it ---
 and consequently, the matrix elements corresponding to the product  $f_{a,b,N}f_{a',b',N'}$  belong
 to $w_{mn}$ with $\eta m, \eta n \ge \lambda + \lambda' $, reproducing \eqref{eq:++} or \eqref{eq:--}.

Similarly,   since for $\psi_{m_1 \lambda_1 n_1}^+ \in {\cal D}_{\lambda_1}^+$ we have $m_1,n_1\ge\lambda_1>1/2$ and for
 $\psi_{m_2 \lambda_2 n_2}^-\in {\cal D}_{\lambda_2}^-$ we have  $m_2,n_2\le- \lambda_2<-1/2$, $m=m_1+m_2, n=n_1+n_2$ can take any value
 in $\mathbb Z$,
the product of matrix elements of
 a representation bounded from below and  a representation bounded from above
belongs to $w_{mn}$ for any values of $m$  and $n$,
and hence belongs to $\LL$.
 Therefore, using the Plancherel Theorem, we reproduce Eq.[\ref{eq:+-3}]. It is straightforward to verify that we reproduce either \eqref{eq:+-1} or \eqref{eq:+-2}, depending on whether $\lambda>\lambda'$ or
$\lambda'>\lambda$ holds. If $\lambda=\lambda'$, the terms appearing in the decomposition of the product ${\cal D}_\lambda^+ \otimes {\cal D}_\lambda^-$ all belong to the continuous series.
Since on the one hands the trivial representation is not normalisable and on the other hands doesn't appear in the
Plancherel Theorem, the trivial representation doesn't appear at all  in the decomposition above.

 To extend this analysis to the case where the product of matrix elements
 of a principal continuous series is involved is more delicate, and we proceed applying the same strategy used for the inspection of the Losert basis (See Sec. \ref{sec:Los}).\\

Thus, at this stage we have
\begin{enumerate}
\item Product of two discrete series bounded from below or above
  (throughout $a=(\rho,\varphi_1,\varphi_2)$ labels a point of the manifold
  $\SL$, thus an element of the group itself):
  \beqa
  \psi^{\pm }_{m_1\lambda_1m'_1}(a)\psi^{\pm }_{m_2\lambda_2m'_2}(a) =
\sum \limits_{\lambda \ge \lambda_1 + \lambda_2}
  C_{\pm \lambda_1, \pm \lambda_2}^{\pm \lambda }{}_{m_1,m_2,m_1',m_2'}
  \psi^{\pm }_{m_1+m_2 \lambda m'_1+m'_2}(a)\nn
  \eeqa
  with
\beqa
& C_{\pm \lambda_1, \pm \lambda_2}^{\pm \lambda }{}_{m_1,m_2,m_1',m_2'}=
\frac{(4 \lambda_1-2)(4 \lambda_2-2)}{4 \lambda-2}
\scriptsize
\bpm\pm;\lambda_1&\pm;\lambda_2&\pm;\lambda\\
m_1&m_2&m_1+m_2\epm
\overline{\bpm\pm;\lambda_1&\pm;\lambda_2&\pm;\lambda\\
m'_1&m'_2&m'_1+m'_2\epm}
\nn
\eeqa
Because of our different normalisation for the matrix element (see \eqref{eq:Dorth}),
  the Clebsch-Gordan coefficients are slightly different from those in determined in \cite{hb1}: for any discrete series
${\cal D}^\pm_\lambda$   in  the LHS (resp. in the RHS) multiply (resp. divide) the RHS by $4 \lambda-2$.

\item Product of  one  discrete series bounded from below and one discrete series bounded from  above:
  \beqa
  \psi^{+ }_{m_1\lambda_1m'_1}(a) \psi^{-}_{m_2\lambda_2m'_2}(a)&=&
  \sum_{\frac12<\lambda\le |\lambda_1 -\lambda_2|}   C_{+\lambda_1,- \lambda_2}^{\eta_{12} \lambda}{}_{m_1,m_2,m_1',m_2'}
  \psi^{ \eta_{12}}_{m_1+m_2 \lambda m_1'+m'_2}(a)  \nn\\
  &&+ \int\limits_0^{+\infty}\text{d} \sigma \; \sigma \tanh \pi(\sigma+i\epsilon_{12}) C_{+\lambda_1,- \lambda_2}^{\epsilon_{12} i\sigma}{}_{m_1,m_2,m_1',m_2'}
  \psi^{\epsilon_{12} }_{m_1+m_2 i\sigma m'_1+m'_2}(a)\nn
\eeqa
($\eta_{12}$ being the sign of $\lambda_1-\lambda_2$ and $\epsilon_{12}=0,1/2$ depending if the RHS is a boson or a fermion),     where
\beqa
 C_{+\lambda_1,- \lambda_2}^{\eta_{12} \lambda}{}_{m_1,m_2,m_1',m_2'}&=&\frac{(4 \lambda_1-2)(4 \lambda_2-2)}{4 \lambda-2}\scriptsize
\bpm +;\lambda_1&-;\lambda_2&\eta_{12};\lambda\\
 m_1&m_2&m_1+m_2\epm
 \overline{\bpm +;\lambda_1&-;\lambda_2&\eta_{12};\lambda\\
 m'_1&m'_2&m'_1+m'_2\epm}\nn\\
 \normalsize
C_{+\lambda_1,- \lambda_2}^{\epsilon_{12} i\sigma}{}_{m_1,m_2,m_1',m_2'} &=&
 \scriptsize
 \bpm +;\lambda_1&-;\lambda_2&\epsilon_{12};i\sigma\\
 m_1&m_2&m_1+m_2\epm
 \overline{ \bpm +;\lambda_1&-;\lambda_2&\epsilon_{12};i\sigma\\
 m'_1&m'_2&m'_1+m'_2\epm}\ . \nn
\eeqa
 Again, note that, when the principal series is involved,  our Clebsch-Gordan coefficients have a slightly different normalisation with respect to \cite{hb1,hb2} (see {\it e.g.} Eq. [2.7] of \cite{hb2}),
in order to be in accordance with the Plancherel Theorem, {\it i.e.},  it has a $\sigma \tanh \pi(\sigma+ i \epsilon_{12})-$term
in the integral (see Eq. [\ref{eq:Plan}]).
\end{enumerate}
The Plancherel Theorem enables to obtain an alternative expression of the coefficients $C$ above. Indeed, in the case of the product of
a discrete series bounded from below by a discrete series bounded from above, by \eqref{eq:PC}  we have that
\beqa
C^{+ \lambda_1,- \lambda_2}_{\eta_{12}\lambda}{}_{m_1,m_2,m_1',m_2'}&=&(\psi^{\eta_{12}}_{m_1+m_2 \lambda m'_1+m'_2},
\psi^{+ }_{m_1\lambda_1 m'_1} \psi^{-}_{m_2\lambda_2 m'_2})\nn\\
C^{+\lambda_1,- \lambda_2}_{\epsilon i \sigma}{}_{m_1,m_2,m_1',m_2'}&=&(\psi^{\epsilon_{12}}_{m_1+m_2 i\sigma m'_1+m'_2},
\psi^{+}_{m_1\lambda_1 m'_1} \psi^{-}_{m_2\lambda_2m'_2}).\nn\
\eeqa

\subsection{Product of vectors of the Losert basis}\label{sec:Los}
In the previous section, we have been able to analyse the expansion of the product of matrix elements of representations bounded from below/above because these matrix elements are square-integrable. A similar analysis does not hold when we are considering at least
a continuous representation, because the corresponding matrix elements are not square-integrable. To perform the analysis in this case, we now use the Losert basis and, in particular, the fact that any matrix element of a continuous representation can be  expanded in the Losert basis, {\it i.e.}, in the elements $e_{m,n,k}^p$
that generate $\LL^{d^\perp}$ (see \eqref{eq:LP}). We already know that the elements $e^d_{m,n,k}$ do satisfy \eqref{eq:convab};  the same holds for the elements $e^p_{m,n,k}$, which belong to $\LL^{d^\perp}$  and also have to satisfy \eqref{eq:convab}. Thus the
products $e^d_{m,n,k} e^p_{m',n',k'}$     and $e^p_{m,n,k} e^p_{m'n',k'}$ are square-integrable, and therefore decompose in the basis ${\cal B}_d \oplus {\cal B}_{d^\perp}$ of $\LL$ (see \eqref{eq:Losert} and \eqref{eq:Hil2}). Note  again those products are also Schwartz functions (see \eqref{eq:asym2}).
By using   \eqref{eq:LP}, we reproduce  the results of \cite{hb2}, {\it i.e.} we express
    the product
    of matrix elements when at least a continuous  principal series occurs.
In particular, this means that the decomposition involves both discrete and principal representations, and thus  \eqref{eq:DC1},
    \eqref{eq:DC2} and \eqref{eq:CC1},     \eqref{eq:CC2}.

\begin{enumerate}
\item Product of a discrete series with a principal continuous series:
    \beqa
    \psi^{\pm }_{m_1\lambda m'_1}(a) \psi^{\epsilon }_{m_2 i\sigma m' _2}(a)&=&
    \sum\limits_{\frac12 <\lambda'\le \text{min}(|m_1+m_2|, |m'_1+m'_2|)}
    C_{\pm \lambda, \epsilon i\sigma}^{\eta \lambda'}{}_{m_1,m_2,m_1',m_2'}
    \psi^{ \eta}_{m_1+m_2 \lambda'm'_1+m'_2}(a)\nn
\nn\\
&&+\int\limits_0^{+\infty}\text{d} \sigma' \;
\sigma'\tanh \pi(\sigma'+i \epsilon')
C_ {\pm \lambda, \epsilon i\sigma}^{\epsilon'i \sigma'}  {}_{m_1,m_2,m_1',m_2'}{}
  \psi^{\epsilon' }_{m_1+m_2 i\sigma' m'_1+m'_2}(a)\nn
  \eeqa
  The sum over the discrete series is possible if $(m_1+m_2)(m'_1+m'_2)>0$ and
  $|m_1+m_2|, |m'_1+m'_2|>1/2$. We have introduced
  $\eta=|m_1+m_2|/(m_1+m_2)$ and $\epsilon'=0$ or $1/2$, depending whether the LHS is a boson or a fermion.
Even if the Plancherel theorem doesn't hold for the product of the matrix elements of a continuous and discrete series, with the purpose of having a unified notation, we consider a different normalisation than that in \cite{hb2} for the Clebsch-Gordan coefficients,
in order to have a term like $\sigma'\tanh \pi(\sigma'+ i\epsilon')$ in the second line of the decomposition above.
We have
\beqa
 C_{\pm \lambda, \epsilon i\sigma}^{\eta \lambda'}{}_{m_1,m_2,m_1',m_2'}&=&{\frac{4 \lambda -2}{4 \lambda' -2}}\;
 \scriptsize \bpm \pm;\lambda&\epsilon;i\sigma&\eta;\lambda'\\
      m_1&m_2&m_1+m_2\epm\;
 \overline{\bpm \pm;\lambda&\epsilon;i\sigma&\eta;\lambda'\\
      m'_1&m'_2&m'_1+m'_2\epm}\nn\\
C_{\pm \lambda, \epsilon i\sigma}^{\epsilon'i \sigma'}  {}_{m_1,m_2,m_1',m_2'}{}&=&(4\lambda-2)
\scriptsize \bpm \pm;\lambda&\epsilon;i\sigma&\epsilon';i\sigma'\\
      m_1&m_2&m_1+m_2\epm\;
 \overline{\bpm \pm;\lambda&\epsilon;i\sigma&\epsilon';i\sigma'\\
      m'_1&m'_2&m'_1+m'_2\epm}\nn
\eeqa

\item Product of  two principal continuous series:
  \beqa
  \label{eq:??}
 & \psi^{\epsilon_1}_{m_1 i\sigma_1 m'_1}(a) \psi^{\epsilon_ 2}_{m_2i \sigma_ 2m'_2}(a)=\scriptsize
 \sum \limits_{\frac12<\lambda\le \text{min}(|m_1+m_2|,|m'_1+m'_2|)} C_{\epsilon_1 i\sigma_ 1, \epsilon_2 i\sigma_2}^{\eta \lambda}{}_{m_1,m_2,m_1',m_2'}\;
 \psi_{m_1+m_2 \lambda m'_1+m'_2}^{\eta}(a)\nn\\
 &+ \int\limits_0^{+\infty}\text{d} \sigma \; \sigma\tanh \pi(\sigma+i\epsilon)\;
C_{\epsilon_1 i \sigma_1, \epsilon_2 i\sigma_2}^{\epsilon i\sigma}{}_{m_1,m_2,m_1',m_2'}
\psi_{m_1+m_2 i\sigma m'_1+m'_2}^{\epsilon}(a)
\eeqa
The  sum over the discrete series is possible if $(m_1+m_2)(m'_1+m'_2)>0$ and
  $|m_1+m_2|, |m'_1+m'_2|>1/2$.
We have introduced $\eta=|m_1+m_2|/(m_1+m_2)$. Note also that both sides have the same spin.
Even if the Plancherel Theorem does not hold for the product of the matrix elements of two continuous series
 (because it is not square-integrable, not being even normalisable), to have  unified notation we consider a different normalisation than that in \cite{hb2} for the Clebsch-Gordan coefficients,
in order to have a term like $\sigma\tanh \pi(\sigma+ i\epsilon)$  in the second line of the decomposition above.
We have
\beqa
C_{\epsilon_1 i\sigma_1, \epsilon_2 i\sigma_2}^{\eta \lambda}{}_{m_1,m_2,m_1',m_2'}&=& \frac 1 {{4 \lambda -2}}
\scriptsize \bpm \epsilon_1;i\sigma_1&\epsilon_2;i\sigma_2& \eta;\lambda\\
m_1&m_2&m_1+m_2\epm
\overline{\bpm \epsilon_1;i\sigma_1&\epsilon_2;i\sigma_2& \eta;\lambda\\
m'_1&m'_2&m'_1+m'_2\epm}\nn\\
C_{\epsilon_1 i \sigma_1, \epsilon_2 i\sigma_2}^{\epsilon i\sigma}{}_{m_1,m_2,m_1',m_2'}&=& \phantom{+}
\scriptsize\bpm \epsilon_1;i\sigma_1&\epsilon_2;i\sigma_2& \epsilon;i\sigma\\
m_1&m_2&m_1+m_2\epm_1
\overline{\bpm \epsilon_1;i\sigma_1&\epsilon_2;i\sigma_2& \epsilon;i\sigma\\
m'_1&m'_2&m'_1+m'_2\epm}_1\nn\\
&&+\scriptsize \bpm \epsilon_1;i\sigma_1&\epsilon_2;i\sigma_2& \epsilon;i\sigma\\
m_1&m_2&m_1+m_2\epm_2
\overline{\bpm \epsilon_1;i\sigma_1&\epsilon_2;i\sigma_2& \epsilon;i\sigma\\
m'_1&m'_2&m'_1+m'_2\epm}_2\nn
\eeqa
where  $\Big( \hskip 1.truecm \Big)_1$ and $\Big( \hskip 1.truecm \Big)_2$  are the two independent solutions  appearing in the decomposition
 ${\cal C}^{i\sigma_1,\epsilon_1} \otimes {\cal C}^{i\sigma_2,\epsilon_2} \supset 2 {\cal C}^{ i \sigma,[\epsilon_1+\epsilon_2]}$ \cite{hb2}.
    \end{enumerate}
\medskip

In order to have a simplified notation,
all cases treated above will be written as
\beqa
\label{eq:matmat}
\psi_{m_1\Lambda_1  m'_1}(a) \psi_{m_2\Lambda_ 2m'_2}(a) &=&   \sum_{\Lambda} \hskip -.6 truecm \int{}\;
{ C}_{\Lambda_1, \Lambda_2}^\Lambda{}_{m_1,m_2, m'_1,m'_2}
\psi_{m_1+m_2 \Lambda m'_1+m'_2}(a)\nn\\
&\equiv&
{C}_{\Lambda_1, \Lambda_2}^\Lambda{}_{m_1,m_2 , m'_1,m'_2}
\psi_{m_1+m_2 \Lambda m'_1+m'_2}
\eeqa
where $\Lambda_1, \Lambda_2= (\lambda,+), (\lambda,-), (i\sigma,0)$ or $(i\sigma,1/2)$, depending on the representation
to which the matrix elements (discrete series bounded from below/above or continuous principal series of bosonic/fermionic nature) belong, and $\Lambda$ acquires one of the allowed  values that have been identified.
The symbol $\sum_\Lambda \hskip -.65 truecm \int{}\; $ indicates a summation over discrete values of $\Lambda$ and  an integration over continuous values. In the last equality we have even skipped the symbol of summation over $\Lambda$.\\

Since the expansion of elements of $\LL$ can  either be obtained by the matrix elements of the $\SL-$representation
(Plancherel Theorem) or
by the Losert basis (Hilbert basis of $\LL$), and since the product of two elements of the Losert basis is square-integrable (this product is even a Schwartz function; see \eqref{eq:asym2}), an alternative
description of the study of coupling of representations {\it via} the traditional Clebsch-Gordan coefficients
would be to express the product of two Losert elements  (which will still be square-integrable) in the Losert basis:
\beqa
\label{eq:ee}
\phi_{m,n,k}(a) \phi_{m',n',k'}(a) = C_{kk'}^{k''}{}_{m m'n n'}\phi_{m+m',n+n',k''}(a).
\eeqa
As convened, we do not write  the summation term $\sum_{k''}$.
The coefficients $C_{kk'}^{k''}{}_{mm'nn'}$ can in principle be deduced by the decomposition of the product of matrix elements given above, by \eqref{eq:PL} and the inverse formula \eqref{eq:LP}.
The advantage of \eqref{eq:ee} with respect to \eqref{eq:matmat} is that we only have a discrete sum and not an integral.  Finally, we mention that in \eqref{eq:ee}, the product
$\phi_{m,n,k} \phi_{m',n',k'}$  can be either only in $\LL^d$ or  only in $\LL^{d^\perp}$  in some cases, as it happens (when certain conditions on the factor series are met; see above) for product of matrix elements of $\SL$ that involves only matrix elements of either discrete or  principal continuous representations.

\section{An infinite dimensional algebra associated to $\SL$}\label{sec:infa}

In \cite{rmm, rmm2} two of us have defined Kac-Moody algebras associated to a  manifold $\cal M$ equal to either  a  compact Lie group $G_c$ or a coset  space $G_c/H$ (with $H\subset G_c$,  and no assumption on the symmetric nature of the coset).  One of the main tools in this  construction was the Peter-Weyl theorem, which enables us to identify  a Hilbert basis of $L^2({\cal M})$.
The purpose of this section is  to extend the results associated to compact manifolds to the case of  the non-compact
(group) manifold  ${\cal M}=\SL$.
\subsection{Construction of the algebra}
In the case of the non-compact manifold $\SL$, as for the extensions of Kac-Moody algebras over compact manifolds $\cal M$, we consider square-integrable functions defined on $\SL$ (more precisely we assume to deal with Schwartz functions), in
order
to associate a corresponding  infinite dimensional Lie algebra.
There are two equivalent ways to define this algebra: either using the Plancherel Theorem, or using the Hilbert basis of $\LL$ we have called the  Losert basis. This algebra is defined in several steps.\\

Let $\g$ be a simple Lie algebra ($\g$ can be a complex or real simple Lie algebra) with basis $\{T^a, a=1,\cdots,\dim \g\}$, Lie brackets
\beqa
\big[T^a, T^b\big] = i f^{ab}{}_c T^c \ , \nn
\eeqa
and Killing form
\beqa
\Big<T^a, T^b\Big>_0=g^{ab} = \text{Tr}\Big(\text{ad}(T^a)\; \text{ad}(T^b)\Big) \ . \nn
\eeqa

Consider now the set of smooth maps
\beqa
\begin{array}{rll}
\SL&\to&G\\
(\rho,\varphi_1,\varphi_2)&\mapsto&g(\rho,\varphi_1,\varphi_2)
\end{array}\nn
\eeqa
where $G$ is the Lie group of $\g$ and $(\rho,\varphi_1,\varphi_2)$ parametrises a point in the manifold $\SL$ (see \eqref{eq:param}). We can thus write
\beqa
g(\rho,\varphi_1,\varphi_2)= e^{i\theta_a(\rho,\varphi_1,\varphi_2) T^a} \nn
\eeqa
(in the case of a non-compact Lie algebra $\g$ we have in fact a finite product of exponentials). In the vicinity of the identity, this reduces to
\beqa
g(\rho,\varphi_1,\varphi_2)\sim 1 + i\theta_a(\rho,\varphi_1,\varphi_2) T^a \ . \nn
\eeqa
We now assume that the functions $\theta_a\in  {\cal S} \subset \LL$. By the Plancherel theorem we have (see \eqref{eq:Plan} and  \eqref{eq:IP})
\beqa
\label{eq:TP}
\theta_a(\rho,\varphi_1,\varphi_2)= \II \theta_{a}^{n\Lambda m} \psi_{m\Lambda n}(\rho,\varphi_1,\varphi_2)
\eeqa
where $\Lambda=(\lambda, +), (\lambda,-), (i\sigma,0)$ and $(i\sigma,\frac12)$ with $\lambda>\frac12, \sigma>0$ and $\psi_{n\Lambda m}$ are the matrix elements
of the discrete series (bounded from below/above) or the principal continuous series (of fermionic/bosonic type).
 Of course, we can also  equivalently expand the $\theta$'s in the Losert basis
\beqa
\label{eq:TL}
\theta_a(\rho,\varphi_1,\varphi_2)= \theta_{a}^{mnk} \phi_{m, n, k}(\rho,\varphi_1,\varphi_2)\ .
\eeqa
Of course \eqref{eq:TP} and \eqref{eq:TL} are equivalent because of \eqref{eq:LP} and \eqref{eq:PL}.

For simplicity in the exposition, from now on we call the expansion \eqref{eq:TP}  the Plancherel basis, and the expansion \eqref{eq:TL} the Losert basis.  The algebra $\mathfrak{g}(SL(2,\mathbb{R}))$ will thus be defined, in the
Plancherel basis, as
\beqa
{\g}(\SL)&=& \Bigg\{T^a \psi^{+}_{n\lambda m}(\rho,\varphi_1,\varphi_2)\ ,\ T^a \psi^{-}_{n\lambda m}(\rho,\varphi_1,\varphi_2), \ \ \lambda>\frac12, \ \  mn>0,\ \  |m|,|n|>\lambda , \nn\\
&&\hskip .35truecm T^a \psi^{\epsilon}_{ni\sigma m}(\rho,\varphi_1,\varphi_2), \epsilon=0,\frac12, \sigma>0, m,n\in \mathbb Z + \epsilon, \ \
a=1,\cdots,\dim \g\Bigg\}\\
&=&\Bigg\{T^{a}_{m\Lambda n}=T^a \psi_{m\Lambda n}, \ \ \Lambda=(\lambda, +), (\lambda,-), (i\sigma,0), (i\sigma,\frac12)\ , a=1,\cdots,\dim \g\ \
\Bigg\}\nn
\eeqa
and in the Losert basis by
\beqa
{\g}(\SL)&=& \Bigg\{T^a_{m n k}=T^a \phi_{m,n,k}(\rho,\varphi_1,\varphi_2), \ \ m,n \in \mathbb Z  + \epsilon, k \in \mathbb N\ , \epsilon=0,\frac12 \Bigg\}
\eeqa
 Of course, we can express the $T^a_{m n k}$ in terms of the $T^{a}_{m\Lambda  n}$ and {\it vice versa}, because of
\eqref{eq:LP} and \eqref{eq:PL}. In the Plancherel basis, the Lie brackets take the form
\beqa
\big[T^{a}_{m\Lambda n}, T^{a'}_{m'\Lambda'n'}\big] =i f^{a a'}{}_{a''} C_{\Lambda, \Lambda'}^{\Lambda''}{}_{m,m',n, n'} T^{a''}_{m+m'\Lambda''n+n'} \ , \nn
\eeqa
with the notations of \eqref{eq:matmat}, while in the Losert basis they adopt the form
\beqa
\big[T^a_{m n k}, T^{a'}_{m'n'k'}\big]=if^{aa'}{}_{a''} C_{kk'}^{k''}{}_{m,m',n, n'} T^{a''}_{m+m'n+n'k''} \nn
\eeqa
by \eqref{eq:ee}. As expected, because of \eqref{eq:LP} and \eqref{eq:PL}, the two presentations of the algebra are equivalent.
Clearly this algebra is a natural extension of the loop algebra  ({\it i.e.} the algebra associated to the set of functions from the circle to the Lie group $G$)
to the  manifold $\SL$.

Finally, the Killing form on ${\g}(\SL)$ is given by
\beqa
\Big<X, Y \Big>_1= \frac 1 {4\pi^2}\int \limits_{0}^{+ \infty} \text{d} \rho \sinh \rho \cosh\rho
\int \limits_0^{2\pi} \text{d} \varphi_1 \int \limits_0^{2\pi} \text{d} \varphi_2 \Big<X,Y\Big>_0 \nn
\eeqa
for $X, Y \in {\g}(\SL)$.
 In the notation above $\big<\, , \,\big>_0$
(resp.$\big<\, , \,\big>_1$) refers to the scalar product in $\g$ (resp. in $\g(\SL)$).
In particular, we have in the Plancherel basis
\beqa
\label{eq:scalP}
\Big<T^{a\eta}_{m\lambda n}, T^{a' \eta'}_{m' \lambda'n'}\Big>_1 &=& g^{aa'} \delta_{\lambda, \lambda'} \delta_{\eta + \eta'} \delta_{m+m'} \delta_{n+n'}\nn\\
\Big<T^{a\epsilon}_{m i\sigma n}, T^{a' \epsilon'}_{m' i\sigma'n'}\Big>_1 &=& \frac 1{\sigma\tanh \pi(\sigma + i\epsilon)} g^{aa'}\delta(\sigma-\sigma') \delta_{\epsilon, \epsilon'} \delta_{m+m'} \delta_{n+n'}\
\eeqa
 (to simplify the reading we henceforth denote the usual Kronecker symbol $\delta_{m,-n}\equiv \delta_{m+n}$) because of \eqref{eq:Dorth} and \eqref{eq:Corth} and the conjugation relations \eqref{eq:cc1} and \eqref{eq:cc2}. In the Losert basis we get
\beqa
\label{scalL}
\Big<T^{a}_{mnk}, T^{a'}_{m'n'k'}\Big>_1 &=& g^{aa'} \delta_{k,k'}  \delta_{m+m'} \delta_{n+n'}
\eeqa
because of \eqref{eq:BHil}  and the conjugation relations \eqref{eq:cc3}.\\

The next step in the construction of the infinite dimensional algebra associated to the manifold $\SL$ is to introduce the maximal set of Hermitean operators which mutually
commute, namely $L_0, R_0$. We thus have
\beqa
\big[L_0,T^{a}_{m\Lambda n}\big]= m T^{a}_{m\Lambda n}\ , \ \ \big[R_0,T^{a}_{m\Lambda n}\big]= n T^{a}_{m\Lambda n}\nn
\eeqa
in the Plancherel basis, and
\beqa
\big[L_0,T^{a}_{mnk}\big]= m T^{a}_{mnk}\ , \ \ \big[R_0,T^{a}_{mnk}\big]= n T^{a}_{mnk}\nn
\eeqa
in the Losert basis.\\

The last step in the construction of the Kac-Moody algebra associated to $\SL$ is the introduction of central extensions.
To centrally extend a Lie algebra, and in particular the algebra obtained so far, is not trivial. Indeed, the central extension has to be, on the one hand, compatible with the Jacobi identity,
and on the other hand, non-trivial, {\it i.e.}, it must not vanish by a change of coordinates.
This problem
was completely solved by Pressley and Segal in \cite{ps} (see Proposition 4.28  therein)
{\it i.e.}, they identified all possible central extensions of the Lie algebra $\widehat{\g}({\cal M})$
where ${\cal M}$ is an $n-$dimensional manifold ${\cal M}$ (compact or  non-compact).
Given a one-chain $C$ ({\it i.e.}, a closed one-dimensional
 piecewise smooth curve)  in ${\cal M}$, the central extension is given by the two-cocycle
\beqa
\omega_C(X,Y) = \oint_C \big<X, \d Y\big>_0  \ , \nn
\eeqa
where $\d Y$ is the exterior derivative of $Y$.
 The two-cocycle can be written in alternative form  \cite{bt}
\beqa
\omega_C(X,Y) = \int_{{\cal M}}  \big<X, \d Y\big>_0 \wedge \gamma \ ,
\eeqa
where $\gamma$ is a closed $(n-1)-$current (a distribution)  associated to $C$.

In the case of the three-dimensional  ($n=3$) manifold $\SL$ we have an infinite number of possible central extensions (associated to any curved ${\cal C}$ or current $\gamma$). In \cite{sorb} all central extensions associated
to the two-torus and the two-sphere were classified. It is a general feature that such algebras admit an infinite
number of central extensions (see also \cite{MRT}).

In our construction we restrict ourselves to the case of
two central charges in duality  (in the sense specified by definitions \eqref{eq:def-1}) with the two Hermitean operators $L_0$ and $R_0$. Indeed, we consider the two two-forms:
\beqa
\gamma_L &=& -\frac i{4\pi^2} k_L\Big(\d \rho \wedge \d \varphi_1+\d \rho \wedge\d\varphi_2\Big) \sinh \rho \cosh \rho\ , \nn \\\label{eq:def-1}
\gamma_R &=&  -\frac i{4\pi^2} k_R\Big(\d \rho \wedge \d \varphi_1-\d \rho \wedge\d\varphi_2\Big) \sinh \rho \cosh \rho\ ,
\eeqa
which of course satisfy $\d \gamma_L = \d \gamma_R =0$.
A direct computation leads to
\beqa
\omega_L(X,Y)&=&-\frac {k_L} {4\pi^2}  \int\limits _0^{+\infty} \d\rho \sinh \rho \cosh \rho \int \limits_0^{2\pi} \d \varphi_1  \int \limits_0^{2\pi}\d \varphi_2
\;\Big<X,L_0 Y\Big>_0\nn\\
\omega_R(X,Y)&=& -\frac {k_R} {4\pi^2}\int\limits _0^{+\infty} \d\rho \sinh \rho \cosh \rho \int \limits_0^{2\pi} \d \varphi_1  \int \limits_0^{2\pi}\d \varphi_2
\;\Big<X,R_0 Y\Big>_0\ . \nn
\eeqa

We now compute explicitly the central extensions in the Losert and Plancherel basis.
In the Plancherel basis we have to use \eqref{eq:cc1} and \eqref{eq:cc2} together with \eqref{eq:Dorth} and \eqref{eq:Corth}:
\beqa
\omega_L(T^{a\eta}_{n\lambda m}, T^{a' \eta'}_{n'\lambda 'm'})&=&  n k_L g^{aa'} \delta_{\lambda,\lambda'} \delta_{\eta+\eta'} \delta_{m+m'} \delta_{n+n'}\nn\\
\omega_L(T^{a\epsilon}_{n i\sigma m}, T^{a'\epsilon'}_{n' i\sigma'm'})&=&nk_Lg^{aa'} \frac{\delta(\sigma-\sigma')}{\sigma\tanh \pi(\sigma+i\epsilon)} \delta_{\epsilon,\epsilon'} \delta_{m+m'} \delta_{n+n'}\nn\
\eeqa
and
\beqa
\omega_R(T^{a\eta}_{n\lambda m}, T^{a' \eta'}_{n'\lambda'm'})&=& mk_R g^{aa'} \delta_{\lambda,\lambda'} \delta_{\eta+\eta'} \delta_{m+m'} \delta_{n+n'}\nn\\
\omega_R(T^{a\epsilon}_{ni\sigma m}, T^{a' \epsilon'}_{n' i\sigma' m'})&=&mk_Rg^{aa'} \frac{\delta(\sigma-\sigma')}{\sigma\tanh \pi(\sigma+i\epsilon)} \delta_{\epsilon,\epsilon'} \delta_{m+m'} \delta_{n+n'}\nn\
\eeqa
Defining
\beqa
\delta(\Lambda,\Lambda') =
\left\{
\begin{array}{cc}
 \delta_{\lambda,\lambda'} \delta_{\eta+\eta'} & \Lambda=(\lambda,\eta)\ ,\ \  \Lambda'=(\lambda',\eta')\\
 \frac{\delta(\sigma-\sigma')}{\sigma\tanh \pi(\sigma+i\epsilon)} \delta_{\epsilon,\epsilon'} &
 \Lambda=(i\sigma,\epsilon)\ , \ \  \Lambda'=(i\sigma', \epsilon')\\
 0&\text{elsewhere}
\end{array}
\right.\nn
\eeqa
we obtain
\beqa
\omega_L(T^{a}_{n\Lambda m}, T^{a'}_{n'\Lambda'm'}) &=& \;n k_L\;g^{aa'} \delta(\Lambda,\Lambda')  \delta_{m+m'} \delta_{n+n'}\nn\\
\omega_R(T^{a}_{n\Lambda m}, T^{a'}_{n'\Lambda'm'}) &=& m k_R \;g^{aa'}\delta(\Lambda,\Lambda')  \delta_{m+m'} \delta_{n+n'}\nn \ .
\eeqa
In the Losert basis we get
\beqa
\omega_L(T^a_{nmk}, T^{a'}_{n'm'k'})&=& n k_L \;g^{aa'}\delta_{k,k'} \delta_{n+n'} \delta_{m+m'} \nn\\
\omega_R(T^a_{nmk}, T^{a'}_{n'm'k'})&=&  mk_R \;g^{aa'}\delta_{k,k'} \delta_{n+n'} \delta_{m+m'} \nn
\eeqa
because of \eqref{eq:cc3} and \eqref{eq:BHil}.\\

 To summarize, the generalised Kac-Moody algebra associated to the non-compact manifold $\SL$,  denoted by $\widehat{\g}\big(\SL\big)$, {\it i.e.}, the algebra  ${\g}\big(\SL\big)$ with the central extension above, is  by generators
\begin{enumerate}[noitemsep]
\item The generators of $\g(\SL)$\;
\item The commuting Hermitean operators $L_0, R_0$;
\item The central charges $k_L,k_R$;
\end{enumerate}
and the Lie brackets whose central extension reads
\beqa
\label{eq:LieP}
\big[T^{a}_{m \Lambda n}, T^{a'}_{m'\Lambda'n'}\big] &=&i f^{a a'}{}_{a''} C_{\Lambda, \Lambda'}^{\Lambda''}{}_{m,m',n, n'} T^{a''}_{m+m'\Lambda''n+n'}
+ (mk_L+nk_R) \delta(\Lambda,\Lambda') \delta_{m+m'} \delta_{n+n'}\ , \nn \\
\big[L_0,T^{a }_{m\Lambda n}\big]&=& m T^{a}_{m\Lambda n}\ , \\
 \big[R_0,T^{a,}_{m\Lambda n}\big]&=& n T^{a}_{m\Lambda n} \nn
\eeqa
in the Plancherel basis, and
\beqa
\label{eq:LieL}
\big[T^{a}_{m n k}, T^{a'}_{m'n' k'}\big] &=&i f^{a a'}{}_{a''} C_{k k'}^{k''}{}_{m,m',n,n'} T^{a'' }_{m+m'n+n' k''}
+ (mk_L+nk_R) \delta_{kk'} \delta_{m+m'} \delta_{n+n'}\ , \nn \\
\big[L_0,T^{a }_{mn  k}\big]&=& m T^{a}_{mn k}\ , \\
 \big[R_0,T^{a}_{mn k}\big]&=& n T^{a}_{mn k} \nn
\eeqa
in the Losert basis. Note that the RHS of the first line in \eqref{eq:LieP} may involves integral and/or summation
in the sense of Eq.[\ref{eq:matmat}].
The two presentations of the algebra \eqref{eq:LieP} and \eqref{eq:LieL} are equivalent because of \eqref{eq:PL} and \eqref{eq:LP}. The former presentation
involves an integral
in the summation (first term in the first line) and the central terms involve a Dirac $\delta-$distribution (second term of the first line) when continuous principal series
are involved  (because they are not normalizable, and thus they do not belong to $%
L^{2}(SL(2,\mathbb{R}))$), whilst the latter presentation involves only summation and  $\delta-$Kronecker terms. This difference is related to the fact that, in the Plancherel basis, the matrix elements are not square-integrable, whereas the Losert basis is an orthonormal Hilbert basis.  Each presentation has its own advantages. In the Plancherel basis, each term
is easily classified in terms of representations of $\SL$, and in the Losert basis, we only have square-integrable terms. Finally, observe that the construction is analogous to the corresponding construction when $\cal M$ is a compact manifold \cite{rmm,rmm2}. However, it is important to realise that in the case of a compact manifold, we do not encounter non-normalisable
terms and $\delta-$distributions.\\

Finally, the algebra can be written using the usual current algebra with a Schwinger term \cite{Scw}. Indeed, if we define $T^a(\rho,\varphi_1,\varphi_2) \in
\LL$ and set
\beqa
\big[T^a(\rho,\varphi_1,\varphi_2), T^{a'}(\rho',\varphi'_1,\varphi'_2)\big]&=&
\Big(if^{a a'}{}_{a''} T^{a''}(\rho',\varphi'_1,\varphi'_2) - g^{ab}(k_L L_0 +k_R R_0)\Big)\times \nn\\
&& \delta(\sinh^2 \rho-\sinh^2\rho') \delta(\varphi_1-\varphi'_1)  \delta(\varphi_2 -\varphi'_2)  \nn
\eeqa
we reproduce either \eqref{eq:LieP} or \eqref{eq:LieL}. Indeed, in the first case we expand $T^a(\rho,\varphi_1,\varphi_2)$  by means of  the Plancherel Theorem and use the expression
 of the $\delta-$distribition
\eqref{eq:delP},  whilst in the second case we expand  $T^a(\rho,\varphi_1,\varphi_2)$ in the Losert basis and use the definition of $\delta$ \eqref{eq:delL}.
This algebra can even be written in a more intrinsic  way, {\it i.e.}, in a base independent manner for any $x,y \in \g$:
\beqa
\big[x(\rho,\varphi_1,\varphi_2),y(\rho',\varphi'_1,\varphi'_2)\big]&=&\Big(\big[x,y\big](\rho',\varphi'_1,\varphi'_2) -\big<x,y\big>_0(k_L L_0 +k_R R_0)\Big)\times\nn \\
&&\delta(\sinh^2 \rho-\sinh^2\rho') \delta(\varphi_1-\varphi'_1)  \delta(\varphi_2 -\varphi'_2) \nn
\eeqa
Thus the Lie brackets involve only  $\g-$tensors and the $\delta-$distribution associated to
SL$(2.\mathbb R)$.
Expanding $x$ and $y$ in various basis leads, after integration over
$\rho,\varphi_1,\varphi_2$ and $\rho',\varphi'_1,\varphi'_2$,
to different presentations of the algebra $\widehat{\g}(\text{SL}(2,\mathbb R))$.
This indeed proves once  again that the presentation in the Plancherel basis \eqref{eq:LieP} and in the Losert basis \eqref{eq:LieL} are equivalent.

\subsection{Root system}
The aim of this section is to identify a root structure of the algebra
$\Kg$. Assume that $\g$ is a rank-$\ell$ Lie algebra.  Let $\{H^i, i=1,\cdots,\ell\}$
be a Cartan subalgebra,  let  $\Sigma$ be the root system  and let $E_\alpha, \alpha
\in \Sigma$  be the corresponding root vector of $\g$.
Correspondingly, the Kac-Moody algebra associated to $\SL$ is generated by
$\Kg=\{H^i_{mnk}, E_{\alpha, mnk}, m,n \in \mathbb Z+ \epsilon, \epsilon=0,\frac12, k\in \mathbb N, L_0, R_0, k_L, k_R\}$ in the Losert basis (it turns out that the analysis of the root system of $\Kg$ is easier in the Losert basis). In this basis, the Lie brackets take the form:

\beqa
\label{eq:CW}
\big[H^i_{ mnk}, H^{i'}_{ m 'n'k'}\big] &=&  (m k_L + n k_R) \delta_{m+m'}\delta_{n+n'} \delta_{kk'}   h^{ii'}   \ , \nn\\
\big[H^i_{m n k }, E_{\alpha, m'n'k'} \big]&=&   \alpha^i C_{kk'}^{k''}{}_{m,m',n,n'}\; E_{\alpha , m+m'  n+n'k'' }\  ,
\nn\\
\big[E_{\alpha, m,n,k}, E_{\beta , m'n'k'}\big] &=&
\left\{
\begin{array}{l}
{\cal N}_{\alpha,\beta} \;  C_{kk'}^{k''}{}_{m,m',n,n'}\; E_{\alpha+\beta, m+m'  n+n' k'' }\ ,
\\ \hskip 2.5truecm \text{if} \ \ \alpha+\beta \in \Sigma \ ,\\[4pt]
  C_{kk'}^{k''}{}_{m,m',n,n'}\ \alpha\cdot H_{ m+m' n+n' k''  }
+ (m k_L + n k_R) \delta_{m+m'}\delta_{n+n'} \delta_{kk'}  \ ,   \\[4pt]\hskip 2.5truecm    \text{if} \ \ \alpha + \beta=0\ ,\\[4pt]
0\ ,
\hskip 2.5truecm   \text{if} \ \ \left\{\begin{array}{l}\alpha + \beta \ne 0\ \ ,\\[4pt] \alpha+\beta \not \in \Sigma \ , \end{array} \right.\
\end{array}
\right.\\
\big[L_0,E_{\alpha, m n k }\big]&=& m E_{\alpha,mnk}\ ,\ \ \big[R_0,E_{\alpha, m n k }\big]= n E_{\alpha,mnk}\ ,  \nn\\
\big[L_0,H^i_{ m n k }\big]&=& m H^i_{mnk}\ ,\hskip .52truecm \big[R_0,H^i_{ m n k }\big]= n H^i_{mnk}\ , \nn
\eeqa
where
\beqa
h^{ij} = \big<H^i,H^j\big>_0 \ ,
 \ \  h_{ij} \equiv (h^{-1})_{ij}\ , \ \  \alpha \cdot H_{mm'k} = h_{ij}
\,\alpha^i H^j_{mm'k}\nn
\eeqa
and the operators associated to roots of $\mathfrak g$ are normalised as
\beqa
\big<E_\alpha, E_\beta\big>_0 = \delta_{\alpha,-\beta} \ . \nn
\eeqa
Finally, the coefficients ${\cal N}_{\alpha,\beta} =-{\cal N}_{\beta,\alpha}$ characterise the Lie algebra $\g$ (and are
equal to $\pm 1$ if $\g$ is simply laced).

The root spaces reduce then to
\beqa
\begin{split}
  \mathfrak{g}_{(\alpha,m,n)} &= \Big\{E_{\alpha,m n k} \ , k\in \mathbb N \Big\}\ , \ \ \alpha \in\Sigma, m,m \in \mathbb Z + \epsilon, \epsilon=0,\frac12\\
  \mathfrak{g}_{(0,m,n)} &= \Big\{H^i_{mnk}\ , i=1,\cdots,\ell\ ,  k\in \mathbb N  \Big\}\ ,  m,n \in \mathbb Z+\epsilon, \epsilon=0,\frac12
  \end{split}\nn
\eeqa
with the obvious commutation relations
\beqa
\Big[\mathfrak{g}_{(\alpha,m,n)},\mathfrak{g}_{(\beta,p,q)} \Big]&\subset &\mathfrak{g}_{(\alpha+\beta,m+p,n+q)}\ , \ \ \text{if} \ \ \alpha+\beta\in \Sigma \cup\{0\} \ \  \text{(and $=0$ otherwise)}\nn\\
  \Big[\mathfrak{g}_{(\alpha,m n)},\mathfrak{g}_{(0,pq)} \Big]&\subset& \mathfrak{g}_{(\alpha,m+p,n+q)}\ .\nn
  \eeqa

Of course each  $E_{\alpha,mnk}$ in  $\g_{(\alpha,m,n)}$ is an eigenvector of $L_0$ and $R_0$ (see \eqref{eq:CW}).
However, as stated previously, the matrix element of the trivial representation ${\cal D}_0$ doesn't appear in the Plancherel Theorem (see Sects. \ref{sec:Mat} and \ref{sec:Plan}). This in particular means that   $E_{\alpha,mnk}$ is not a eigenvector of $H^i_{pqk'}$ for any values of $i,p,q,k'$.
Consequently, we cannot find an operator in $\Kg$ and an  element in $\g_{(\alpha,m,n)}$ with eigenvalue $\alpha$.
This situation is very different to the corresponding situation
when $\cal M$ is related to a compact Lie group, since the mode expansion of square-integrable functions involves the trivial representation, and thus there exists an operator such that all elements in the analogue of $\g_{(\alpha,m,n)}$  have  eigenvalue $\alpha$.

In spite of the non-existence of an operator with eigenvalue $\alpha$, the algebra $\Kg$ has a very interesting property. Indeed, the set
\beqa
\widehat{\mathfrak{h}}=\Big\{H^i_{00k}\ , i=1,\cdots, \ell \ ,  k\in\mathbb N\ ,  L_0\ ,  R_0\ ,  k_L\ ,  k_R\Big\} \nn
\eeqa
 consists of simultaneously commuting operator. This  implies that the rank of the algebra $\Kg$ is infinite. This observation opens immediately one question.
As said previously the trivial representation of $\s$ is not square integrable and  $\g \not \subset \Kg$. So in spite of the non-existence of an operator with eigenvalue $\alpha$,
one may wonder  whether a set $\widehat{\mathfrak{h'}} \subset \widehat{\mathfrak{h}}$ exists which enables an explicit diagonalisation  of the root space.

\section{From SL$(2,\mathbb R)$ to SL$(2,\mathbb R)/U(1)$}
In this section we extend the previous results to the
symmetric space  SL$(2,\mathbb R)/U(1)$. This study is motivated by the
fact that such  a space naturally appears in physical theories, such as supergravity, as the target space of scalar fields.

\subsection{On SL$(2,\mathbb R)/U(1)$ as a Scalar Manifold in $D=3+1$ Supergravity}\label{sec:sugra}

By confining ourselves to $D=3+1$ space-time dimensions, let us now consider
locally supersymmetric theories of (Einstein) gravity (usually named \textit{%
supergravity} theories). The bosonic sector of such theories naturally
contains scalar (\textit{i.e.}, spin-0) fields, which can be regarded as
maps from space-time to a certain target space, named scalar manifold. This
latter is thus coordinatized by the scalar fields. Generally, in low
space-time dimensions many scalar fields appear, resulting from the \textit{%
moduli} of the geometry of the internal manifold in superstring or M-theory
compactifications.

We here want to focus on the simplest case in which only one (complex%
\footnote{%
The complex nature of the scalar field is imposed by ($\mathcal{N}=1$ or $%
\mathcal{N}=2$-extended) supersymmetry (see below).}) scalar field occurs,
and it coordinatizes a scalar manifold which is (locally) isomorphic to the K%
\"{a}hler symmetric coset\footnote{%
The notation \textquotedblleft $\simeq $\textquotedblright\ denotes local
isomorphism of symmetric spaces.}%
\begin{equation}
\frac{SU(1,1)}{U(1)}\simeq \frac{SL(2,\mathbb{R})}{U(1)}\simeq \frac{Sp(2,%
\mathbb{R})}{U(1)}.  \label{m}
\end{equation}%
In presence of $\mathcal{N}\geqslant 2$ supercharges, the non-compact,
Riemannian, locally symmetric, constant curvature, rank-1 space (\ref{m})
occurs three times as scalar manifold of a supergravity theory in $D=3+1$%
\footnote{%
In passing, we should mention that the space (\ref{m}) can also be regarded
as the scalar manifold of $D=9+1$ type IIB non-chiral supergravity \cite{IIB}%
.}. These theories are :

\begin{enumerate}[noitemsep]
\item \textquotedblleft pure\textquotedblright $\mathcal{N}=4$ supergravity
(only containing the gravity multiplet);

\item $\mathcal{N}=2$ supergravity \textit{minimally} coupled to $1$ vector
multiplet;

\item $\mathcal{N}=2$ supergravity obtained as dimensional reduction from $%
\mathcal{N}=2$ \textquotedblleft pure\textquotedblright\ (minimal)
supergravity in $D=5$ (the so-called $T^{3}$ model, with $1$ vector
multiplet).
\end{enumerate}

Since $\mathcal{N}=1$ (local) supersymmetry constrains the geometry of the
(chiral multiplets') scalar manifold to be K\"{a}hler, any $\mathcal{N}=1$
supergravity model with only one chiral multiplet could possibly have the
space (\ref{m}) as target space; however, in most of such models the
electric-magnetic ($U$-)duality\footnote{%
In this paper, $U$-duality is referred to as the \textquotedblleft
continuous\textquotedblright\ symmetries of \cite{CJ}, namely to the
classical (\textquotedblleft continuous\textquotedblright ) limit of
non-perturbative string theory symmetries introduced by Hull and Townsend
\cite{HT}. In supergravity, the alternative naming of \textit{%
electric-magnetic duality} is also used.} group would be trivial. Thus, in
the following treatment we will consider the occurrence of (\ref{m}) as
scalar manifold in those $\mathcal{N}=1$ supergravity models which share the
very same bosonic sector with the $\mathcal{N}\geqslant 2$ supergravities
mentioned at points 1-3 above.

\subsubsection{\label{N=4-pure}\textquotedblleft Pure\textquotedblright\ $%
\mathcal{N}=4$}

In \textquotedblleft pure\textquotedblright\ $\mathcal{N}=4$ supergravity
\cite{Scherk} (\textit{i.e.}, in absence of $\mathcal{N}=4$ vector
multiplets), the $U$-duality group is $G_{\mathcal{N}=4~\text{"pure"}%
}=SO(6)\times SU(1,1)$; indeed, the manifold (\ref{m}) should in this case
be properly written as%
\begin{equation}
\mathcal{M}_{\mathcal{N}=4~\text{"pure"}}:=\frac{SU(1,1)}{U(1)}\times \frac{%
SO(6)}{SO(6)}.  \label{MM}
\end{equation}%
Remarkably, in this theory the kinetic vector matrix is a holomorphic
function \cite{Scherk} (see also Sec. 7 of \cite{Gnecchi-1}, and references
therein) of the unique complex scalar field $u$ of the gravity multiplet
(see (\ref{s}) below), and so this theory is \textquotedblleft
twin\textquotedblright \footnote{%
For a recent account on \textquotedblleft twin\textquotedblright\
supergravities in all dimensions, see \textit{e.g.} \cite{twinsugras}, and references therein.} to (namely, shares the same bosonic sector with) $\mathcal{N}%
=1,D=4$ supergravity with $U$-duality group $SO(6)\times SU(1,1)$,
consisting of the $\mathcal{N}=1$ gravity multiplet coupled to $n_{c}=1$
chiral (Wess-Zumino) multiplets and $n_{v}=6$  ($\mathcal{N}=1$) vector multiplets.

In this case, the space (\ref{MM}) is locally coordinatised by the complex
scalar field%
\begin{equation}
u:=a+ie^{-2\phi },  \label{s}
\end{equation}%
where $a\in \mathbb{R}$ and $\phi \in \mathbb{R}$ are named the axion and
dilaton fields, respectively. The metric of (\ref{MM}) reads \cite{Scherk}
(see also Secs. 5 and 6 of \cite{Gnecchi-2})%
\begin{equation}
ds_{\mathcal{N}=4~\text{\textquotedblleft pure\textquotedblright }}^{2}=g_{u%
\overline{u}}dud\overline{u}=e^{4\phi }da^{2}+d\phi ^{2}.
\end{equation}%
For what concerns the Riemann scalar curvature, the general formula for $1_{%
\mathbb{C}}$-dimensional K\"{a}hler manifolds (see e.g. \cite{KG}) yields%
\begin{equation}
R=-g^{u\overline{u}}\overline{\partial }_{\overline{u}}\partial _{u}\text{ln}%
\left( \text{det}\left( g\right) \right) =-1/2.
\end{equation}

\subsubsection{\label{N=2-mc}$\mathcal{N}=2$ \textit{axion-dilaton}}

The $\mathcal{N}=2$ theory \textit{minimally coupled} to one vector
multiplet \cite{Luciani} is usually named the $\mathcal{N}=2$ \textit{%
axion-dilaton} model; it has $n_{V}=1$ and $n_{h}=0$, where $n_{V}$ and $%
n_{h}$ respectively denote the number of ($\mathcal{N}=2$) vector multiplets
and hypermultiplets. In this model, the $U$-duality group is $G_{\mathcal{N}%
=2\text{,~mc,~}n_{V}=1}=U(1,1)\simeq SU(1,1)\times U(1)$; indeed, the
manifold (\ref{m}) should in this case be written as%
\begin{equation}
\mathcal{M}_{\mathcal{N}=2\text{,~mc,~}n_{V}=1}\simeq \frac{SU(1,1)}{U(1)}%
\times \frac{U(1)}{U(1)}\simeq \overline{\mathbb{C}P}^{1},  \label{D}
\end{equation}%
where $\overline{\mathbb{C}P}^{1}$ denotes  the non-compact form of the $1_{\mathbb{C}}$-dimensional
projective space. This may be endowed with a \textit{projective} \textit{%
special K\"{a}hler} geometry, as consistent with $\mathcal{N}=2$ local
supersymmetry (see e.g. \cite{Strominger, N=2-Big}, and references therein). The
global $U(1)$ factor is a relic of the compact symmetry of the Maxwell
theory of the lone graviphoton, to which the electromagnetic sector reduces
if the vector multiplet is truncated; this $U(1)$ acts only on the Maxwell
vector fields, and not on the scalar fields; see  \textit{e.g.} the discussion in \cite{FMO-mc}.

The kinetic vector matrix of the $\mathcal{N}=2$ axion-dilaton model is a
holomorphic function of the complex scalar field, and so this theory is
\textquotedblleft twin\textquotedblright\ to $\mathcal{N}=1,D=4$
supergravity with $U$-duality group $U(1,1)$, consisting of the $\mathcal{N}%
=1$ gravity multiplet coupled to $n_{c}=1$ chiral multiplets and $n_{v}=2$
vector multiplets.

Note that this model can be obtained from the $\mathcal{N}=4$
\textquotedblleft pure\textquotedblright\ theory considered in Sec. \ref%
{N=4-pure}; in the bosonic sector, this amounts to removing four
graviphotons out of six (see \textit{e.g.} the discussion in \cite{Hayakawa}%
). In this way, the $\mathcal{N}=2$ axion-dilaton model is obtained in a
manifestly $SO(1,1)$-invariant symplectic frame, in which the special
geometry holomorphic prepotential reads%
\begin{equation}
F=-iX^{0}X^{1},  \label{F1}
\end{equation}%
where the $X^{\Lambda }$'s ($\Lambda =0,1$) are the (complex) contravariant
sections of the symplectic bundle over  the manifold (\ref{D}). By projectivizing, one can
define $Z:=X^{1}/X^{0}$, and the resulting K\"{a}hler potential \footnote{%
In (curved) special geometry, in a symplectic frame in which the holomorphic
prepotential $F\left( X\right) $ exists, given the holomorphic symplectic
sections $X^{\Lambda }$ and $F_{\Lambda }=\partial F/\partial X^{\Lambda }$,
the K\"{a}hler potential is given by the formula $K=-\ln \left[ i\left(
\overline{X}^{\Lambda }F_{\Lambda }-X^{\Lambda }\overline{F}_{\Lambda
}\right) \right] $ (see \cite{CDF}, and  references therein), with $\Lambda
=0,1,..,n_{V}$. The index $0$ pertains to the graviphoton, and the section $%
X^{0}$ is nowhere vanishing (and it usually set to $1$ by fixing the K\"{a}%
hler gauge freedom), and it gives rise to the projective nature of the
special K\"{a}hler space in presence of local supersymmetry.} (see \textit{%
e.g.} \cite{Chimento-Klemm-Petri} and  references therein; see also \cite%
{Dall'Agata-Ceresole-First})%
\begin{equation*}
K=-\text{ln}\left( 4\text{Re}Z\right)
\end{equation*}%
along with the constraint%
\begin{equation*}
\text{Re}Z>0,
\end{equation*}%
yields the metric
\begin{equation}
ds_{\mathcal{N}=2~\text{axion-dilaton}}^{2}=g_{Z\overline{Z}}dZd\overline{Z}=%
\frac{1}{4\left( \text{Re}Z\right) ^{2}}\left\vert dZ\right\vert ^{2}.
\end{equation}

Another widely used symplectic frame for this model is the manifestly $%
SU(1,1)$-invariant symplectic frame (also named Fubini-Study symplectic
frame), which provides a model for the \textit{Poincar\'{e} disk}\footnote{%
Recently, the Poincar\'{e} disk received much attention in $\mathcal{N}=1$
supergravity, in relation to the so-called \textquotedblleft $\alpha $%
-attractors\textquotedblright\ in inflationary cosmology \cite%
{Kallosh-Linde, Shahbazi-Lazaroiu}.  Such models are based on the metric given \textit{e.g.} by [6.10] of \cite{Kallosh-Linde-2}, which reduces to \eqref{eq:tt} by setting $\alpha=1/3$. An intriguing interpretation of  $\alpha$-attractors is based on a geometric moduli space with a boundary: a Poincar\'{e} disk model of a hyperbolic geometry with radius $\sqrt{3\alpha}$, beautifully represented by Eschers picture Circle Limit IV (for a computer generated version, see
\href{http://bulatov.org/math/1201/}{http://bulatov.org/math/1201/}).
In these models, the amplitude of the gravitational waves is proportional to the square of the radius of the Poincar\'{e} disk.} (which we will henceforth denote as $%
\mathcal{D}$) : in such a frame, the special geometry holomorphic
prepotential reads
\begin{equation}
F=-\frac{i}{2}\left[ \left( X^{0}\right) ^{2}-\left( X^{1}\right) ^{2}\right]
.  \label{F2}
\end{equation}%
By projectivizing, one can define $z=X^{1}/X^{0}$, and the K\"{a}hler
potential (see e.g. \cite{Gnecchi-1} and  references therein)
\begin{equation*}
K=-\text{ln}\left( 2\left( 1-\left\vert z\right\vert ^{2}\right) \right) ,
\end{equation*}%
along with the Poincar\'{e} disk constraint%
\begin{equation}
1-\left\vert z\right\vert ^{2}>0\ , \label{eq:Poin}
\end{equation}%
gives rise to the following metric of (\ref{D}), providing a realization of
the Poincar\'{e} disk $\mathcal{D}$ in the Argand-Gauss plane $\mathbb{C}$ :
\begin{equation}
\label{eq:tt}
ds_{\mathcal{D}}^{2}=g_{z\overline{z}}dzd\overline{z}=\frac{1}{\left(
1-\left\vert z\right\vert ^{2}\right) ^{2}}\left\vert dz\right\vert ^{2}.
\end{equation}
It is worth remarking that the two symplectic frames (\ref{F1}) and (\ref{F2}%
) are related through a $Sp(4,\mathbb{R})$ transformation, discussed \textit{%
e.g.} in \cite{FMO-mc} (see also \cite{Sabra, Gnecchi-Halmagy,
Cacciatori-et-al-Duality}).

For what concerns the Riemann scalar curvature, again, the general formula
for 1$_{\mathbb{C}}$-dimensional K\"{a}hler manifolds yields%
\begin{equation}
R=-g^{z\overline{z}}\overline{\partial }_{\overline{z}}\partial _{z}\text{ln}%
\left( \text{det}\left( g\right) \right) =-2,
\end{equation}%
of course regardless of the symplectic frame under consideration.

\subsubsection{\label{N=2-T^3}$\mathcal{N}=2$ Kaluza-Klein ($T^{3}$)}

In the so-called $\mathcal{N}=2$ $T^{3}$ model, no global compact factors
are present in the $U$-duality group, which is simply $G_{T^{3}}=SU(1,1)$;
thus, $n_{V}=1$, $n_{h}=0$. In this model, the manifold (\ref{m}) should be
written as%
\begin{equation}
\mathcal{M}_{T^{3}}=\frac{SU(1,1)}{U(1)}.  \label{M-T^3}
\end{equation}%
This space is an isolated case in the classification of symmetric special K%
\"{a}hler spaces \cite{c-map,dWVVP}. Also, this model is not \textquotedblleft twin\textquotedblright\ to any
known supergravity theory; in particular, it is not twin to any $\mathcal{N}%
=1$ supergravity, because its kinetic vector matrix is not holomorphic \cite%
{0707.0964v3}.

Again, local $\mathcal{N}=2$ supersymmetry constrains the space (\ref{M-T^3}%
) to be a projective special K\"{a}hler manifold. Usually, this model is
considered in the symplectic frame in which the special geometry holomorphic
prepotential reads\footnote{%
This is the so-called \textquotedblleft $4D/5D$ special coordinates'
symplectic frame\textquotedblright\ \cite{0707.0964v3}. Another symplectic
frame often used is the one in which $F(X)=\sqrt{X^{0}\left( X^{1}\right)
^{3}}$ (cf. e.g. (2.38) of \cite{dWVP3}, as well as the $T^{3}$-degeneration
of Sec. 3.3 of \cite{BFMY-FI}).}%
\begin{equation*}
F=\frac{\left( X^{1}\right) ^{3}}{X^{0}}.
\end{equation*}%
By projectivizing and defining $T:=X^{1}/X^{0}$, the K\"{a}hler potential
(see \textit{e.g.} \cite{Saraikin-Vafa, 0707.0964v3})%
\begin{equation}
K=\text{ln}\left( \left( \text{Im}T\right) ^{3}\right) ,
\end{equation}%
along with the metric constraint%
\begin{equation*}
\text{Im}T>0,
\end{equation*}%
gives rise to the metric
\begin{equation}
ds_{T^{3}}^{2}=g_{T\overline{T}}dTd\overline{T}=\frac{4}{3}\left( \text{Im}%
T\right) ^{2}\left\vert dT\right\vert ^{2}.
\end{equation}

For what concerns the Riemann scalar curvature, again, the general formula
for 1$_{\mathbb{C}}$-dimensional K\"{a}hler manifolds yields%
\begin{equation}
R=-g^{T\overline{T}}\overline{\partial }_{\overline{T}}\partial _{T}\text{ln}%
\left( \text{det}\left( g\right) \right) =-\frac{2}{3},
\end{equation}%
again regardless of the symplectic frame under consideration.

\subsection{\label{applappl}An infinite dimensional algebra associated to  the Poincar\'{e} disk $\mathcal{D}$}
Another natural system of coordinates on the Poincar\'e disk $\cal D$ can be introduced by
  parametrising $z = r e^{i \varphi}, r \in [0,1[, \varphi \in [0,2 \pi[$, thus
recasting the metric \eqref{eq:tt}
into the form (see also \cite{Hel})
\beqa
\text{ d}^2 s = \frac{1}{(1-r^2)^2}\Big(\text{d}^2 r + r^2 \text{d}^2 \varphi\Big) \ . \label{eq:PDm}
  \eeqa
We denote $g_{ij}$  the tensor metric, $g^{ij}$ its inverse and $g =\det(g_{ij})$.
Introducing the  measure of integration
  \beqa
\sqrt{g}\text{d} r \text{d}\varphi= \frac{r \text{d} r \text{d} \varphi}{(1-r^2)^2}\nn
  \eeqa
  for $f,g \in L^2({\cal D})$ the scalar product is defined by
  \beqa
  \label{eq:scal-D}
(f,g) = \frac 1 { \pi} \int_{\cal D} \frac{r \text{d} r \text{d} \varphi}{(1-r^2)^2}\bar f(r,\varphi) g(r,\varphi) \ .
  \eeqa

The Plancherel  theorem is known to hold for symmetric spaces, as well \cite{Wal, Ba-Ra}.
 Assume that
$H \in U(1)\subset SL(2,\mathbb R)$ is given by
\beqa
H= e^{i \theta R_0} \ , \nn
\eeqa
{\it i.e.},  that $U(1)$ is associated to the   right action.
 Since  the harmonic analysis in a $G/H$ coset (and the related matrix elements) must be $H-$invariant,
 the only matrix elements  which appear in  the Plancherel theorem are those which are $U(1)-$invariant,  thus corresponding to the matrix
 elements  with $m=0$
 (see Sec. \ref{sec:Mat}). The only matrix elements which satisfy this assumption are associated  to the bosonic principal
 continuous series ${\cal C}^{i\sigma,0}$.  By introducing $\varphi= \varphi_1 -\varphi_2, \theta= \varphi_1+\varphi_2$, the matrix elements  with $m=0$  only depend on $\varphi$. Thus,  $\mathfrak{sl}(2,\mathbb R)$
 generators of the left action reduce to (omitting the $\theta-$part)
\beqa
L_\pm= \frac 12 e^{\pm i\varphi} \Big(\pm \partial_\rho +2  i \coth 2 \rho  \partial_\varphi\Big)
\  , \ \ L_0= -i \partial_\varphi \nn \ .
\eeqa
With the variable $x =\cosh 2 \rho$ the matrix elements  can be rewritten as
(we have unified the expression of the matrix element for $n\ge0$ and $n\le 0$
using \eqref{eq:n-n})
\beqa
  \psi_{n,i\sigma,m= 0}^{\epsilon=0}(x,\varphi)&=&
  \frac {(-1)^{\frac12(n+|n|)}} {2^{|n|}\;|n|!}\
  \sqrt{\frac{\Gamma(|n|+\frac 12 +\frac 12 i\sigma)\Gamma(|n|+\frac 12 -\frac 12 i\sigma)}
                              {\Gamma(\frac 12 +\frac 12 i\sigma)\Gamma(\frac 12 -\frac 12 i\sigma)}}
  e^{in \varphi} (x+1)^{\frac{|n|}2}  (x-1)^{\frac{|n|}2}\times \nn \\
  &&{}_2 F_1(|n|+\frac12 +\frac12 i\sigma,|n|+\frac12 -\frac12 i\sigma;|n|+1;- \frac12(x-1)) \ ,
  n \in \mathbb Z\ , \sigma >0 \ .  \nn
\eeqa
If we define the new variable $y$ by $x = \frac{y+1}{1-y}$, we have a   bijection from $[1,+\infty[$ to $[0,1[$. If we now
set        $y=r^2$ the generators $\mathfrak{sl}(2,\mathbb R)$ reduce to
\beqa
L_\pm=\frac 12 e^{\pm i \varphi} \Big(\pm (r^2-1) \partial_r -i \frac{r^2+1}r \partial_ \varphi\Big) \ , \ \
L_0= -i\partial_\varphi
\eeqa
and the matrix elements to
\beqa
 \Psi_{ni\sigma}(r,\varphi)&=&
  \frac {(-1)^{\frac12(n+|n|)}} {|n|!}\
  \sqrt{\frac{\Gamma(|n|+\frac 12 +\frac 12 i\sigma)\Gamma(|n|+\frac 12 -\frac 12 i\sigma)}
                              {\Gamma(\frac 12 +\frac 12 i\sigma)\Gamma(\frac 12 -\frac 12 i\sigma)}}
  e^{in \varphi} \frac {r^{|n|}}{(1-r^2)^{|n|}}\times \nn \\
  &&{}_2 F_1(|n|+\frac12 +\frac12 i\sigma,|n|+\frac12 -\frac12 i\sigma;|n|+1; -\frac{r^2}{1-r^2}) \nn
\eeqa
Furthermore   the scalar product becomes (see \eqref{eq:sp})
\beqa
(f,g)=\frac 1{ 2\pi}\frac 1 2 \int\limits_1^{+\infty} \text{d} x\int \limits_0^{2 \pi} \text{d} \varphi \bar f(x,\varphi)
g(x,\varphi)= \frac 1 { \pi}
\int_{\cal D} \frac{r \text{d} r \text{d} \varphi}{(1-r^2)^2} \bar f(r,\varphi) g(r,\varphi)
\eeqa
and  the Casimir operator turns out to reduce to the Laplace-Beltrami operator of $\cal D$
\beqa
Q&=& L_ 0^2 -\frac 12 L_+ L_- -\frac12 L_- L_+ =\frac 14\frac {(r^2 -1)^2}{r^2}\Big(r \partial_r r \partial_r  + \partial^2_\varphi\Big)
\nn\\
&=&\frac 14 \frac1 {\sqrt{g}} \partial_i \sqrt{g} g^{ij} \partial_j \ .
\eeqa

Let $\Phi \subset {\cal D} \subset \Phi'$ be a Gel'fand triple, then for $f \in \Phi$
the Plancherel theorem takes the form (\cite{Hel} p. 9; see also \cite{Knapp} and
\cite{Ba-Ra} Theorem 1 p. 441, where the harmonic analysis on $G/H$ is studied in a more general context, with the second reference using the
language of Gel'fand triples)
\beqa
f(r,\varphi) &=& \int \limits_0^{+\infty} \text{d} \sigma \sigma \tanh \pi \sigma \sum \limits_{n\in \mathbb Z} f^n(\sigma)
\psi_{n i \sigma}(r,\varphi) \nn\\
f^n(\sigma)&=&(\psi_{n i \sigma}, f) \ . \nn
\eeqa

The elements of the Losert basis, which are neutral with respect to the charge $U(1)$, correspond to the elements of $W_{0n}$ for which $\epsilon=0$
(see \eqref{eq:ed})
\beqa
\Phi_{0,n,k}(x,\varphi)=2^{k} \sqrt{2k+|n|+1} e^{i n \varphi} (x-1)^{\frac{|n|}2} (x+1)^{-\frac{|n|} 2 -k-1}
  P_k^{(|n|,-|n|-2k-1)}(x) \   , \ \ n\in \mathbb Z\ , k \in \mathbb N \ .\nn
\eeqa
(Note that the normalisation differs by a factor two, since previously the functions were normalised by $\int  \text{d} x$ and
now are normalised with $ 2\int \text{d} x$). As it is the case for the Losert basis of SL$(2,\mathbb R)$, these
functions also belong to the Schwartz space $\Phi$.
With the variables $(x,\varphi)\to (r,\varphi)$ the functions adopt the form
\beqa
\Phi_{n,k}(r,\varphi)=\frac 12\sqrt{2k+|n|+1}  e^{i n \varphi} r^{|n|}(1-r^2)^{k+1} P_k^{(|n|,-|n|-2k-1)}\Big(\frac {1+r^2}{1-r^2}\Big) \ . \nn
\eeqa
Then the set
\beqa
\label{eq:Hil-D}
\{\Phi_{n,k}, n \in \mathbb Z, k \in \mathbb N\}
\eeqa
constitutes a Hilbert basis of $L^2({\cal D})$ and we have for
any square-integrable function $f$:
\beqa
f(r,\varphi) &=& \sum \limits _{k \in \mathbb N}\sum \limits _{n \in \mathbb Z} f^{nk} \Phi_{nk}(r,\varphi)\ , \nn\\
 f^{nk} &=&(\Phi_{nk},f) \ . \nn
 \eeqa

The results of Section 4 simplify and lead to
 \beqa
 \label{eq:intrin} 
 \Psi_{ni \sigma}(r,\varphi) \Psi_{n' i \sigma'}(r,\varphi)=
     \int \limits _0^{ + \infty} \text{d} \sigma'' \; \sigma''\tanh \pi \sigma'' C_{i\sigma, i\sigma', n ,n'}^{i \sigma''} \Psi_{n+n', i \sigma''}(r,\varphi) \ ,
 \eeqa
 and
 \beqa
\Phi_{n,k}(r,\varphi) \Phi_{n',k'}(r,\varphi)=C_{k k'}^{k''}{}_{nn'} \Phi_{n+n',k''}(r,\varphi) \ ,\nn
\eeqa
with respectively the notations of \eqref{eq:??} and \eqref{eq:ee}, with the notations $C_{i\sigma i\sigma'}^{i \sigma''}{}_{nn'} =C_{i\sigma i\sigma'}^{i\sigma''}{}_{nn' 0 0}$ (resp.
$C_{k k'}^{k''}{}_{nn'}=C_{k k'}^{k''}{}_{nn'00}$). For
the first equation above note however that the product  decomposes only onto
continuous bosonic series.

Having introduced the Losert and Plancherel basis (in the terminology of Section 5.1), it is immediate to define
the algebra $\widehat{\mathfrak g}({\cal D})$, {\it i.e.}, the centrally extended algebra
associated to the Poincar\'e disk  $\mathcal{D}$, defined in the complex plane by the constraint \eqref{eq:Poin}. The generators of $\mathfrak g({\cal D})$ are given by
\beqa
\mathfrak g({\cal D})=\Bigg\{T^a_{n i \sigma}=T^a \Psi_{n i\sigma}(r,\varphi)\ , \ \ n\in\mathbb Z\ , \sigma>0\Bigg\}\nn
\eeqa
in the Plancherel basis, and by
\beqa
\mathfrak g({\cal D})=\Bigg\{T^a_{n k}=T^a \Phi_{n,k}(r,\varphi)\ , \ \ n\in\mathbb Z\ , k \in \mathbb N \Bigg\}\nn
\eeqa
in the Losert basis.
We have only one differential operator $L_0$ and one central extension associated to the one-form
\beqa
\gamma=-\frac i\pi k \frac {r}{(1-r^2)^2} \text{d} r \nn
\eeqa
leading to the two-cocycle
\beqa
\omega(X,Y) = - \frac k \pi \; \int \limits_{\cal D} \frac{r \text{d}r\text{d} \varphi}{(1-r^2)^2} \Big<X,L_0 Y\Big>_0 \ ,\nn
\eeqa
with the notations of Section 5.1.
The centrally extended Lie brackets take the form:
\beqa
\big[T^{a}_{n i\sigma}, T^{a'}_{n' i\sigma'}\big] &=&i f^{a a'}{}_{a''} \int \limits_0^{+\infty}
\text{d} \sigma''  \; \sigma''\tanh \pi \sigma''
C_{i\sigma, i\sigma'}^{i\sigma''}{}_{n, n'} T^{a''}_{n+n' i \sigma''}
+ nk \delta(\sigma-\sigma')  \delta_{n+n'}\ , \nn \\
\big[L_0,T^{a}_{n i\sigma}\big]&=& nT^{a}_{n i\sigma} \ ,\nn
\eeqa
in the Plancherel basis, and
\beqa
\big[T^{a}_{n k}, T^{a'}_{n' k'}\big] &=&i f^{a a'}{}_{a''} C_{k k'}^{k''}{}_{n,n'} T^{a'' }_{n+n' k''}
+ nk \delta_{kk'}  \delta_{n+n'}\ , \nn \\
\big[L_0,T^{a}_{n, k}\big]&=& nT^{a}_{n, k}\nn
\eeqa
in the Losert basis.

\bigskip
The coset  $SL(2,\mathbb R)/U(1)$ has been endowed with the metric \eqref{eq:PDm} (or, equivalently, \eqref{eq:scal-D}),
thus providing a model for the Poincar\'e disk \eqref{eq:Poin}.  So, the resulting algebra should be relevant for the $\mathcal{N}=2$
axion-dilaton model treated in Section \ref{N=2-mc}.
However, other metrics, for instance those relevant to other models of supergravity in $D=3+1$, could be defined (see Section \ref{sec:sugra}). Thus, we now  briefly address
the possibility to endow   $SL(2,\mathbb R)/U(1)$ with a metric $h_{ij}$ and a corresponding algebra
$\widehat{\mathfrak g}(\text{SL}(2,\mathbb R)/U(1), h_{ij})$. Since this construction proceeds through either the Plancherel theorem or  the identification of a Hilbert basis of $L^2(\text{SL}(2,\mathbb R)/U(1), h_{ij})$ we shall follow here the second path, namely we will identify a Hilbert basis. Two possible approaches  are mentioned below.

\bigskip
Firstly, consider a manifold $\cal M$ endowed with two different scalar products associated to two integration
measures $\text{d} \alpha$ and $\text{d} \beta$ respectively (\cite{Mc}, p. 100):
\beqa
\begin{array}{ll}
  ({\cal M}, \text{d} \alpha):& (f,g)_\alpha= \int_{\cal M} \text{d}\alpha \bar f(m) g(m)\\[.2cm]
  ({\cal M}, \text{d} \beta):& (f,g)_\beta= \int_{\cal M} \text{d}\beta \bar f(m) g(m)
  \end{array}\nn
\eeqa
with $m \in {\cal M}$. We assume further that there exists a mapping $T_{\beta \alpha}$:
\beqa
T_{\beta \alpha\cal}: L^2({\cal M},\text{d} \beta) \to  L^2({\cal M}, \text{d} \alpha) \ ,\nn
\eeqa
such that
\beqa
\int_{\cal M} \text{d} \alpha = \int_{\cal M} \text{d} \beta T_{\beta \alpha}. \ \nn
\eeqa
For instance, for a $n$-dimensional  Riemannian manifold ${\cal M}$ parametrised by $m_1,\cdots,m_n$ with metric $g_\alpha$ (resp. $g_\beta$) we have $\text{d}\alpha = \sqrt{\big|\det g_\alpha\big|} \text{d}^n m$
 (resp. $\text{d}\beta = \sqrt{\big| \det g_\beta\big|} \text{d}^n m$) and thus  $T_{\beta \alpha} = \sqrt{\big|\det g_\alpha\big|/\big|\det g_\beta\big|}$.
Thus if $\{f^\beta_i, i\in \mathbb N\}$ is a Hilbert basis of $L^2({\cal M},\text{d} \beta)$, then
$\{f^\alpha_i = \frac{f^\beta_i }{\sqrt{T_{\beta \alpha}}}, i\in \mathbb N\}$ is a Hilbert basis for  $L^2({\cal M},\text{d} \alpha)$
  and we obviously have
  \beqa
(f^\beta_i,f^\beta_j)_\beta = \delta_{ij}\ \  \Longleftrightarrow \ \ (f^\alpha_i,f^\alpha_j)_\alpha = \delta_{ij} \  \nn
  \eeqa
  and the map $T_{\beta\alpha}$ is unitary.

  Starting from the Poincar\'e  disc $\cal D$  with scalar product \eqref{eq:scal-D} and Hilbert basis \eqref{eq:Hil-D},
  one can introduce a metric $h_{ij}$ whose associated  scalar product reads
  \beqa
(f,g)_h = \int_{\text{SL}(2,\mathbb R)/U(1)} \sqrt{\det(h)} \text{d} r \text{d} \varphi \bar f(r,\varphi) g(r,\varphi) \ . \nn
    \eeqa
    Then, this is related to \eqref{eq:scal-D} by the mapping
\footnote{
In the \textquotedblleft generalized special geometry\textquotedblright\ of $%
\mathcal{N}>1$-extended supergravity in $D=3+1$ space-time dimensions \cite%
{ADF-Gen,FK-N=8}, the (unitary) mapping $T$ does not have to be necessarily
realized through a symplectic transformation of the (generalized) symplectic
sections. When it is not, the mapping $T$ also yields a change of the
geometric properties of the coset itself, thus implying a change of the
supergravity model (as well as, possibly, of the total amount of local
supersymmetry).
}
$T= \sqrt{ \det(h)}(1-r^2)^2/r $ and the procedure above leads naturally to a Hilbert basis
    of  $L^2(\text{SL}(2,\mathbb R)/U(1))$ with metric $h$, which we denote
as $L^2(\text{SL}(2,\mathbb R)/U(1),h)$ and
we explicitly realise as $\{\Phi^h_{nk}, n \in \mathbb Z, k \in \mathbb N\}$
    . 
      In order to apply  this procedure, we should however warn the reader that
       the product of two arbitrary elements
      $\Phi^h_{nk}\Phi^h_{n'k'}$ should be  square-integrable.
    The corresponding Lie algebra $\widehat{\mathfrak g}({\cal M}, h)$ follows.

\bigskip
Secondly, we recall that   Bargmann \cite{bar}
showed  that for any $\lambda \in \mathbb N \setminus\{0\}$ or $\lambda \in \mathbb N \setminus\{0\} + \frac12$, one can
realise the discrete series bounded from below ${\cal D}_\lambda^+$ as holomorphic functions on  the coset SL$(2,\mathbb R)/U(1)$ with
metric
\beqa
\label{eq:scal-lambda}
(f,g)_{\lambda }=\frac{1}{2\pi i}\int_{|z|<1}\text{d}z\text{d}\bar{z}\frac{\bar{%
f}(\bar{z})g(z)}{(1-|z|^{2})^{2-2\lambda }}\ .
\eeqa
Let ${\bf D}_\lambda$ be the the coset SL$(2,\mathbb R)/U(1)$  with the  $\lambda-$dependent scalar product above (not to be confused with ${\cal D}_\lambda^+$, the representation
bounded from below). Note that since $\lambda>1/2$ we cannot reproduce the metric of the Poincar\'e disk \eqref{eq:scal-D} in this way.
The generators of $\mathfrak{sl}(2,\mathbb R)$ are given by
\begin{equation*}
K_{0}=z\partial _{z}+\lambda \ ,\ \ K_{+}=z^{2}\partial _{z}+2\lambda z\ ,\
\ K_{-}=\partial _{z}\ ,
\end{equation*}%
and ${\cal D}_\lambda=\{f_{n,\lambda}, n \in \mathbb N\}$  defined  by
\begin{equation*}
f_{n,\lambda }(z)=\sqrt{\frac{(2\lambda +n-1)!}{%
n!(2\lambda -2)!}}z^{n}\ ,n\in \mathbb{N}\ .
\end{equation*}%
The action of $K_\pm, K_0$  onto $f_{\lambda,n}$ reproduces  \eqref{eq:D+}
 with $n \to n+\lambda$.
Since ${\cal B}_\lambda=\{f_{n,\lambda}, n \in \mathbb N\}$ constitutes a Hilbert basis of ${\bf D}_\lambda$ with scalar product \eqref{eq:scal-lambda}, it enables us to define the algebra $\widehat{\mathfrak g}({\bf D}_\lambda)$.  However, in this case this algebra
coincides with the usual affine extension of $\g$, namely $\widehat{\mathfrak g}({\bf D}_\lambda) \cong \g^+$,  for any value of $\lambda>1/2$.

\section{Conclusion and perspectives}
In the present paper we extended the results obtained in \cite{rmm},
concerning compact manifolds, to the case of \textit{non-compact} manifolds,
by explicitly considering the case of $SL(2,\mathbb{R})$ and the related
Riemannian symmetric coset $SL(2,\mathbb{R})/U(1)$. This generalization has
proved to be highly non-trivial, as resulting from the novel results which
we have obtained on the harmonic analysis over $SL(2,\mathbb{R})$. Our
approach has been twofold : on the one hand, we have been exploiting  the Plancherel
theorem, while on the other hand we identified a Hilbert basis on the space
of square-integrable functions $L^{2}\left( SL(2,\mathbb{R})\right)$.

The appearance of $SL(2,\mathbb{R})/U(1)$ as a target space of one complex
scalar field in the bosonic sector of some Maxwell-Einstein supergravity
theories in $D=3+1$ space-time dimensions, hints at an application of the
infinite-dimensional Lie algebra $\widehat{\mathfrak{g}}\left( SL(2,\mathbb{R%
})/U(1)\right) $ in the context of supergravity  (in particular, as resulting from the treatment given in Section \ref{applappl}, for the $%
\mathcal{N}=2$ axion-dilaton supergravity). However, this may be
complicated by the fact that the rank of this algebra is
infinite. Nevertheless, it is here worth listing at least a
couple of remarks, which may hint to further future developments.

First of all, one may choose $\mathfrak{g}=\mathfrak{so}(3,1)$, thus
obtaining $\widehat{\mathfrak{so}}(3,1)\left( SL(2,\mathbb{R})/U(1)\right) $%
. In the approach based on the Plancherel theorem, this
infinite-dimensional, centrally-extended Lie algebra would be generated by a
suitable merging the usual Lorentz generators with matrix elements of the
bosonic continuous principal series of $SL(2,\mathbb{R})$, which are
functions of the manifold $SL(2,\mathbb{R})/U(1)$ itself. On the other hand,
by employing the square-integrable basis constructed by V. Losert, the
generators of this algebra would instead be resulting from a merging of the
usual Lorentz generators with the functions $W_{0n}$ given by (\ref{eq:ed}).
The fact that $SL(2,\mathbb{R})/U(1)$ is coordinatized in a
Lorentz-independent way (\textit{i.e.}, by the would-be spin-$0$ fields in
the corresponding $D=3+1$ space-time) is pointing to an inherent consistency
of this application, whose relevance to four-dimensional supergravity we
hope to investigate in future work.

Finally, we should point out that our approach is in many respects
alternative to the `Kac-Moody perspective' on the symmetries of supergravity
theories, leading to affine ($\mathfrak{u}^{+}$), Kac-Moody ($\mathfrak{u}%
^{++}$), and generalized Kac-Moody ($\mathfrak{u}^{+++}$)  extensions of the
(simple) $U$-duality Lie algebra $\mathfrak{u}$ in $D=2+1$ space-time
dimensions (see \textit{e.g.} \cite{West, Nicolai, West-VP-Riccioni}, and  references therein). Therefore, it would be interesting to investigate the
relationship between the two infinite-dimensional Lie algebras $\widehat{%
\mathfrak{u}}\left( SL(2,\mathbb{R})/U(1)\right) $ and $\mathfrak{u}^{+++}$;
nevertheless, we can anticipate that such algebras exhibit fundamentally
different features, because the former is of infinite-rank, but all its roots
and the corresponding generators are explicitly known, whereas the latter
has finite rank $rk(\mathfrak{u})+3$, but its generators are only
iteratively known in terms of the Chevalley-Serre relations.

\appendix

\section{Hypergeometric functions}\label{app:hyper}
 The  second order differential equation (hypergeometric equation \cite{Ince})
 \beqa
 \label{eq:hyper}
z(1-z) \frac{\text{d}^2 F(z)}{\text{d} z^2} +(\gamma-(\alpha+\beta+1)z)\frac{\text{d} F(x)}{\text{d} z} -\alpha \beta F(z)=0 \ ,
 \eeqa
  admits the two independent solutions (see {\it e.g.} \cite{vk}, p. 381):
 \beqa
 F_1(z) &=& {}_2F_1(\alpha,\beta;\gamma;z)\ ,\nn\\
 F_2(z) &=&z^{1-\gamma} {}_2F_1(1+\alpha-\gamma,1+\beta-\gamma;2-\gamma;z) \ , \nn
 \eeqa
 where ${}_2F_1(\alpha,\beta;\gamma;z)$ is the Euler hypergeometric function.  If $\gamma$ is not an integer, the second solution is easily seen to be linearly independent of ${}_2F_1(\alpha,\beta;\gamma;x)$.

If $|z|<1$, the  hypergeometric function can be defined by the convergent series \cite{hyper}
\begin{equation*}
_{2}F_{1}(\alpha ,\beta ;\gamma ;z)=1+\sum\limits_{n>0}\frac{1}{n!}\frac{%
\alpha (\alpha +1)\cdots (\alpha +n-1)\beta (\beta +1)\cdots (\beta +n-1)}{%
\gamma (\gamma +1)\cdots (\gamma +n-1)}z^{n}.
\end{equation*}%
Note that, if $\gamma$ is a negative integer,
the hypergeometric series is indeterminate.

If $\alpha$ or $\beta$ is a negative integer number, the infinite series ends and ${}_2F_1(\alpha,\beta;\gamma;z)$ reduces to a hypergeometric polynomial. Thus we have ($n \in \mathbb N$)
\beqa
\label{eq:poly}
{}_2 F_1(-n,\beta;\gamma;z) = \sum \limits_{k=0}^n (-1)^k \begin{pmatrix} n\\k\end{pmatrix} \frac{(\beta)_k}{(\gamma)_k} z^k \ ,
\eeqa
with
\beqa
(\alpha)_k= \alpha(\alpha+1)\cdots(\alpha+k-1) \ , \nn
\eeqa
the Pochhammer symbol. Using a result of Courant and Hilbert (\cite{cr}, p.  90), we have
\beqa
\label{eq:CH}
&(1+z)^{-m-n} z^{m-n}{}_2F_1(-n+\lambda,-n-\lambda+1;1+m-n;-z)\nn\\[0.3cm]
&=\frac{(m-n)!}{(m-\lambda)!}
\frac{{\rm d}^{n-\lambda}}{{\rm d} z^{n-\lambda}}\Big(z^{m-\lambda} (1+z)^{-m -\lambda}\Big) \ .
\eeqa
From  the identity
\beqa
\label{eq:ab}
\int \limits_0^{+ \infty} \text{d} \rho \cosh^{2a+1} \rho \sinh^{2b+1}\rho= \frac 12 \frac{\Gamma(1+b)\Gamma(-a-b-1)}{\Gamma(-a)}
\ \ \text{if} \ \ a+b<-1\ , b>-1,
\eeqa
 and using \eqref{eq:ab} for the lowest and highest degree of ${}_2 F_1(-n,\beta;\gamma;-\sinh^2 \rho)$,
we have for $m\ge n \ge 0$ that
\beqa
\label{eq:cv}
\int\limits  \text{d} \rho \cosh^{-2(n+m)+1}  \rho\sinh^{2(m-n)+1}\rho {}_2 F_1(-n  +\lambda,\beta;\gamma;-\sinh^2 \rho)^2
\ \ \text{converges} \ \ \textit{iff} \ \ ,  \lambda > \frac 12\
\eeqa
and if $P_N$ is a degree $N$ real polynomial in $-\sinh^2 \rho$, then
\beqa
\label{eq:convP}
\int\limits  \text{d} \rho \cosh^{2a+1} \rho\sinh^{2b+1} \rho P_N(-\sinh^2 \rho)^2
\ \ \text{converges} \ \ \textit{iff} \ \ a+b+2 N<-1 \ \text{and} \ b>-1.
\eeqa

We finally  enumerate some relations:
\beqa
\label{eq:Lhyper}
&\Big(m-n+ 2nz + z(1-z)\frac{\text{d}}{\text{d} z}\Big){}_2F_1(\alpha,\beta;\gamma;z)=(m-n)
{}_2F_1(\alpha-1,\beta-1;\gamma-1;z)\nn\\
&\frac{{\rm d}}{{\rm d} z} {}_2F_1(\alpha,\beta;\gamma;z)=\frac{n(n-1)-q}{m-n+1} {}_2F_1(\alpha+1,\beta+1;\gamma+1;z)\nn\\
&\Big(m+n + (1-z)\frac{\text{d}}{\text{d} z} \Big){}_2F_1(\alpha,\beta;\gamma;z)=
\frac{m(m+1)-q}{m-n+1}{}_2F_1(\alpha,\beta;\gamma+1;z)\\
&\Big(m-n+ z\frac{\text{d}}{\text{d} z}\Big){}_2F_1(\alpha,\beta;\gamma;z)=(m-n){}_2F_1(\alpha,\beta;\gamma-1;z)\nn
\eeqa
with $\alpha= 1/2(-2n+1 + \sqrt{4q+1}), \beta=1/2(-2n+1 - \sqrt{4q+1}), \gamma= m-n+1$.

\section{Jacobi Polynomials}\label{app:J}

Let $\alpha,\beta >-1$ (in fact $\alpha$ and $\beta$ can be complex, and we have $\text{Re}(\alpha),  \text{Re}(\beta)>-1$). A Jacobi polynomial $P_n^{(\alpha,\beta)}$ of degree $n$  satisfies the second-order ordinary differential equation \cite{be, nu}
\beqa
(1-x^2) \frac{\text{d}^2P_n^{(\alpha,\beta)} (x)}{\text{d} x^2} + \big[\beta-\alpha-(\alpha+\beta+2)x\big] \frac{\text{d} P_n^{(\alpha,\beta)}(x)}{\text{d} x}+n(n+\alpha+\beta+1) P_n^{(\alpha,\beta)}(x)=0 \ \nn
\eeqa

The Jacobi polynomials, as a particular case of the general hypergeometric functions, belong to the class of classical orthogonal polynomials and satisfy the condition   (with
$\text{Re}(\alpha),  \text{Re}(\beta)>-1$)
\beqa
\label{eq:N1}
\int \limits_{-1}^1 \text{d} x (1-x)^\alpha(1+x)^\beta P_m^{(\alpha,\beta)}(x)P_n^{(\alpha,\beta)}(x)
=\delta_{mn} \frac{2^{\alpha +\beta+1} \Gamma(n+\alpha+1)\Gamma(n+\beta+1)}{n! (2n+\alpha+\beta+1)\Gamma(n+\alpha+\beta+1)} \ .
\eeqa
 It follows that $\{P^{(\alpha,\beta)}_n, n\in \mathbb R\}$ is a Hilbert basis of $L^2([-1,1])$ with the weight function
$ (1-x)^\alpha (1+x)^\beta$ (see \cite{be} for details).
Note that we have also the relation  (see again \cite{be} p. 285)
\beqa
\label{eq:N2}
\int \limits_{-1}^1 \text{d} x (1-x)^{\alpha-1}(1+x)^\beta P_m^{(\alpha,\beta)}(x)^2
= \frac{2^{\alpha +\beta } \Gamma(n+\alpha+1)\Gamma(n+\beta+1)}{n!  \alpha\Gamma(n+\alpha+\beta+1)} \
\ , \ \text{Re}(\alpha)+1,  \text{Re}(\beta)>-1 \ . \ .
\eeqa

We  have the following relationship between hypergeometric and Jacobi polynomials   (from \cite{vk} p. 289)
   \beqa
   \label{eq:HJ}
P^{(\alpha,\beta)}_n(x)= \frac{\Gamma(n+\alpha+1)}{n! \Gamma(\alpha +1)}\; {}_2 F_1(-n,n+\alpha+\beta+1; \alpha+1; \frac12(1-x)) \ .
\eeqa
Thus, the orthogonality property   of the matrix elements of
the series bounded from below/above (see (\ref{eq:matD+}-\ref{eq:matD-})), {\it i.e.}
$\int \limits _0^{+_ \infty} \cosh \rho \sinh \rho \text{d} \rho
\left| \psi^\eta_{n\lambda m}(\rho)\right|\left| \psi^\eta_{n\lambda' m}(\rho)\right| = \delta_{\lambda \lambda'}
$ with $\eta= \pm 1$
reduces to:
\beqa
\label{eq:N3}
  \int \limits_{1}^{+\infty}  (x-1)^p \;(x+1)^q\; P_r^{(p,q)}(x) \;P_{s}^{(p,q)}(x)\; \text{d} x =
  \delta_{rs}\; 2^{p+q+1} \frac{(p+r)!\;(-p-q-r-1)!}{r!\; (-q-r-1)!\; (-p-q-2r-1)} \ \ ,\nn\\
  \eeqa
where, $p, q\in \mathbb R$,  $r, s \in \mathbb N$  with $p > -1$ , $r, t \ge 0$ and
$p + q + r + s < -1$. (See also \cite{vk} p. 485 --- with a misprint corrected.)

The mapping
\beqa
\label{eq:bij}
x \mapsto y = \frac{3+x}{1-x}
\eeqa
is a bijection from $[-1,1 [$ to $[1,+\infty  [$. Using (\cite{vk} p. 291)
\beqa
\label{eq:this2}
P_r^{(\alpha,\beta)}(x)= \Big(\frac{x-1}2\Big)^r P_r^{(\beta,-\alpha-\beta-2r-1)}\Big(\frac{x+3}{1-x}\Big)
\eeqa
we obtain from \eqref{eq:N2}
 \beqa
 \label{eq:N4}
  \int\limits_1^{+\infty}
  (x-1)^p (x+1)^{q-1} P_r^{(p,q)}(x)^2 \text{d} x=
    2^{p+q} \frac{(p+r)!(-p-q-r-1)!}{r! (-q-r-1)! (-q)}\  ,
    \eeqa
where $p, q\in \mathbb R$,  $r \in \mathbb N$  with $p > -1$ , $r \ge 0$ and
$p + q + 2r  < 0$.

\bigskip \noindent \textbf{Acknowledgements}
  Viktor Losert is gratefully acknowledged for the construction of the Hilbert basis of $\LL$.
We would like to thank
Hubert Rubenthaler, Marcus Slupinski, Robert Stanton, Valdemar Tsanov for enlightening discussions.
The work of AM is supported by a \textquotedblleft Maria Zambrano\textquotedblright\ distinguished researcher fellowship at the
University of Murcia, Spain, financed by the European Union within the NextGenerationEU programme.

\bibliographystyle{utphys}
\bibliography{ref}

\end{document}